\documentclass[english]{iopart}

\usepackage[dvips]{graphicx}

\usepackage{iopams}
\usepackage{amsgen,amsfonts,amsbsy,amssymb}

\begin{document}

\title[Globally nilpotent operators]
{\Large Globally nilpotent differential operators and the square Ising model.}

\author{ A. Bostan$^\P$, S. Boukraa$^\dag$, S. Hassani$^\S$,
J.-M. Maillard$^\ddag$, \\
 J-A. Weil$^\pounds$ and N. Zenine$^\S$}

\address{$^\P$ \ INRIA Rocquencourt, 
Domaine de Voluceau, B.P. 105
78153 Le Chesnay Cedex, France} 
 
\address{$^\S$ \  Centre de Recherche Nucl\'eaire d'Alger, \\
2 Bd. Frantz Fanon, BP 399, 16000 Alger, Algeria}

\address{$^\dag$ LPTHIRM and D\'epartement d'A{\'e}ronautique, \\
Universit\'e de Blida, Algeria}

\address{$^\ddag$\ LPTMC, CNRS, Universit\'e de Paris, Tour 24,
 4\`eme \'etage, case 121, \\
 4 Place Jussieu, 75252 Paris Cedex 05, France} 

\address{$^\pounds$ \  XLIM, Universit\'e de Limoges, 
123 avenue Albert Thomas,
87060 Limoges Cedex, France} 
 
\ead{alin.bostan@inria.fr, boukraa@mail.univ-blida.dz, maillard@lptmc.jussieu.fr, 
jacques-arthur.weil@unilim.fr, njzenine@yahoo.com}

\begin{abstract}

We recall various multiple integrals with one parameter, related to the 
isotropic square Ising model, and 
corresponding, respectively, to the  $n$-particle 
contributions of the magnetic susceptibility, to the  
(lattice) form factors, to the two-point
 correlation functions and to their
 $\, \lambda$-extensions. The univariate analytic functions defined 
by these integrals are holonomic and 
even  {\em G}-functions: they satisfy Fuchsian linear
differential equations with polynomial coefficients
 and have some arithmetic properties. We recall 
the explicit forms, found in previous work, of these Fuchsian
equations, as well as their russian-doll and direct sum
 structures. These differential 
operators are very selected Fuchsian linear
differential operators, and their remarkable 
properties have a deep geometrical origin: 
they are all globally nilpotent,
 or, sometimes, even have zero $\, p$-curvature.
We also display miscellaneous examples of
globally nilpotent operators emerging from enumerative
combinatorics problems for which no integral
representation is yet known. 
Focusing on the factorised parts of all these operators, we find out that
the global nilpotence of the factors 
(resp.  $\, p$-curvature nullity)
corresponds to a set of selected structures of algebraic geometry: 
elliptic curves, modular curves, curves of genus five, six, \ldots , 
and even a remarkable weight-1 modular form emerging
 in the three-particle contribution
$\, \chi^{(3)}$ of the magnetic susceptibility of the square 
Ising model. Noticeably,
this associated  weight-1 modular form
is also seen in the factors of the differential operator for another
$\, n$-fold integral of the Ising class, $\, \Phi_H^{(3)}$, for
the staircase polygons counting, and in Ap\'ery's study of $\, \zeta(3)$.
 {\em G}-functions naturally occur as solutions of
 globally nilpotent operators. 
 In the case where we do not have  {\em G}-functions,
but Hamburger functions (one irregular 
singularity at $\, 0$ or $\, \infty$)
that correspond to the 
confluence of singularities in the scaling limit,
the $\, p$-curvature is also found to verify new structures
associated with simple deformations
of the nilpotent property.

\end{abstract}


\noindent {\bf AMS Classification scheme numbers}: 34M55, 
47E05, 81Qxx, 32G34, 34Lxx, 34Mxx, 14Kxx

\vskip .5cm
 {\bf Key-words}: Globally nilpotent operators, $\, p$-curvature, 
$\, G$-functions,  
 arithmetic Gevrey series,
Form factors of the square Ising model, 
susceptibility of the Ising model,
 Fuchsian linear differential equations, 
moduli space of curves,   two-point 
correlation functions of the lattice Ising model,   
 complete elliptic integrals, scaling limit of the Ising
model, 
 apparent singularities, modular forms, Atkin-Lehmer involutions, 
Fricke involutions,
Dedekind eta functions, 
Weber modular functions, Calabi-Yau manifolds,
 three-choice polygons, enumerative combinatorics.

\section{Introduction}
\label{intro}

Generating large series expansions of physical quantities
that are quite often defined as $\, n$-fold integrals is
the bread and butter of lattice statistical mechanics, 
enumerative combinatorics, and more generally theoretical physics.
The $\, n$-fold integrals considered in theoretical physics
are integrals of some more or less simple 
{\em algebraic integrands}:
they are therefore holonomic~\cite{Griffiths,Kawai}. We
 actually found explicitely~\cite{ze-bo-ha-ma-04,ze-bo-ha-ma-05}
 the (highly non-trivial) Fuchsian linear ODEs satisfied by the first 
 $\, n$-particle contribution $\, \chi^{(n)}$
 of the magnetic susceptibility of
the isotropic square Ising model for $n =3, 4$ (and $n =5$ 
modulo a prime~\cite{Experimental}).
 Mathematicians use to say of such $\, n$-fold integrals
of algebraic integrands that they are ``{\em derived from geometry}'' (DFG),
which means that they can be interpreted as 
periods\footnote[4]{See Picard-Fuchs operators~\cite{pic} and
{\em Gauss-Manin connection}~\cite{man}.} of
some algebraic variety~\cite{Wall} closely related to the
 algebraic integrand\footnote[2]{These $\, n$-fold integrals 
can also be seen as the ``diagonal'' 
of an algebraic expression closely linked to 
the algebraic integrand~\cite{Chris}.}. 
Leaving behind all the cohomology
that can be done on these algebraic varieties and other mixed Hodge
structures~\cite{Bertin}, let us just remind that
 such DFG quantities are remarkably 
selected and structured. 
Considering the roots of the
indicial polynomials of these 
 ODEs (the critical
exponents), one finds out 
that {\em all} the critical exponents of {\em  all} the 
singularities of the corresponding {\em minimal order}
 Fuchsian linear ODEs are necessarily
 {\em rational numbers}~\cite{Griffiths,Andr}. 
Coming back to series expansions, these DFG  $\, n$-fold integrals
necessarily correspond to, not only Gevrey 
series, but {\em convergent series} and, often,
to {\em arithmetic Gevrey 
series}~\cite{Andr,preuve2,Andre5,resconj,Andr,Garou,Naga}.
There is a notion of order of Gevrey 
series: order zero corresponding
to {\em G}-functions~\cite{Andr,Gerotto,Andre2,Andre3,Andre4},
 and order
 one to ``Hamburger''~\cite{Hamburg} functions\footnote[3]{In 
that case the integrand in these
$\, n$-fold integrals is not an algebraic function anymore.}, that is to 
say ODEs with an {\em irregular singularity only}\footnote[5]{The 
irregular singularity is at $\, \infty$ or $\, 0$, but not $\, 0$ and
$\, \infty$. Mathematicians do not like to consider direct sums of 
ODEs corresponding to $\, G$-functions 
and  ``Hamburger'' functions~\cite{Hamburg}: the sum
of Gevrey series of different order is {\em not summable} and requires
the multi-summability introduced 
by \'Ecalle~\cite{Ecalle,Ecalle2}.} {\em at $\, \infty$}. 
These $\, n$-fold integrals, corresponding to  Fuchsian linear ODEs, 
are thus necessarily  {\em G}-functions i.e. solutions 
of linear differential equations with
 arithmetic properties~\cite{Andr,Andre2,Andre3}. In a series of
 papers~\cite{Chud,Chud0,Chud1,Chud2,Chud3,Chud9,Lucia}, the Chudnovskys 
underlined the crucial role of this fundamental class of
 functions\footnote[1]{First introduced by Siegel~\cite{Siegel}.}.
They proved in~\cite{Chud1} that solutions of linear 
differential equations satisfying an {\em arithmetic} growth property,
the $\, G$-property\footnote[8]{The  $\, G$-property is an 
arithmetic growth property, on the coefficients
of a solution-series.}, have special {\em geometric} properties in the 
sense that the corresponding minimal linear differential operators are
{\em globally nilpotent}\footnote[9]{In particular, all solutions 
of such globally nilpotent operators with algebraic 
initial conditions are {\em G}-functions.}. 
From an arithmetic, as well as effective (i.e. computational) 
 view-point, 
globally nilpotent linear differential
 operators~\cite{Schmitt,Turrittin} correspond to 
highly selected structures with a large number of
remarkable properties. In particular, their wronskians
 are $\, N$-th roots
of rational functions. In fact, this property 
holds as soon as the indicial polynomials for each singularity
have {\em integer} coefficients\footnote[3]{Generically
 the indicial polynomials 
of Fuchsian ODEs, with coefficients in $\mathbb{C}[x]$, 
do not have integer coefficients. In our examples
originating from a lattice problem we do have integer
coefficients (see \ref{fuchindi}).}. 
A much more selected property is that
{\em all} the critical exponents
 of {\em  all} the regular singularities of these ODEs 
are necessarily {\em rational numbers}\footnote[8]{In this
 arithmetic framework, this can be seen as a consequence of 
the Kronecker's theorem. When the polynomial coefficients of
 Fuchsian ODEs have integer coefficients the critical exponents
are algebraic numbers. In a globally nilpotent context the critical exponents
necessarily reduce to integers 
modulo every prime. Algebraic integers
reducing to integers  modulo every prime are necessarily
rational numbers (Kronecker's
 theorem~\cite{Kronecker,Kronecker2,Kronecker3}).}. 
Recall, however, that
the  rationality of critical exponents is a consequence
of the DFG structure.

Unfortunately this Grothendieck's geometry view-point 
is not very well-known in physics, and thus, like Monsieur
 Jourdain\footnote[5]{Le bourgeois gentilhomme (Moli\`ere).},
theoretical physicists often study series expansions 
(low or high temperature series expansions, generating 
functions, \ldots ), or divergent series, without knowing that 
they are Gevrey series (and {\em arithmetic Gevrey series})
 and often $\, G$-functions~\cite{Andre2,Andre3}. In particular, they 
take for granted the rational character of the
critical exponents, or the algebraic simplicity of the wronskians
of the ODEs they encounter. Many simple 
remarkable results on form factors,
or non-trivial identities on some well-poised hypergeometric series
and other Bayley pairs~\cite{Berkovic,Ole}, are
 not sufficiently recognized
as a straight consequence of the fact
that {\em G}-functions naturally occur.
Along this hypergeometric line\footnote[1]{Which is not a surprise
 for Yang-Baxter
specialists, see for instance~\cite{Perk}.}, the paradigm of functions
that can be interpreted as periods on an algebraic variety
are the {\em hypergeometric function} $_2F_1$
 (or sometimes complete elliptic
 integrals of the first or second kind $\, K$, $\, E$), 
or, more generally,
hypergeometric functions $_{n+1}F_n$ for some 
selected arguments. Dwork conjectured~\cite{Dwork} that
globally nilpotent linear differential operators of second order,
are necessarily reducible, up to
a rational pull-back and up to the $N$-th root of a rational function,
to the hypergeometric functions $_2F_1$. This
 initial conjecture was ruled-out
by Krammer\footnote[9]{In fact, this operator was first introduced
by the Chudnovskys.} 
who provided a counterexample~\cite{Krammer}
 which comes from the periods
 of a family of abelian surfaces over a 
{\em Shimura curve} (``wrong'' elliptic curves)
$\,  {\mathbb P}^{1}\,  \setminus \{0, 1, 81,  \, \infty \}$. 
Later other examples that are not even associated
 with arithmetic Fuchsian lattices,
or Shimura curves, were also found~\cite{Detweil}. 
Now the conjecture is 
rephrased to embed  hypergeometric functions $_2F_1$ 
and such counterexamples.
In this paper, we will call ``Dworkian'' a 
globally nilpotent linear differential operator 
of second order corresponding to this ``extended'' conjecture. 

We will try to promote a DFG Grothendieck's viewpoint 
 using a {\em learn-by-example approach} which focuses on the
(quite arithmetic) notion of {\em global nilpotence} of
the linear differential operators of various holonomic
quantities we already encountered in physics. Alternatively, one
 could also imagine to perform systematic 
analysis of the differential Galois group of the  corresponding 
linear differential operators in order to obtain a deeper understanding
of these operators. To some extent differential Galois group 
analysis and $\, p$-curvature calculations (see below) are very close. 
In pratice, the analysis of the differential Galois group requires
 much more time and becomes {\em very difficult to perform for 
linear differential operators of order larger than four}, which 
is the vast majority of the operators we encounter in physics. 
In contrast, $\, p$-curvature calculations provide more partial
 information (only a finite amount of primes can be checked), but 
are easy to perform, simple, 
and effective. 

We will consider many $\, n$-fold integrals, or generating functions, 
that originate from the square Ising model or from 
enumerative combinatorics. Some quantities are not naturally 
expressed as $\, n$-fold integrals of an algebraic integrand: the
discovery of the global nilpotence of the corresponding (minimal
 order Fuchsian linear differential) operator has to be seen 
as a strong indication that they are DFG (can be expressed as
$\, n$-fold integrals of an algebraic integrand). Other quantities,
like noticeably the $\, \chi^{(n)}$ contributions 
of the magnetic susceptibility of the
 square Ising model~\cite{ze-bo-ha-ma-04,ze-bo-ha-ma-05}, are defined as
$\, n$-fold integrals of an algebraic integrand\footnote[3]{The 
corresponding linear differential operators
are thus holonomic~\cite{Kawai,Griffiths2}, because the 
integral of a holonomic $\, D$-module is  necessarily holonomic.}. 
The integrand 
of the algebraic function\footnote[9]{In the integration-variables
 $\, e^{2 i \pi \phi_n}$ and
in the ``parameter'' $\, w$, see (\ref{Gn}) below.}
 can be chosen continuous and single-valued
on the torus of integration: {\em they are indeed}~\cite{KatzIHES} a
 {\em family of periods} (in other words DFG).

In this last case, our purpose is not to
give another proof of this global nilpotence\footnote[1]{The
 $\, n$-fold integrals  (over a closed $\, n$-cycle) 
of rational expressions are 
necessarily DFG. More generally, $\, n$-fold integrals 
(over a closed $\, n$-cycle)
of rational expressions on some algebraic
 variety~\cite{Griffiths,FritsBeukers,Saito} 
are necessarily DFG.}, but 
to see {\em how} these linear differential operators manage to be 
globally nilpotent. The corresponding
minimal order differential operators can be of a quite high 
order (for example order 33 for $\, \chi^{(5)}$, see~\cite{Experimental}),
 but are always
 factorised\footnote[2]{As direct sum factorisations 
or straight factorisations.}
into linear differential operators
 of smaller orders. Necessarily, all these factors  
{\em have to be} globally nilpotent\footnote[4]{The
product of globally nilpotent operators is necessarily globally 
nilpotent. More precisely, the characteristic 
polynomial of the $\, p$-curvature 
of the product operator is the product of the characteristic
 polynomial of the $\, p$-curvatures
of the factors (see Theorem 5.
in~\cite{Schmitt} or Corol. 2.1.3. in~\cite{Dwork}). }. The global
nilpotence of order-one linear differential operators is easy to see:
their wronskians are $N$-th roots of  rational functions.  The global
nilpotence of order-two linear differential operators is much more
interesting: are they ``Dworkian'' operators 
(see before), and, more specifically, do they correspond to $_2F_1$
functions, or do they correspond to
counterexamples similar to Krammer's, 
 where Heun\footnote[5]{They are
straight generalisations of hypergeometric ODEs, four singular
points $[0, 1, \alpha,  \infty]$ replacing the three
points $[0, 1,  \infty]$ of  $_2F_1$ (see the Heun functions
 generalizing $_2F_1$).} functions occur, or to the more general
counterexamples of Dettweiler and Reiter~\cite{Detweil}? 
We will also display many
 globally nilpotent linear differential operators 
of order three, four, etc.  Are they reducible to 
the global nilpotence of the previous order-two operators, because
they are equivalent\footnote[8]{In the sense
of the equivalence of linear differential
 operators~\cite{PutSinger}. We refer to this (classical) notion 
of  equivalence of linear differential operators everywhere 
in this paper.},
to symmetric squares, symmetric cubes, \ldots  of  globally
nilpotent linear differential operators of order two? 
Do they correspond to selected  $_{n+1}F_n$?
We will see that the answers to these questions are quite
 non-trivial, and shed an interesting light on
the very nature of the globally nilpotent operators emerging from
physics. 

From an ``experimental mathematics'' viewpoint, checking the 
global nilpotence of linear differential operators
 amounts to studying these operators {\em modulo as
 many primes as possible}~\cite{vdP1,vdP2}, more precisely by calculating
the $\, p$-curvature of these differential linear operators 
mod prime for different primes (see below).
Along this line it is worth recalling
that, in a previous paper~\cite{Experimental}, we 
performed massive calculations 
on series expansions of many $\, n$-fold integrals. 
These massive calculations 
were performed modulo various prime numbers and enabled
to get many highly non-trivial exact results on these
physical quantities, thus showing that modulo prime calculations
are not artificial or academic:
they are actually a very powerful, and efficient, tool to get
highly non-trivial exact results and they are possibly the only way to
get some ``extreme'' results in physics.
The $\, p$-curvature calculations performed here 
are a natural extension of the series and ODE mod
prime calculations performed
 in~\cite{Experimental}. Let us recall that the 
linear differential operators that annihilate our  $\, n$-fold integrals,
 factorise in operators of much smaller order. 
In this paper, we will systematically calculate the   
$\, p$-curvature of these linear differential operators (for moderate $\, p$)
but also of each of the differential factors in their factorisation 
(in direct sums and in products of differential operators).
Except in the examples for which we do not have
an $n$-fold integral representation of the holonomic function, these two set
of $p$-curvature calculations are not performed 
to check a global nilpotence that we know to be a simple consequence 
of the integral (of an algebraic integrand) representation,
but to get a {\em global understanding} of these ODEs {\em beyond the 
usual local analysis} (singularities, exponents, formal series expansions, 
see the notion of 
``accessory parameters''~\cite{Messing,Splendid,Accessory2,Accessory} below)
 and get more precise details on these 
operators\footnote[9]{Like the characteristic and minimal polynomial
 of the $p$-curvature, the Jordan-block reduction of the $p$-curvature, 
 hopefully in order to get some
hint on the factorisation or direct-sum decompositions of these
operators.}. 

\vskip .1cm 
 The paper is organised as follows: we first recall
 a few $\, n$-fold integrals
and some basic facts on global nilpotence. 
The factorization of the linear differential operators annihilating these
holonomic $\, n$-fold integrals will provide a bunch
 of non-trivial examples of 
globally nilpotent operators of growing 
orders. We successively consider the 
global nilpotence of such operators
 of order two, three and four. This study 
will provide a deep understanding
 of these global nilpotence from the discovery
 of the {\em underlying structures of selected algebraic 
varieties} (hypergeometric functions
with a Hauptmodul pull-back, various modular structures, etc.). 
We will finally show that there is clearly a 
generalization of global nilpotence  
to be discovered in the scaling limit of our lattice models. 
We will conclude with a systematic program of analysis 
of  $\, n$-fold integrals
in theoretical physics.

 In our learn-by-examples approach the variable 
in the ODEs\footnote[5]{To avoid multiplying
the notations, we will sometimes use
the same notations for different operators
when there is no possibillity of confusion.} is generally 
called $\, x$, except when we need to 
recall previous results on the Ising model, where the variables were 
called $\, w$ or $\, t$, or when we need to introduce
 some change of variables. 

\vskip .2cm 

\section{Recalls on Fuchsianity and global nilpotence.}
\label{recallsonFuchsglob}

\subsection{Recalls on Fuchsianity for lattice problems.}
\label{recallsonFuchs}

 When a (minimal order) linear differential operator
 with polynomial coefficients
is discovered for series expansions 
in lattice statistical mechanics, or enumerative combinatorics
on a lattice, one always finds out that
 it is a Fuchsian differential operator.
This {\em lattice} property is not true in the scaling limit
 (see section (\ref{beyondglob}) below). The 
regularity of two singular points 
of the ODE, $\, 0$ and $\, \infty$,
is a simple consequence of the well-known existence of various kinds 
of series expansions (low-temperature,
high-temperature, high-field, large $\, q$ expansions, \ldots ).
The fact that the other singularities are regular 
may seem more mysterious at first, except if one 
remarks that these series with rational number
 coefficients, or even integer coefficients,
have a finite radius of convergence, and, in fact, are $\, G$-functions 
(see section (\ref{notDFG}) below).

It is important to remark that the Fuchsian ODEs
one  encounters in lattice statistical mechanics, 
or enumerative combinatorics
are {\em not} the most generic Fuchsian ODEs, but very
{\em selected} ones.
One inherits from their
{\em  lattice origin} the fact that their polynomial coefficients 
have {\em integer coefficients}, that all the
{\em  indicial polynomials of all the singular points have 
integer coefficients}, and, thus, that all the
 critical exponents are {\em algebraic numbers, 
but not necessarily rational ones} (see \ref{fuchindi}). 

Let us consider such an  order $\, q$ Fuchsian linear operator.
Denote  $x_k$  the $n$ regular singularities, including the apparent ones
and excluding the point at infinity, and $\rho_k^{(j)}$ the local
exponents corresponding to the singularity $x_k$. It is straithforward 
(using Fuchs' relations) 
to write the rational coefficient in front of
the $(q-1)$-derivative  in terms of the local exponents 
of the various singularities:
\begin{eqnarray}
 D_x^q \, \,  \,  
+ \, \sum_{k=1}^n {\frac{q \cdot (q-1)/2\, -\sum_{j=1}^q\rho^{(j)}_k }{x-x_k}}
\cdot  D_x^{q-1}
 \, \, + \, \, \cdots , \qquad  D_x \, = \, \, {{d} \over {dx}}. 
\nonumber 
\end{eqnarray}
The local exponents at the regular singular point $x_k$ are roots
of an indicial equation which is a polynomial
in $\rho$ with integer coefficients, $\, a_0^{(k)} + a_1^{(k)}\, \rho 
+ \cdots \,+ \, a_{q-1}^{(k)}\, \rho^{q-1}\,
 + a_{q}^{(k)}\, \rho^{q}\,=\,0$, 
and, thus, the sum 
$\sum_{j=1}^q\rho^{(j)}_k\, = \, \,  -a^{(k)}_{q-1}/a^{(k)}_q$, 
associated to the singular point $\, x_k$,
 is necessarily a rational number.
One sees, as a consequence, that there exists 
an integer $\, N$ such that the 
$\, N$-th power of the wronskian
\begin{eqnarray}
W(x) \, = \, \, \, 
  \prod_{k=1}^n \, (x-x_k)^{\sum_{j=1}^q \, \rho^{(j)}_k-q(q-1)/2},
\end{eqnarray}
 is a rational function.

All the examples of Fuchsian ODEs displayed in this paper
 (see in particular (\ref{wronsk}) below)
have wronskians that are $\, N$-th roots of 
rational functions, and, as we just saw it, this is  straightforwardly
inherited from the underlying lattice. Lattice statistical mechanics and
enumerative combinatorics naturally 
provide Fuchsian ODEs with $\, N$-th roots of 
rational function wronskians, and algebraic numbers critical exponents.

We will show, in the following, that {\em an even more selected 
set of Fuchsian linear differential operators naturally occurs
 in theoretical physics, the globally nilpotent
 operators}: the previous, at first sight, 
 algebraic numbers critical exponents
 have {\em necessarily} to be rational 
critical exponents\footnote[5]{Stricto sensu
 the rationality of all the critical exponents
of a Fuchsian ODE is {\em not sufficient} to have
 the global nilpotence property, see~\cite{Honda}.
Global nilpotence is stronger than the rationality 
of all the critical exponents for
our ``lattice'' Fuchsian ODEs. One has conditions on the so-called
{\em  ``accessory 
parameters''}~\cite{Messing,Splendid,Accessory2,Accessory}.}.

\subsection{Recalls on global nilpotence}
\label{recallsonglob}

A linear differential homogeneous equation of order $\, q$, with
polynomial coefficients in $\, \mathbb{Q}[x]$,
can always be written as a first order system of homogeneous 
linear differential equations :
\begin{eqnarray}
\label{firstorder}
Y' \, = \, \, A \cdot Y,  \quad \, \,  
Y \, = \, \,   \left[ \begin {array}{c} y\\
\noalign{\medskip} y' \\
\noalign{\medskip} \vdots  \\
 \noalign{\medskip} y^{(q-1)}
\end {array} \right] \quad \, \,  \hbox{or:} \, \, 
  \quad   \, \,  
\Bigl( {{d} \over {dx}} \, - \, A \, \Bigr) \cdot Y
 \, = \, \, 0, \quad 
\end{eqnarray}
where the entries of the matrix $\, A$ are rational functions of $\, x$.
Instead of studying the connection
 $\, d/dx \, - \, A$, one can, alternatively, 
consider for almost any\footnote[2]{Almost all,
 here, and in the following, means for all primes except 
a {\em finite set of} primes.} prime $\, p$, 
 its $\, p$-iterate modulo $\, p$:
\begin{eqnarray}
\label{psip}
\psi_p \, = \, \, \Bigl( {{d} \over {dx}} \, - \, A \, \Bigr)^p, 
\qquad  \quad
 mod \quad  p.
\end{eqnarray}
This $\, \psi_p$ is called the $\, p$-curvature, and {\em it turns out 
that this $\, p$-curvature for any prime number $\, p$, is
  a $\mathbb{F}_p(x)$-linear operator, so that it can be 
represented by a matrix whose entries
are rational functions of $\, x$, rather than a linear differential
operator}. The differential system $Y' \, = \, \, A \cdot Y$
 yields for the $i$-th derivative of $Y$:
\begin{eqnarray}
\label{Lie}
Y^{(i)} \, = \, \, A_i \cdot Y, \quad  \hbox{with:} \quad 
A_{i+1} \, = \, \, \, \,   {{d\, A_i} \over {dx}} \, + \, \, A_i \cdot A, 
\quad  A_1\, = \, A.
\end{eqnarray}
Katz shows~\cite{Katz1982} that computing the $p$-curvature
 amounts to calculating $\, A_p$ 
modulo $\, p$ from the Lie-sequence (\ref{Lie}). 
This can be done by performing $p$  products of $\, q \times q$ matrices
whose entries are rational
 functions in  $\,\mathbb{F}_p(x)$ (i.e. rational
 functions with coefficients  
 in $\,\mathbb{F}_p$ where 
 $\,\mathbb{F}_p \,=\, \, \mathbb{Z}/p\mathbb{Z}$).
These $\, p$-curvatures were introduced in the framework of the 
Grothendieck conjecture, to provide ``algebraic criteria'' 
for the monodromy group of (\ref{firstorder}) 
to be finite\footnote[3]{The Grothendieck-Katz
 $\, p$-curvature conjecture is a problem 
on linear ordinary differential equations, related 
to differential Galois theory.
It is a conjecture of A. Grothendieck 
from the late 1960s, and apparently 
not published by him in any form; it has been publicised, reformulated
 and in some cases related to deformation theory 
proved by N. Katz in a series of papers~\cite{Katz1,Katz11,Katz2,Katz3}. 
The question is to give an arithmetic criterion for when there is
 a full set of algebraic function solutions.}.

In the case of order-one linear differential operators, the 
Grothendieck's conjecture was proved by Honda~\cite{Honda}. The  
fact that the exponents of the various regular singularities are
 {\em rational numbers} can be seen as a consequence of the 
 {\em Kronecker's theorem which says that any algebraic
 number which reduces to integers modulo 
almost every prime is necessarly a
 rational number}~\cite{Kronecker,Kronecker2,Kronecker3}.
The conjecture can also be proved in some particular 
cases: by Dwork for ordinary
hypergeometric equations~\cite{Dwork}, and by Katz for
 Gauss-Manin differential equations 
(see~\cite{PutSinger,vdP1} for more details on the
 second order linear differential equations
when one has only three regular singular points,
like $\, 0, \, 1, \, \infty$
for hypergeometric functions). The conjecture is still 
open for general second order operators.

Rather than the Grothendieck-Katz
 $\, p$-curvature conjecture, one can consider various 
theorems by Katz~\cite{Katz1982},
in particular Proposition 9.3 in~\cite{Katz1982}, which 
shows that the reductions modulo $\, p$
of the Lie algebra of the differential Galois group 
contain the $\, p$-curvatures $\, \psi_p$.

Beyond the situation of the Grothendieck's
 conjecture where these $\, p$-curvatures
vanish, another highly selected situation corresponds to 
the case where these $\, p$-curvatures 
{\em are nilpotent modulo $\, p$, for almost all primes $\, p$}
(for all primes except a {\em finite set of} primes). In 
that case, the linear
 differential operator is called {\em globally nilpotent}.
A globally nilpotent linear differential operator is necessarily 
a {\em Fuchsian} linear differential operator, but it has many more
strong remarkable structures. For instance, 
all the exponents of  all its various regular singularities are
 {\em rational}, but the reciprocal statement is not true:
a  Fuchsian linear differential operator with
 rational exponents is {\em not necessarily 
globally nilpotent}. Global nilpotence is a
 stronger structure than 
having regular singularities with rational exponents~\cite{Honda}.
It  is a very strong arithmetic
 property with a large number of 
remarkable consequences:  for instance modulo any prime $\, p$
the Fuchsian linear differential operator
factorizes, and {\em for almost all primes, it factorizes
 into linear differential operators of  order one},
 each  operator of order one 
having {\em rational solutions} modulo  $\, p$. 
Such a property is quite well illustrated\footnote[5]{Note
 a misprint in~\cite{Singul} one should read
  $\ln A_i$, instead of  $\, A_i$,
in the equations defining the $\, A_i$ after equation
 (H.2) in~\cite{Singul,Kiev}.} 
in Appendix H of ~\cite{Singul,Kiev}  on 
 $\, n$-fold integrals
 related to Ap\'ery's analysis of $\, \zeta (3)$. In that case, 
we even have a factorisation
 into order-one linear differential operators 
on the rationals $\,\mathbb{Q}$ and not only modulo (almost all) primes.
The fact that the solutions of these
 order-one linear differential operators
are actually rational solutions modulo primes, is clearly
 reminiscent of the occurrence
of wronskians that are $\, N$-th roots of rational expressions.

Global nilpotence is often said to suggest a ``deep 
geometrical interpretation'', namely that the solutions
of a globally nilpotent linear differential operator
can be interpreted {\em as periods} of some (hidden \ldots) 
algebraic variety, suggesting more or less a {\em Gauss-Manin 
connection}~\cite{man} interpretation for 
these linear differential operators.

Beyond the linear differential operators associated
with Ap\'ery's analysis of $\, \zeta (3)$, almost all
 examples of globally nilpotent 
linear differential operators correspond to hypergeometric 
functions, and other 
Katz's rigid local systems~\cite{Katz0}, for which
 an interpretation of the solutions as periods of an
algebraic variety plays a central role. Within the known examples,
the overlap between hypergeometric functions (and their 
simple generalisations) and global nilpotence was so large 
that Dwork proposed a conjecture~\cite{Dwork} 
that all the globally nilpotent linear 
differential operators correspond to 
{\em hypergeometric functions up 
to simple transformations}. This conjecture
 was ruled out by Krammer~\cite{Krammer}. Therefore, at the present 
moment, beyond the fact that 
it is a highly remarkable arithmetico-geometric selected property, 
 one can say that one does not have
 a complete understanding of global nilpotence.

In the following we are going to find globally
 nilpotent linear differential
operators corresponding to various 
$\, n$-fold integrals that occur naturally
in the case of the off-critical lattice Ising model, or corresponding to
enumerative combinatorics for which no $\, n$-fold integral
representation is yet known.
We will also explore situations that are
 precious to understand namely other 
 $\, n$-fold integrals that naturally occur
 in particle physics (Feynman diagrams~\cite{Kreimer2000}),
like, for instance, some selected scaling limits.

\subsection{Krammer's counterexample.}
\label{Krammer}

Let us recall briefly Krammer's counter-example\footnote[9]{The
 uniformizing linear differential equation
of an arithmetic Fuchsian lattice~\cite{Krammer}.} to Dwork's 
conjecture~\cite{Krammer,Detweil} which comes from the periods
 of a family of {\em abelian surfaces} over a {\em Shimura curve} 
 $\, \mathbb{P}^{1}\,  \setminus \, \{0, 1, 81,  \, \infty\}$:
\begin{eqnarray}
\label{Kram}
&&Y' \, = \, \, \Bigl({{A_0} \over {x} } \, + {{A_1} \over {x-1} } \, 
+ {{A_{81}} \over {x-81} } \Bigr)\cdot Y,  \,\,\,\qquad \,\, \, 
A_0 \, = \, \,
 \left[ \begin {array}{cc} 0&0\\
\noalign{\medskip}-1/2&-1/2\end {array} \right],\nonumber \\
&&\quad A_1 \, = \, \,
\, 
\left[ \begin {array}{cc} 0&0\\
\noalign{\medskip}4/9&-1/2\end {array}  \right],
\quad \quad \quad A_{81} \, = \, \,
\left[ \begin {array}{cc} 0&1\\
\noalign{\medskip}0&1/2\end {array}
 \right], 
\end{eqnarray}
yielding the second order operator
 (on the first component of the vector $\, Y$):
\begin{eqnarray}
\label{O1}
&& O_1 \, \, \, = \, \,  \, \, D_{x}^2 \, 
  +  {{1} \over {2}} \cdot \Bigl({{1} \over { x}} \, 
+{{1} \over {x-1}} \, \, 
+{{1} \over {x-81}}\Bigr)
 \cdot D_x   \\
&& \qquad \qquad \qquad \, +\,{\frac {x-9}{ 18 \, (x-81) 
\, (x-1) \cdot  x}}, 
\qquad D_x \, = \, \, {{d} \over {dx}}, \nonumber 
\end{eqnarray}
or the second order operator (on the second component of the vector $\, Y$):
\begin{eqnarray}
\label{O2}
&&O_2\, = \, \,
 18\,x \cdot  (x-1) \, (x-9) \, (x-81)^2 \cdot  D_x^2\, \,  \\
&&\qquad +27\,(x-81) \cdot (x^3-123\,x^2+1491\,x-729) 
\cdot  D_x \nonumber \\
&&\qquad 
+(x^3+549\,x^2 \,+13203\,x \, -1003833). \nonumber 
\end{eqnarray}
The two linear differential operators $\, O_1$ and $\, O_2$
are, of course, equivalent 
and the squares of their wronskians are simple rational functions. 
We have calculated their $\, p$-curvatures
and confirmed that they are globally nilpotent. 

The general solution of  $\, O_1$ 
reads in terms of Heun functions~\cite{Kuiken}:
\begin{eqnarray}
&& \mu \cdot Heun(81, 1/2; 1/6, 1/3, 1/2, 1/2; \,  x) \, \nonumber \\
&&\qquad \qquad +\, \lambda  \cdot 
x^{1/2} \cdot Heun(81, 21; 2/3, 5/6, 3/2, 1/2; \, x), \nonumber 
\end{eqnarray}
where $\, \mu$ and  $\, \lambda$ are two constants. 
The differential  Galois group of (\ref{Kram}) (or (\ref{O1}), (\ref{O2}))
is a central 
extension of $\,  SL(2, \, C)$.
Calculating the indicial polynomials of $\, O_1$
 at the various singularities, one finds the indicial polynomial
$\, (6\,r-1)  \cdot (3\,r-1)$
for  $\, t \, = \, \, \infty$, and 
$\, r \cdot (2\,r-1)$
for the singularities $\,  t \, = \, \, 0,\,  1, \, 81$.  
These Heun functions {\em cannot be reduced}~\cite{Krammer} 
 to hypergeometric functions  $\, _2F_1$ (up to 
multiplication and some pull-back). Along this line,
it is important to note that, generically, a Heun function 
{\em does not} correspond to a globally nilpotent second order
 differential operator. For instance
we calculated the $\, p$-curvature of a second order operator
very similar to (\ref{O1}):
\begin{eqnarray}
\label{nonnilp}
D_x^{2}\, +\, {{1} \over {2 }} \, \left( {{1} \over {x}}
+  {{1} \over {x-1}}
+  {{1} \over {x-81}} \right) \cdot  D_x\, 
+ {{1} \over {2 }} \,{\frac
{81 -28\,x}{(x -81)\, (x-1)\, x }},  \qquad 
\end{eqnarray}
which has as solutions the Heun functions
\begin{eqnarray}
&&\, \mu \cdot Heun(81,\,  -81/2, \, -7/2,\, 4, \, 1/2, \, 1/2; \,  x)
 \nonumber \\
&&\,\qquad \qquad  + \, \lambda \cdot x^{1/2} \cdot  
 Heun(81,\,  -20, \, 9/2,\, -3, \, 3/2, \, 1/2; \,  x), \nonumber 
\end{eqnarray}
and we found that (\ref{nonnilp}) is {\em not} globally nilpotent.

\vskip .1cm

\section{Global nilpotence of a few $n$-fold integrals of the Ising class.}
\label{isingclass}

\subsection{Recalls of a few $n$-fold integrals of the Ising class.}
The susceptibility of the Ising model can be 
written~\cite{wu-mc-tr-ba-76} as an infinite sum
of $n$-fold integrals.
These $n$-particle contributions $\, \chi^{(n)}$
 are given by $(n-1)$-dimensional 
integrals~\cite{nickel-99,nickel-00,yamada-84} that read
\begin{eqnarray}
\label{chi3tild}
\tilde{\chi}^{(n)}(w)\,\,=\,\,\,\, {\frac{1}{n!}}  \cdot 
\Bigl( \prod_{j=1}^{n-1}\int_0^{2\pi} {\frac{d\phi_j}{2\pi}} \Bigr)  
\Bigl( \prod_{j=1}^{n} y_j \Bigr)  \cdot   R^{(n)} \cdot
\,\, \Bigl( G^{(n)} \Bigr)^2, 
\end{eqnarray}
where\footnote[1]{The fermionic term $\,G^{(n)}$ has several 
representations~\cite{nickel-00}.} 
\begin{eqnarray}
\label{Gn}
G^{(n)}\,=\,\, \prod_{1\; \le\; i\;<\;j\;\le \;n} \, h_{ij}, \,  \quad 
h_{ij}\,=\,\,
{\frac{2\sin{((\phi_i-\phi_j)/2) \cdot \sqrt{x_i \, x_j}}}{1-x_ix_j}}, 
\end{eqnarray}
and
\begin{eqnarray}
\label{Rn}
R^{(n)} \, = \,\,\, \,  {\frac{1\,+\prod_{i=1}^{n}\, 
x_i}{1\,-\prod_{i=1}^{n}\, x_i}}, 
\end{eqnarray}
with
\begin{eqnarray}
\label{thex}
&&x_{i}\, =\,\,\,  \,  \frac{2w}{1-2w\cos (\phi _{i})\, 
+\sqrt{\left( 1-2w\cos (\phi_{i})\right)^{2}-4w^{2}}},  
  \\
\label{they}
&&y_{i} \, = \, \, \,
\frac{2w}{\sqrt{\left(1\, -2 w\cos (\phi_{i})\right)^{2}\, -4w^{2}}}, 
\quad
 \quad  \quad  \quad \sum_{j=1}^n \phi_j=\, 0  
\end{eqnarray}
valid for small $w$ and, elsewhere, by analytical continuation.
We actually 
found~\cite{ze-bo-ha-ma-04,ze-bo-ha-ma-05,ze-bo-ha-ma-05b,ze-bo-ha-ma-05c} 
the linear ODEs for some of these
holonomic  $n$-particle contributions namely 
$\, \chi^{(3)}$, $\, \chi^{(4)}$ and,  {\em modulo a prime},
 for $\, \chi^{(5)}$.
From a arithmetic Gevrey series and {\em G}-function viewpoint it is 
worth noticing that the series
 expansion of the $\, \tilde{\chi}^{(n)}$, 
in the variable $\, w$, are series expansions 
with {\em integer} coefficients:
\begin{eqnarray}
\label{closedn}
&&\tilde{\chi}^{(n)}(w) \, = \, \, 2^n \cdot w^{n^2} \cdot 
\Bigl(1 \, + \, 4 \, n^2 \cdot w^2 \,  
+ \, 2 \cdot (4\, n^4 \, +13\, n^2 \, +1)\cdot w^4 \, \, \nonumber \\
&& \qquad \qquad + \,
{{8} \over {3}} \cdot (n^2+4)\, (4\, n^4\, +23\, n^2+3) \cdot w^6
\, + \,  \, \, \cdots  \,  \, \, \Bigr),
\end{eqnarray}
where the $\, w^2$ coefficient is valid for $\, n\, \ge \, 3$,
 the $\, w^4$ coefficient is valid 
for $\, n\, \ge \, 5$, and the $\, w^6$ coefficient
 is valid for $\, n\, \ge \, 7$.
Note that the  $\, w^6$ coefficient is {\em always 
an integer}\footnote[2]{It would 
be interesting to get much longer series 
expansion like (\ref{closedn}), valid for arbitrary $\, n$,
to see if these successive
 rational functions of $\, n$ are actually 
functions of $\, n^2$.}. 

In previous publications~\cite{Singul,Landau}, 
we also introduced
 some integrals of the so-called 
``Ising class''\footnote[8]{The terminology 
{\em integral of the Ising class} has 
been proposed by Bailey, Borwein and Crandall in~\cite{crandall}.}.
We considered several kinds of integral representations (one-dimensional 
and multidimensional) of these holonomic functions 
which belong to the Ising class~\cite{crandall}.
Again we obtained the linear ODEs of these sets of integrals
for the first values of $\, n$, through series
 expansions~\cite{Landau,Singul}. 
In~\cite{Singul} a detailed analysis of 
the multiple integrals
$\Phi_H^{(n)}$ was performed. These $\, n$-fold integrals amount
to removing the Fermionic factor
 $\, G^{(n)}$ in (\ref{chi3tild}), so that
one introduces:
\begin{eqnarray}
\label{In}
\Phi_H^{(n)}(w) \, \,= \,\,\, \,\, {\frac{1}{n!}}  \cdot 
\Bigl( \prod_{j=1}^{n-1}\int_0^{2\pi} {\frac{d\phi_j}{2\pi}} \Bigr)  
\Bigl( \prod_{j=1}^{n} y_j \Bigr)  \cdot  
 {\frac{1\,+\prod_{i=1}^{n}\, x_i}{1\,-\prod_{i=1}^{n}\, x_i}}.
\end{eqnarray}

Even simpler integrals (over a {\em single} variable), 
were also introduced
and denoted~\cite{Landau} $\, \Phi_{D}^{(n)}$:
\begin{eqnarray}
\label{chinaked}
\Phi_{D}^{(n)}(w) \,=\,\,\,\,  
 -{{1} \over {n!}}\,\,  \,+ {{2} \over {n!}} \, \int_0^{2\pi} 
{\frac{d\phi}{2\pi}}
 \,   \,   {\frac{1}{1\, -x^{n-1}(\phi)  \cdot  x ((n-1)\phi)}}, 
\end{eqnarray}
where $\, x(\phi)$ is given by (\ref{thex}). 

\subsection{Results on global nilpotence of these
 $n$-fold integrals of the Ising class.}
\label{globising}

Let us display here our results for
 the calculations of $\, p$-curvatures
 for the {\em minimal order} ODEs of 
$n$-fold integrals (\ref{chi3tild}),
 (\ref{In}), (\ref{chinaked})
of the Ising class.

We have calculated  (modulo the first
thousand primes) the $\, p$-curvature of 
 the order-six linear differential 
operator $\, L_6$ occurring in $\, \chi^{(3)}$ 
(see (\ref{l6}) in section 
(\ref{revisitglob}) and~\cite{ze-bo-ha-ma-05c}),
as well as the  Jordan block reduction 
of the $\, 6 \times 6$ $\, p$-curvature
matrix, and found that the characteristic
 polynomial of the $\, p$-curvature
reads $\, T^6$. This  $\, 6 \times 6$ Jordan block  reduction
can be compared with the two $\, 3 \times 3$ Jordan block  reductions 
corresponding to (the $p$-curvature of)
an order-three operator $\, Z_2 \cdot N_1$ that 
right-divides $\, L_6$, and another order-three operator
$\, Y_3$ that left-divides $\, L_6$
 (see section (\ref{revisitglob}) below). They read
respectively (in block form): 
\begin{eqnarray}
\label{Jordan}
\left[ \begin {array}{cc} 
                  A&0\\
\noalign{\medskip}0&B
\end {array} \right],
\quad \quad 
A \, = \, \, \left[ \begin {array}{ccc} 
                  0&0&0\\
\noalign{\medskip}0&0&1 \\
\noalign{\medskip}0&0&0
\end {array} \right], \quad B \, = \, \,
\left[ \begin {array}{ccc} 
                  0&1&0\\
\noalign{\medskip}0&0&1 \\
\noalign{\medskip}0&0&0
\end {array} \right].
\nonumber 
\end{eqnarray}
The explicit Jordan-block form of the $\, p$-curvature 
is quite reminiscent of the factorisation 
of the operator $\, L_6$
(see section (\ref{revisitglob}) below)
in an order-three linear
differential operator,  and another 
order-three operator, itself product of
an order-two  operator
and  an order-one  operator.
Note, however, that one should not extrapolate 
beyond simple product factorisations: 
the Jordan-block form of the $\, p$-curvature 
gives systematically a misleading prejudice
of direct-sum structures that do not exist. 

 The global nilpotence 
of the order-ten Fuchsian linear differential
 operator~\cite{ze-bo-ha-ma-05b}
 for  $\chi^{(4)}$, is confirmed by the calculation of the 
$\, p$-curvature for
all the primes up to $\, p \, \le \,  809$. The $\, p$-curvature 
{\em has been found to be nilpotent for all these primes}.

The global nilpotence of the order 5 and 6  Fuchsian
linear differential operators for $\Phi^{(3)}_H$ and  $\Phi^{(4)}_H$
is confirmed by the calculation of the 
$\, p$-curvature: we have calculated the $\, p$-curvature for
all the primes up to $\, p \, \le \,  809$ and it
{\em has been found to be nilpotent for all these primes}.
 The characteristic polynomial
 of the $\, p$-curvature of the (globally nilpotent)
linear differential operator of $\, \Phi_H^{(3)}$
 has been found to be $\, T^5$ (its  minimal polynomial being $\, T^3$).
The characteristic polynomial
 of the $\, p$-curvature of the (globally nilpotent)
linear differential operator of $\, \Phi_H^{(4)}$
 has been found to be $\, T^5$ (its  minimal polynomial being $\, T^4$).
For  $\, \Phi_H^{(5)}$ the calculations are drastically larger, but,
from a {\em probabilistic algorithm}, we found that the
 characteristic polynomial of the $\, p$-curvature of the
linear differential operator of $\, \Phi_H^{(5)}$ is $\, T^{17}$.  

The minimal polynomial of  $\, p$-curvature
of $\, \Phi^{(3)}_D$ and $\, \Phi^{(4)}_D$ is $\, T^{4}$.
The characteristic polynomial of the $\, p$-curvature 
of $\, \Phi^{(5)}_D$ and $\, \Phi^{(6)}_D$ is $\, T^{5}$.
The minimal  polynomial of  $\, p$-curvature
 of $\, \Phi^{(8)}_D$  is $\, T^{6}$.
Recall that
the characteristic  polynomial
 of  the $\, p$-curvature of a globally nilpotent operator 
of minimal order $\, N$ equals $\, T^N$.

\vskip .2cm 

\subsection{Other $n$-fold integrals of the Ising class.}
\label{other}

Other $\, n$-fold integrals (corresponding to 
the susceptibility of a square Ising model for which a magnetic
field is located only on spins on a particular diagonal of
the square lattice) were introduced 
in~\cite{Diag}. For instance, for $\, T<T_c$,
they read:
\begin{eqnarray}
\label{d2n}
&&{\tilde \chi}^{(2n)}_{d-}(t)\,\, = \,\,\,  \, 
{{  t^{n^2}} \over {
  (n!)^2 }} \, {{1 } \over {\pi^{2n} }} \cdot  
\int_0^1 \cdots \int_0^1\prod_{k=1}^{2n}\,  dx_k  
\cdot   {1\, +t^n\, x_1\cdots x_{2n}\over
  1\,-t^n\, x_1 \cdots x_{2n}}\nonumber \\ 
&&\quad \quad \times\prod_{j=1}^n\, 
\left({x_{2j-1}(1-x_{2j})(1-tx_{2j})\over 
x_{2j}(1-x_{2j-1})(1\, -t\, x_{2j-1})}\right)^{1/2}\nonumber\\
&&\quad \quad \times \prod_{1 \leq j \leq n} \, \, 
\prod_{1 \leq k \leq n}(1\, -t\, x_{2j-1}\, x_{2k})^{-2} \\
&&\quad\quad  \times
\prod_{1 \leq j<k\leq n}(x_{2j-1}-x_{2k-1})^2\,
 (x_{2j}-x_{2k})^2. \nonumber
\end{eqnarray}
and another similar formula for $\, {\tilde \chi}^{(2n+1)}_{d+}(t)$.
They are holonomic functions and their corresponding
Fuchsian linear differential operators were
 given in~\cite{Diag}. Again the calculations of the $\, p$-curvature 
of the corresponding linear differential equations 
of minimal order for $\, {\tilde \chi}^{(3)}_{d+}$
 and $\, {\tilde \chi}^{(4)}_{d-}$ confirmed their global nilpotence
(see (\ref{kidiag}) below and  \ref{diag}).
  
\subsection{ODEs for two-point correlation functions and form factors.}
\label{2point}

Many simple linear ODEs of various orders were obtained
 for the two-point\footnote[3]{Or could have been
obtained for any $\, N$-point correlation functions.}
correlation functions of the square Ising model~\cite{PainleveFuchs}.
The two-point correlation functions were found to be polynomials (with 
rational function coefficients) of complete elliptic integrals
of the first and second kinds: their
 global nilpotence is, thus, a
 straight consequence of their hypergeometric nature. 

Along this correlation function line, 
we can recall the linear differential operators $\, F_j(N)$ we obtained 
for the {\em form factors}\footnote[9]{Coefficients in $\, \lambda^j$
 of $\, C(N, \, N; \, \lambda)$, the
 $\, \lambda$-extension~\cite{Holo} of  the two-point 
correlation function $\, C(N, \, N)$.} $ \, f^{(j)}_{N,N}$
of the (off-critical) square Ising model~\cite{Holo}
and, in particular, their russian-doll structure.

The linear differential operators $F_{2n+1}(N)$, which annihilate
the form factors $f^{(2n+1)}_{N,N}$ 
have a ``russian-doll'' structure. They
are such that: 
\begin{eqnarray}
\label{F7531}
&&F_1(N) \, = \, \,    L_2(N).  \nonumber \\
&&F_3(N) \, = \, \,    L_4(N) \cdot  L_2(N),  \\
&&F_5(N) \, = \, \,   L_6(N) \cdot  L_4(N) \cdot  L_2(N),
 \,\,\,\,\,\, \cdots
 \nonumber 
\end{eqnarray}
where the differential operators  $\, L_r(N)$ are of order $\, r$.
The first one reads:
\begin{eqnarray}
 L_2(N)  \, = \,\,\,   \, D_t^2\,\, 
 + {\frac {2\,t-1}{ \left( t-1 \right) t}} \cdot D_t \, \, 
-{{1} \over {4 \, t}} +{{1} \over {4 \, (t-1) }}\, 
   - \,{\frac {{N}^{2}}{ 4 \, {t}^{2}}},
\end{eqnarray}
and the expressions of $\, L_4(N)$,
 $\, L_6(N)$, $\, L_{8}(N)$ 
and $\,   L_{10}(N)$ are given in~\cite{Holo}.  

Thus we see that the linear differential operator 
for $\, f^{(2n-1)}_{N,N}$
{\em rightdivides} the differential operator 
for $\, f^{(2n+1)}_{N,N}$,  $\, n\, \leq 3$. 
Similar relations occur for the $\, F_{2\, n}(N)$'s. 
We conjectured~\cite{Holo}
 that this property holds for all values of $n$.
We thus have a  ``russian-doll'' (telescopic) structure of 
these successive linear differential operators. 

Again, these form factors were found to be polynomial (with 
rational function coefficients) of complete elliptic integrals
of the first and second kinds: the 
 global nilpotence of the corresponding operators
is, again, a straight consequence of their hypergeometric 
nature. 

\subsection{Modular ODEs for lattice form factors.}
\label{formfact}

Along this correlation function line, it is also worth
 recalling the Fuchsian linear ODEs 
we found~\cite{Holo} for some $\, \lambda$-extensions
$\, C(N, \, N; \, \lambda)$
of two-point correlation functions of the (off-critical)
lattice Ising model for selected values 
of the parameter $\, \lambda$.
As examples of these Fuchsian linear differential 
operators, we found, for instance, that 
$\, C_{-}(N,N;\cos(\pi/4))$, for
 $\, N\, =\, 0,\, 1,\, 2, \, \cdots $, are annihilated,
respectively, by\footnote[9]{Note two misprints in~\cite{Holo}
 for $\,L^{[1/4]}_1$
and $\,L^{[1/4]}_2$, corresponding to the $\, D_t$ coefficient.}
\begin{eqnarray}
\label{1over4}
&&L^{[1/4]}_0\,=\,\,\, \, (t-1)^2\,t \cdot D_t^2\,\,
+{3\over 8}(t-1)(3t-2)\cdot Dt\,\,\,\,
-{15\, t\over  256}\,\,+{3\over 32}, \nonumber \\
&&L^{[1/4]}_1\,=\,\, \, (t-1)^2\,t\cdot D_t^2\,+{{(t-1)(5t-2)} 
\over {8}} \cdot Dt\,\,
 -{7\, t \over  256}\,+{1\over 16}, \nonumber\\
&&L^{[1/4]}_2\,=\, \, 
(t-8)(t-1)^2\,t \cdot D_t^2\,\, +{{ 7}\over {8}} \cdot
 (t-1)(t^2-2t+16) \cdot D_t
\nonumber\\
&& \quad\quad \quad \quad \quad +{209\, t^2 \over 256}\,
 -{25 t\over 16}\,+{1\over 2}, 
\end{eqnarray}
One has homomorphisms between the linear 
differential operators with same parity:
 the  $\, L^{[1/4]}_n$ for $\, n$ odd
 (resp. even) are equivalent~\cite{vdP1}.
We also have higher order ODEs like, for instance, the 
order-four linear differential operators~\cite{Holo}, 
denoted $\, L^{[1/3]}_N$,  
corresponding to $\, C_{-}(N,N;\cos(\pi/3))$.

We have calculated the $\, p$-curvatures of 
all these irreducible linear differential operators and
seen that they  have {\em zero $\, p$-curvatures}. Not 
surprinsingly the corresponding wronskians associated to these
$\, \lambda$-extensions
of two-point correlation functions are $\, N$-th root
 of rational functions and read:
\begin{eqnarray}
\label{wronsk}
&&W(L_0^{[1/4]}) \, = \, \, \,
(1-t)^{-3/8} \cdot  t^{-3/4}, \quad \quad  \quad 
  W(L_1^{[1/4]} ) \, = \, \,
(1-t)^{-3}\, t^{-2}, \nonumber \\
&&W(L_2^{[1/4]}) \,  \, = \, \, \,
(t-8)^{-8/7} \cdot (1-t)^{-15/7} \cdot  t^{-2/7},
 \quad \quad \cdots   \\
&&W(L_0^{[1/3]})  \, = \, \, 
(1-t)^{-11/3} \cdot  t^{-11/3},  \quad  
\,  W(L_1^{[1/3]}) \, = \, \,
 t^{-14/3} \cdot  (1-t)^{-8/3}, \nonumber \\
&&W(L_2^{[1/3]}) \, = \, \,\,
\left( 11+21\,t \right) \cdot (1-t)^{1/3} \cdot  t^{-20/3}, 
\quad \quad \cdots  \nonumber 
\end{eqnarray}

The fact that the 8-th power
(instead of the square in most of the examples of this paper) 
of $\, W(L_0^{[1/4]})$, or the 7-th power of  $\, W(L_2^{[1/4]})$,
is rational is in agreement with the interpretation 
of $\, \lambda$ we gave in~\cite{Holo}.
 These Fuchsian linear ODEs actually correspond 
to {\em  algebraic functions}\footnote[1]{Note that
maple also solves these ODEs in terms of (algebraic) 
hypergeometric or Heun functions
(see (\ref{alghyp}), (\ref{alghyp2}) below).}, 
 and are often, explicitely,
associated with {\em modular curves}. 

\vskip.3cm

\subsection{More zero $\, p$-curvatures: Joyce's Fuchsian ODEs.}
\label{moreJoyce}

Finally, along this zero $\, p$-curvature line, it is also 
worth recalling the large set of 
higher order Fuchsian ODEs obtained 
by  Joyce~\cite{Joyce1,Joyce2,Joyce3,Joyce4,Joyce5,Joyce6}.
There are not so many examples of Fuchsian linear ODEs
of high order in the litterature.  Joyce has been
 one of the few authors
to provide such non-trivial examples. 
We have calculated the $p$-curvature of 
a large set of these Fuchsian linear
differential operators, namely  (42) 
in~\cite{Joyce2}, (85-86) in~\cite{Joyce3},
 (5.22) of~\cite{Joyce1}, (2.16) of~\cite{Joyce4},  \ldots
We  found that they are more than globally nilpotent:
their $p$-curvature is {\em zero for
 almost every prime} (for all primes except 
a  finite set of primes).
They have a basis of {\em algebraic
 solutions}
 that can be expressed in terms of simple 
Legendre-P functions and simple  algebraic functions.
 These Fuchsian linear ODEs actually correspond 
to {\em  algebraic functions}, and are often, explicitely,
associated with {\em modular curves}. 

\vskip .1cm 

\section{Global nilpotence from the 
global nilpotence of the factors.}
\label{revisit}

\subsection{Revisiting the global nilpotence of $\, \chi^{(3)}$}
\label{revisitglob}
The minimal linear differential operator for $\, \chi^{(3)}$
is an order-seven operator $\, L_7$ which
 can be written as the direct sum of the
order-one linear differential operator for $\, \chi^{(1)}$
and an order-six linear differential operator for 
$\, 2 \, \chi^{(3)} - \chi^{(1)}$,
namely $\, L_6$ which factorizes into an order-three 
linear differential operator $\, Y_3$,  
an order-two linear differential operator 
$\, Z_2$, and an order-one linear differential operator $\, N_1$:
\begin{eqnarray}
\label{l6}
L_6 \,\, = \, \,\, Y_3 \cdot  Z_2 \cdot N_1. 
\end{eqnarray}
The explicit expressions of $\, Y_3$ and $\, Z_2$ are given
 in Appendix A of~\cite{ze-bo-ha-ma-05c}.

We have the following, almost obvious remark\footnote[5]{If one 
takes as a definition of global nilpotence the 
factorization modulo (almost all) primes
of the operator in order-one operators with rational function solutions
 modulo primes (see~\cite{Schmitt} 
and Lemma 0.6.2 in~\cite{Dwork}).}, that {\em the product of  
globally nilpotent operators is necessarily globally nilpotent}.
The global nilpotence of the order-one linear
 differential operator $\, N_1$ is 
obvious. Furthermore, we found the remarkable result that the solutions
of $\, Y_3$ are quadratic expressions of the complete elliptic integrals 
of the first and second kind $\, K$ and $\, E$ 
(see appendix B of~\cite{ze-bo-ha-ma-05c}). From a differential algebra 
viewpoint this amounts to saying that $\, Y_3$ is {\em equivalent
to the symmetric square} of the second order 
linear differential operator $\, L_E$ corresponding to  $\, E$
(see~\cite{Fuchs}). Since hypergeometric functions 
correspond to  globally nilpotent
operators, $\, Y_3$ {\em is therefore globally nilpotent}.

The global nilpotence of the linear differential
 operator for $\, \chi^{(3)}$
{\em thus reduces to the global nilpotence of the second order
linear differential operator} $\, Z_2$. 
The linear differential operator $\, Z_2$, is  {\em an example
 of globally nilpotent operator
which does not straightforwardly reduce to hypergeometric functions}
up to change of variables (pull-back) and multiplications.

In fact, a simple right-multiplication of $\, Z_2$
by $\, h(w)\, = \, 1/(1\, +4\, w)/(1\, -4\, w)^2$, 
 enables one to get rid
of the singularity $\,w \, = \, -1/4$.
Instead of the solutions $\, F(w)$ of $\, Z_2$,
this amounts to considering the second order linear differential operator
 $\, \tilde{Z}_2 $ 
with solutions  $\, F(w)/h(w)$. Denoting
 $\, D_w \, = \, d/dw$, this linear  differential operator reads:
\begin{eqnarray}
 \tilde{Z}_2 \, \, = \,  {{1} \over {h(w)}} \cdot Z_2 \cdot h(w)
\, = \, \,\, \, \,q_2 \cdot q_{app} \cdot D_w^2 \, 
+q_1 \cdot D_w\, + \, 24\,w \cdot q_0, \nonumber 
\end{eqnarray}
with:
\begin{eqnarray}
&&q_2 \, \, = \, \,\, \,
w \cdot (1-w)  \left( 1-4\,w \right)  \, (1+2\,w) \,
(1+3\,w+4\,{w}^2),  \, \\
&&q_1 \, \, = \, \,\, \,
1-2\,w+{w}^{2}-216\,{w}^{3}-336\,{w}^{4}\, 
+1656\,{w}^{5}+1040\,{w}^{6}\nonumber \\
&& \qquad 
-2560\,{w}^{7}-6400\,{w}^{8}-6144\,{w}^{9}, 
\nonumber \\
&&q_0 \, \, = \, \,\, \,
 1-7\,w-4\,{w}^{2}-47\,{w}^{3}+36\,{w}^{4}
+280\,{w}^{5}+160\,{w}^{6}+256\,{w}^{7},\nonumber \\
&&q_{app} \, \, = \, \,\, \,
1 \, -3\,w \, -18\,{w}^{2}\, +104\,{w}^{3} \,+96\,{w}^{4}.
\nonumber 
\end{eqnarray}
The polynomial $\, q_{app}$ corresponds to {\em apparent} singularities.
All the other singularities are regular singularities
{\em  remarkably with integer exponents
and all yielding logarithmic behaviours}. This can simply be seen 
from their corresponding indicial polynomials and 
the formal series around these singularities. 
The differential Galois group 
 of $\,\tilde{Z}_2$ is $\, SL(2, \, C)$,
 a consequence of the existence of logs in
the formal solutions around singular points, together
 with a wronskian being rational,
and the operator being irreducible~\cite{Boucher}. 

Calculating the indicial polynomials at the
 various singularities, one finds
 the following critical exponents: 
\begin{eqnarray}
\label{indici}
&&\rho=\, 0,\,0,  \quad \quad  \qquad \hbox{for:}
\quad  \qquad  w=\, 0, \,\,\,
 \, 1/4,\,\,\,\,\, {{-3\, \pm i\sqrt {7}} \over {8}}, \nonumber \\
&&\rho=\, 0,\, 2,  \quad\quad  \qquad  \hbox{for:} \quad \qquad 
 w=\,1, \,\,  \,-1/2,\nonumber \\
&& \rho=\, -1,\, -2  \qquad \quad \hbox{for:} 
\qquad \quad  w \, = \, \infty.
\end{eqnarray}
If one wants to see the solution as hypergeometric 
functions up to a pull-back,
 the change of variables to be done must try to 
``wrap'' all these seven singularities 
onto only three ones: $\, 0$, $\, 1$, $\, \infty$.
Naively one can thing of wrapping the singularities 
according to the previous roots
 of indicial polynomials (\ref{indici}), namely 
$\, w=\, 0, \,\,\, w \, = \, 1/4,  \,\,\, (-3\, \pm i\sqrt {7})/8
\,  \rightarrow \,  \, 0$,
$\, w\,=\,\,1, \,\,\, w \, = \,-1/2, \, \rightarrow \,  \, 1/4$,
$\, w=\, \infty \,  \rightarrow \, \, \infty$, however, since the 
critical exponents are all integers, 
 these seven singularities have, in fact, to 
be considered {\em on the same footing}. This will be discussed
in more details in section (\ref{nondwork}), where
 the explicit solution of $\, Z_2$ 
will be given {\em in term of a modular form of weight one}.  

\subsection{Revisiting the global nilpotence of $\, \chi^{(4)}$}
\label{revisitglobbis}

For the order-ten linear differential operator of $\, \chi^{(4)}$ 
we have similar calculations. The  
order-ten linear differential operator
 $\, L_{10}$ of $\, \chi^{(4)}$ 
is the direct sum of the order-two linear
 differential operator for $\, \chi^{(2)}$
and of an order-eight linear differential operator 
which factorizes~\cite{ze-bo-ha-ma-05b,ze-bo-ha-ma-05c}
 into an order-four operator 
and four order-one operators (see (F.4) in~\cite{ze-bo-ha-ma-05b}):
\begin{eqnarray}
\label{cal8}
{\cal L}_8 \, \, = \, \, \, 
M_2 \cdot L_{25} \cdot L_{12} \cdot L_3 \cdot L_0,
\end{eqnarray}
The global nilpotence of the order-one operators 
$\, L_{25}$, $\,  L_{12}$, $\,  L_3$ and
 $\,  L_0$ is a simple consequence of the fact that all these 
operators are of the form $\,  D_x \,\, + \, \, R'/R$,
where $\, R$ denotes a rational function, or the square of
a rational function (simply
 related to the wronskian of these operators), and $\, R'$
 its first derivative with respect to $\, x$.
 These $\, R$ 
functions read respectively for $\, L_{25}$, $\,  L_0$,
 $\,  L_{12}$ and $\,  L_3$:
\begin{eqnarray}
&&R_{25}\, = \, \, \, {\frac { (5\,x+7)  \, (4-x)^{7/2}}
{ (14-12\,x+9\,{x}^{2}-5\,x^3) 
\cdot {x}^{2} \left(1-x \right)^2}}, 
\quad \quad R_{0}\, = \, \,  \, 1, \\
&&R_{12}\, = \, \, \,
{\frac { (14-12\,x+9\,{x}^{2}-5\,x^3)\cdot  x}
{ (2+ 3\,{x}^{2}+{x}^{3})  \, (1-x)^{7/2}}}, 
\qquad \quad R_{3}\, = \, \,  \,
{\frac {2+3\,x^2+x^3}{ (1-x)^2 \cdot x^{3/2}}}.\nonumber
\end{eqnarray}
The order-two linear differential operators 
$\, N_i$, $\, i \, = \, 1, \, \cdots \,, \,  9$,
sketched in~\cite{ze-bo-ha-ma-05c} and which happen 
in other factorizations of $\, L_{10}$,
are all equivalent to the order-two
linear differential operator~\cite{ze-bo-ha-ma-05c} $\, N_0$ 
\begin{eqnarray}
N_0 \, = \, \,  \,  \,  \,
D_x^2 \, \, \, \,
 - \, \, {\frac { 1+\,x}{\, 1-\,x}} \cdot {{ D_x} \over {x}}
 \, \, \,  \,
+ \, {{1} \over {4\,x}} \cdot {\frac {1}{ 1-x  }},
\end{eqnarray}
having hypergeometric solutions (corresponding to $\, \chi^{(2)}$):
\begin{eqnarray}
\chi^{(2)} \, = \, \, {x}^{2}  \, \cdot  \, \,
_2F_1 \left( [3/2,5/2],[3],x \right).
\end{eqnarray}
Most of the ``complexity'' of $\, \chi^{(4)}$ is 
thus ``encapsulated'' in the order-four
linear differential operators $\, M_2$  (or 
equivalently $\, M_1$ of~\cite{ze-bo-ha-ma-05c}). 

In a remark (in Appendix B of~\cite{ze-bo-ha-ma-05c} page 27),
we mentioned the fact that the solutions of $\, M_2$
can be expressed as linear combinations of products
 of complete elliptic integrals. 
One can make this statement more precise. Let
 us introduce the {\em symmetric cube}
of the linear operator~\cite{ze-bo-ha-ma-05c}
  $\, N_0$ (associated to $\, \chi^{(2)}$),
the order-four linear differential operators $\, M_2$ (or 
equivalently $\, M_1$) is {\em equivalent} to the symmetric cube 
$\, Sym^3(N_0)$. There exists
two order-three interwinners $\, I_1$ and $\, I_2$ such that:
\begin{eqnarray}
 I_1 \cdot Sym^3(N_0) \,\, = \, \,\,\, \, M_2 \cdot I_2. 
\end{eqnarray}

In other words, the solutions of $\, M_2$
 are cubic (homogeneous) polynomials~\cite{ze-bo-ha-ma-05c}
of the two solutions of $\, N_0$ (hypergeometric functions). 
Therefore  $\, M_2$ {\em is globally nilpotent},
 and consequently the order-ten linear differential
 operator for $\, \chi^{(4)}$
is also globally nilpotent.
 One sees that, paradoxically, the global nilpotence of 
the linear differential operator for $\, \chi^{(4)}$
is much simpler to understand than the global nilpotence of 
the linear differential operator for $\, \chi^{(3)}$,
which is a consequence of the, after first sight,
subtle global nilpotence of $\, Z_2$.

\subsection{Revisiting the global nilpotence of $\, \Phi^{(3)}_H$.}
\label{subrevi21}
\vskip .1cm 
In order to revisit the global nilpotence of $\, \Phi^{(3)}_H$
 (see (\ref{In})), 
let us study the factorisation of the corresponding order-five 
Fuchsian linear operator given in~\cite{Singul}.  This order-five 
Fuchsian linear operator factorizes into 
an order-three and an order-two 
Fuchsian linear operators:
\begin{eqnarray}
\label{Phi3Hrevim3m2}
L_{\Phi^{(3)}_H} \, =\, \, \, M_3 \cdot M_2, 
\end{eqnarray}
where the second order operator $\, M_2$ is given
 in \ref{m2}. The wronskian of the order-two operator
 $\, M_2$ is a simple rational function.
The indicial polynomials for $\, M_2$ yield integer critical exponents,
when the formal series solutions show logarithms
 for all the singularities, the second order operator $\, M_2$
 being irreducible.
Consequently~\cite{Boucher}, the 
differential Galois group of $\, M_2$ is $\, SL(2, \, C)$.

We have calculated the $\, p$-curvature of $\, M_3$ and $\, M_2$ 
and found that their characteristic polynomial equals 
their minimal polynomials,
being respectively $\, T^3$ and $\, T^2$ for almost all primes.
These two linear differential operators
 are thus  globally nilpotent. 

{\em More remarkably, we actually found
that $\, M_2$ is  
equivalent}\footnote[1]{Up to the
 change $\, x \,= \, \,  4 \, w$.} to
 $\, Z_2$ of $\,\chi^{(3)}$, (see (\ref{40}) below). 
{\em We also found that $\, M_3$ is  
equivalent to}  $\, Y_3$ of $\,\chi^{(3)}$, or equivalently 
$\, Sym^2(Q_E)$, the symmetric square of 
the linear differential operator $\, Q_E$ 
corresponding to the complete elliptic integral $ \, {\it E}(4\, x)$:
\begin{eqnarray}
\label{32}
Q_E \, = \, \,  \,\, D_x^{2} \, \, +{\frac {D_{x}}{x}} \, \, 
+{\frac {16}{ ( 1-4\, x)  \, (1 \, +4\, x) }}.
\end{eqnarray}
Therefore $\,\chi^{(3)}$ and $\, \Phi^{(3)}_H$ have extremely close 
structures.

 As usual the order-two intertwinners $H_1$ and
 $\, H_2$ (or $\, H'_1$ and $\, H'_2$) in the equivalence of 
$\,M_3$ and $\, Y_3$ 
 \begin{eqnarray}
  M_3 \cdot H_1 \, = \, \, H_2 \cdot Sym^2(Q_E), \qquad  
 H'_1 \cdot  M_3  \, = \, \,  Sym^2(Q_E) \cdot H'_2,
 \nonumber
\end{eqnarray}
are themselves equivalent and, again, there exist two 
order-one intertwinners $K_1$ and $\, K_2$
(resp.  $K'_1$ and $\, K'_2$) such that 
$\,  K_1 \cdot H_1 \, = \, \, H_2 \cdot K_2$, and  
$\,K'_1 \cdot H'_1 \, = \, \, H'_2 \cdot K'_2$, which
 are again equivalent. One thus has a ``tower''
 of equivalent differential operators.
 It is important to note that the 
``intertwinning'' operators in such 
a ``tower'' of equivalences 
 {\em are not necessarily globally nilpotent!}

\subsection{Revisiting the global nilpotence of $\, \Phi^{(4)}_H$.}
\label{subrevi22}
\vskip .1cm 
The global nilpotence of $\, \Phi^{(4)}_H$ can
 be understood from the factorisation of 
the corresponding order-six 
Fuchsian linear differential operator given in~\cite{Singul}. It 
factorises as follows:
\begin{eqnarray}
\label{Phi4H}
L_{\Phi^{(4)}_H} \, = \,  \, \,  \,
M_4 \cdot P_1 \cdot Q_1, 
\end{eqnarray}
where the order four operator $\, M_4$ is given in \ref{m4} and 
the two order-one operators $\, P_1$ and $\, Q_1$ read:
\begin{eqnarray}
\label{Phi4Hfact}
&&P_1  \, = \, \,\,  \,  D_x \, \,\,
 +\, {{1} \over {2}} \,{{d} \over {dx}}
 \ln \Bigl((x-4)\,  (x-1)^2 \cdot   x^2  \Bigr),
 \nonumber \\
&& Q_1 \, = \, \, \, \,  D_x \, \, \,
+\, {{1} \over {2}} \,{{d} \over {dx}} 
\ln \Bigl( (x-1)^2 \cdot   x  \Bigr). 
\nonumber 
\end{eqnarray}
The global nilpotence of $\, P_1$ and $\, Q_1$
corresponds to the fact that the square of the wronskians of 
these first order operators are {\em simple rational functions}.

The order-four operator $\, M_4$ is irreducible but
{\em it is actually equivalent to} 
$\, Sym^3(L_E)$, the symmetric cube of
the linear differential operator $\, L_E$ having
${\it E}(x^{1/2})$ (complete elliptic integral) as solution: 
\begin{eqnarray}
\label{LE}
 L_E \, = \,\, \, \,\, D_x^2 \, \, \, +{\frac { D_x}{x}}  \, \,
\, +{{1} \over {4}}\,{\frac {1}{ (1-x)\, x }}.
\end{eqnarray}

The global nilpotence of $\, \Phi^{(4)}_H$
is clearly a straight consequence of this last equivalence
 with a symmetric cube of a hypergeometric operator.
Comparing the factors (and their equivalence) for $\, \Phi^{(4)}_H$ 
 one finds out, after the remarkable identification
 of structure  between $\, \Phi^{(3)}_H$ 
 and  $\, \chi^{(3)}$,
 that one has exactly the same identification of structures
  for $\, \Phi^{(4)}_H$ 
 and for $\, \chi^{(4)}$. Again, slightly surprisingly, the global nilpotence 
of  $\, \Phi^{(4)}_H$ (or $\, \chi^{(4)}$) is
 straightforward to understand
compared to the global nilpotence 
of  $\, \Phi^{(3)}_H$ (or  $\, \chi^{(3)}$)
 which amounts to understanding
the more subtle global nilpotence of the second order operator $\, Z_2$.

\subsection{Revisiting the global nilpotence of the $\, \Phi^{(n)}_D$'s.}
\label{revi20}
Similar (detailed factorisations) calculations can be performed for 
$\, \Phi^{(3)}_D$, $\, \Phi^{(4)}_D$, $\, \Phi^{(5)}_D$,
and $\, \Phi^{(6)}_D$ defined by (\ref{chinaked}).  They
 are displayed in \ref{revin}. 
Again one finds out that the global nilpotence of  many of the  
factors occurring in the factorisations can be explained 
by the occurrence of  complete elliptic integrals
 of the first or second kind.  The example of $\, \Phi^{(6)}_D$ 
in \ref{revin} corresponds to a much more interesting situation, 
also occurring with $\, \Phi^{(5)}_D$, where we contemplate
in the factorisation of the operator for $\, \Phi^{(6)}_D$,
 a remarkable combination of a
 globally nilpotent order-two
operator associated with complete elliptic integrals
 of the first or second kind,
{\em together with}  a second order operator of 
{\em zero $\, p$-curvature} associated with highly non-trivial (genus 
five) algebraic curve (with a group of quaternion as its differential 
Galois group). 
Let us illustrate a similar structure focusing on 
the global nilpotence of  $\, \Phi^{(5)}_D$.

\subsection{Revisiting the global nilpotence of  $\, \Phi^{(5)}_D$.}
\label{revi20}
The linear differential operator for $\, \Phi^{(5)}_D$
is a Fuchsian linear differential operator of order five 
which factorises in three different ways ($N_4,\,N_3, \, N_2, \, N_1$
are second order linear differential operators, $ \,L_1$ and  $ \,L_2$ 
first order):
\begin{eqnarray}
L_{\Phi^{(5)}_D} \, = \, \, \,
 L_1 \cdot  N_2 \cdot N_1
 \, = \, \, \,N_3 \cdot L_2 \cdot N_1
 \, = \, \, \,N_3 \cdot N_4 \cdot D_x.
\end{eqnarray}

This means that it factorises 
into the direct sum
of two order-two linear differential operators
and the operator $\, D_x$:
\begin{eqnarray}
&&L_{\Phi^{(5)}_D} \, = \, \, \,  M_4 \oplus D_x, \qquad
 \hbox{ with:}
\qquad \quad    M_4 \, = \, \, \,  
 N_2 \cdot N_1,  \qquad
\nonumber 
\end{eqnarray}
where the second order operator $\, N_1$ is 
given in \ref{n1}.
The square of the wronskian of $\, N_1$ is a simple rational function.

The $\, p$-curvature of $\, N_1$
{\em is equal to zero for any prime} $\, \ge 2$, 
therefore, modulo Grothendieck's conjecture\footnote[3]{Namely that
zero $\, p$-curvatures yield~\cite{resconj,Katz1,Katz11,Katz2} a basis 
of algebraic solutions for the 
linear differential operator.},
 {\em it admits a basis of algebraic solutions}. These 
algebraic solutions can be written as 
hypergeometric functions up to a pull-back by a rational function
and multiplication
by a $\, N$-th root of a rational function.
They are of the form~\cite{JAW,Berkenbosch}:
\begin{eqnarray}
\label{alghyp}
R_1(x)^{1/4} \cdot H[R_2(x)],  \qquad \hbox{where:} \qquad  
P(H[R_2(x)], \, x) \, = \, \, 0,  
\end{eqnarray}
where $\,R_1(x)$ and  $\,R_2(x)$ are rational functions,
$\, H$ is an {\em algebraic hypergeometric function}
 and $\, P(y, \, x)\, = \, \, 0 $ is an {\em algebraic curve} of
degree 4 and {\em of genus} $\,6$:  
\begin{eqnarray}
\label{Z4}
&&(1-x-3\,{x}^{2}+4\,{x}^3)^{2} \, (1+x)^2
 \, (1+8\,x+20\,{x}^{2}+15\,{x}^{3}+4\,{x}^4) \nonumber \\
&& \quad \times  \, 
(1+2\,x-4\,{x}^2)  \,(1-3\,x+{x}^2)  \,
(1+2\,x)  \, (1-x) \cdot  y^4\nonumber \\
&&\quad -2\, (1+x)  \, (1-x-3\,{x}^{2}+4\,{x}^3) 
 \cdot g_1 \cdot y^2  \\
&&\quad + \, (1\, + 3\,x\, + 3\,{x}^{2}\, -4\,{x}^{3})^2 \, 
(1+x-6\,{x}^{2}-{x}^{3} +{x}^{4})^2 \, = \, \, 0, \nonumber
\end{eqnarray}
where:
\begin{eqnarray}
&&g_1\, = \, \, -3-24\,x
-20\,{x}^{2}+255\,{x}^{3}
+484\,{x}^{4}-800\,{x}^{5}-1729\,{x}^{6}\nonumber \\
&& \qquad +1296\,{x}^{7}+2236\,{x}^{8}
-1035\,{x}^{9} -1004\,{x}^{10}
+232\,{x}^{11}+160\,{x}^{12}. \nonumber
\end{eqnarray}
Note that the vanishing of the  $\, y^4$ 
coefficient in (\ref{Z4}) corresponds
to singularities of $\, \Phi^{(5)}_D$.

 This {\em genus six} algebraic 
curve should not be confused\footnote[8]{They can, however,
correspond to highly non-trivial relations
expressing such  {\em genus six} algebraic solutions as
linear combinations of complete elliptic
integrals of the third kind with a ``characteristic'' (first argument
of the complete elliptic integral) associated with a {\em genus three}
curve.} with genus three curves
that will be mentioned at the end of section (\ref{towards1}). 

Using the maple command $\texttt{kovacicsols}$
 one can write one of the solutions
 in terms of hypergeometric functions as follows:
\begin{eqnarray}
\label{alghyp2}
\Bigl(R_{\Phi_D^{(5)}}(x) \Bigr)^{1/12} \cdot  \, \, \, 
 _2F_1([-1/4,1/4],[1/2],\, {\cal M}), 
\end{eqnarray}
where the argument in the hypergeometric function 
reads:
\begin{eqnarray}
&&{\cal M} \, = \, \, {{{\cal N}} \over {{\cal D}}},
 \qquad \qquad \hbox{where:} \\
&&{\cal N}\, = \, \, \left( x-1 \right)  \, (1+2\,x) 
 \, (1 -3\,x +{x}^2)  \, (1+2\,x-4\,{x}^2) 
 \nonumber \\
&& \qquad \quad \times  
 (1+8\,x+20\,{x}^{2}+15\,{x}^{3}+4\,{x}^4)
  \, (1+3\, x \, +3\,{x}^2\, -4\,x^3)^{2}
 \nonumber \\
&&\qquad  \quad \times 
 \, (1+x-6\,{x}^{2}-{x}^{3} +{x}^{4} )^2, 
\nonumber \\
&&{\cal D} \, = \, \, 8\, \, (1+x) 
 \, (1-x-3\,{x}^{2}+4\,{x}^3 ) \cdot d_{10}^2, \nonumber \\
&&  d_{10} \, = \, \, 1+8\,x+8\,{x}^{2}-73\,{x}^{3}-148\,{x}^{4}
+144\,{x}^{5}+325\,{x}^{6}\nonumber \\
&&  \qquad \quad \quad 
-152\,{x}^{7}-168\,{x}^{8}+39\,{x}^{9}+28\,{x}^{10}.
\end{eqnarray}
and the rational function $\, R_{\Phi_D^{(5)}}(x)$ reads:
\begin{eqnarray} 
&&R_{\Phi_D^{(5)}}(x)  \, = \, \, \, 
- \, {{n} \over {  g_{1}^2 \cdot {\cal N}  }}, 
\qquad \quad \quad \hbox{where:} \\
&&n\, = \, \, 16 \cdot  (x+1)^2 \,
 (1\, -x\, -3\,{x}^{2}\, +4\,{x}^3)^2
 \cdot d_{10}^4.  \nonumber 
\end{eqnarray}

 The linear differential 
operator $\,N_2$ (or equivalently $\,N_3$) {\em has 
solutions in terms of the complete 
elliptic integrals $\, E$ and $\, K$}.
The $\, p$-curvatures of the linear differential
operator $\,N_2$ (resp. $\,N_3$)
correspond to a globally nilpotent operator. 

\subsection{Towards a geometric interpretation of the $\, \Phi_D^{(n)}$.}
\label{towardsageom}
The DFG (global nilpotence) structure corresponds to the fact
that a holonomic function has an interpretation
as a {\em period of some algebraic variety}.
Along this line, it is worth noting that
 in the case of these $\, \Phi_D^{(n)}$,
 {\em some closed exact expressions} 
for these integrals $\, \Phi_D^{(n)}$ can be
 obtained which give explicit
 examples of such an interpretation as a period. Actually 
the simple integrals  $\, \Phi_D^{(n)}$ can 
all be expressed
as sums of {\em complete elliptic integrals of
 the third kind}, where the characteristic 
$\, y= \, y(w)$ corresponds to some selected, 
and highly non trivial, algebraic curves (genus zero, three, 
ten, \ldots). The results are displayed in \ref{towards1}.

\section{The second order operator $\, Z_2$ and weight-1 modular forms.}
\label{Z2}

The previous section was dedicated to understand the global nilpotence
of linear differential operators of quite high orders
from the global nilpotence of their factors of smaller orders.
To sum up all the previous results, the global nilpotence of the
factors fall in three categories: 

- firstly, a straightforward global nilpotence
of order-one factors (which amounts to saying that
their wronskian is a $\, N$-th root of a rational function), 

- secondly, global nilpotence straightforwardly associated 
with complete elliptic integrals
of the first (or second) kind, or equivalently $_2F_1$
 hypergeometric functions,
or global nilpotence corresponding to symmetric powers of the previous 
second order hypergeometric operators, or to zero curvature operators
with their solutions being a basis of algebraic functions (often
 modular curves, such algebraic functions can be written as pull-backs
 of $_2F_1$ hypergeometric functions),

-  finally, a set of operators
of order two, three, \ldots that do not have a basis
 of algebraic solutions
and that we have not been able to immediately reduce to
$_2F_1$ hypergeometric functions, or products 
of hypergeometric functions.

The second order linear differential
operator $\, Z_2$, occurring in the factorisation of 
the linear operator for $\, \chi^{(3)}$ (or equivalently $\, M_2$ for 
$\,\Phi_H^{(3)}$), is a perfect illustration of this last situation. 
The order two and three operators occurring in 
the operator factorization 
for the three-choice polygons perimeter generating function (see below)
are other examples that do not 
immediately fit in a  $_2F_1$  ``Dworkian'' framework. 

\subsection{Reducing   $Z_2$ to Heun functions}
\label{nondwork}

We found that the $\, Z_2$ of $\, \chi^{(3)}$, and its equivalent 
for $\, \Phi^{(3)}_H$, namely $\, M_2$ 
in (\ref{Phi3Hrevim3m2}) and (\ref{Phi3Hrevi}),  are
 homomorphic\footnote[5]{Up to the
 change $\, x \,= \, \,  4 \, w$ in $\, M_2$.} ($U_1$ and $U_2$
are two order-one intertwinners):
\begin{eqnarray} 
\label{40}
&& Z_2 \cdot U_1   \, = \, \,  U_2 \cdot M_2, \quad   \quad 
U_1   \, = \, \,{\frac {    (1+2\,x)  \, (1-x) }
{ (1-2\,x)  \, (1+4\,x)  \, (1-4\,x)}} \cdot u_1,   \\
&&u_1 \, = \, \,x \cdot  (1-4\,x)  \, (1+3\,x+4\,{x}^{2}) \cdot D_x 
\, + \, \,  (1 \,+\,x \, +12\,{x}^{2} +48\,{x}^{3}),  \nonumber 
\end{eqnarray}

On the other hand, let us introduce the order-two Heunian operator
whose solution is $\, Heun(8/9, 2/3, 1, 1, 1, 1; t)$:
\begin{eqnarray}
\label{lets}
&& {\cal H} \, = \, \, \, D_t^2 \, \, 
+\Bigl( {{1} \over {t}} \, \, 
  +{{1} \over {t-1}} \, + {{9} \over {9\, t\, -8}} \Bigr) \cdot D_t \, \\
&& \qquad  \qquad \quad
 + \, 3\,{\frac {3\,t-2}{ (9\,t-8)\, (t-1)\, t  }},\nonumber 
\end{eqnarray}
a simple change of variable:
\begin{eqnarray}
\label{simplechange}
t \, = \,\,  \, \, \,{\frac {-8\, x}{ (1-4\,x) 
 \, (1-x) }},
\end{eqnarray}
 transforms (\ref{lets}) into the order-two 
linear differential operator:
\begin{eqnarray}
\label{Heunx}
&&  {\cal H}_{x}\, = \, \, D_x^2 \\
&&\, \,\qquad   {\frac {1 -10\,x+19\,{x}^{2}-92\,{x}^{3} 
 +12\,{x}^{4}\,+224\,{x}^{5}-64\,{x}^{6}  }
{ ( 1+3\,x+4\,{x}^2)  \, (1-2\,x)  \, (1+2\,x)\, 
  (1-4\,x)
  \, (1-x) \cdot x  }} \cdot D_x 
\nonumber \\
&&\qquad \, \, + \, 6\,{\frac { (1+7\,x+4\,{x}^{2}) 
 \, (1-2\,x)^2}{ (1+ 3\,x+4\,{x}^2) \,
\, (1-4\,x)^{2} \,(1-x)^2 \cdot x}}.
\nonumber
\end{eqnarray}
This operator (\ref{Heunx}) is just
 the conjugate of $\, M_2$ by the multiplication by 
a simple polynomial function $\, h(x) \, =\, $$ (1-x) \, (1-4\, x)^2$:
\begin{eqnarray}
\label{conj}
h(x) \cdot M_2 \, = \, \, {\cal H}_{x} \cdot h(x). 
\end{eqnarray}

Together with (\ref{40}), this means that the 
second order linear differential
operator $\, Z_2$, corresponding to 
$\, \chi^{(3)}$, reduces to 
a Heun operator given by (\ref{lets}),
to be compared with Krammer's~\cite{Krammer} counterexample (\ref{O1}) 
to Dwork's conjecture.

One then gets the (selected) solution for $Z_2$ in terms of Heun 
functions ($w$ replaced here by $\, x$,
 and $\, t$ is given by (\ref{simplechange})):
 \begin{eqnarray}
 \label{Hg}
 &&r(x)
 \cdot  \Bigl( (1- 9\,x)  \,  (1-4\,x) \, (1-x) \cdot Hg(t) \, \,\,  \nonumber \\
&& \qquad \qquad \qquad  +\, 8  \cdot  (1+3\,x+4\,{x}^{2})  \cdot  x
 \cdot Hg'(t) \Bigr),    \\
&&\hbox{where:}\quad  \quad   \quad r(x)\, = \, {{ (1\, +2\, x)^2 }
 \over { (1-x)^2 \, (1\, +4\, x)\, (1\, -\, 4\, x)^2}},
\quad \quad  \quad  \hbox{and:}  \nonumber \\
&& Hg'(t)\, = \,
 {{d Hg(t)} \over {dt}}, \qquad  \quad 
Hg(t)  \, =\, \, Heun \Bigl(8/9, 2/3, 1, 1, 1, 1, t  \Bigr).
 \nonumber 
\end{eqnarray}

\subsection{Reducing   $Z_2$ to weight-1 modular forms}
\label{weight1}

Recall~\cite{Maier} that
 the {\em fundamental weight-1 modular form}\footnote[2]{
The modular form $\, h_6$ 
is also combinatorially significant: the perimeter generating 
function of the three-dimensional staircase polygons~\cite{Prell} 
can be expressed in terms of $\, h_6$
 (see section (\ref{beyondDwork1}) below). The modular form $\, h_6$ 
also occurs in Ap\'ery's study of $\, \zeta(3)$ (see \ref{fin}). }
  $\, h_N$ for the modular group $\, \Gamma_0(N)$
for $\, N\, = \, 6$,  can 
be expressed as a simple Heun function, 
$\, Heun(9/8, 3/4, 1, 1, 1, 1, - t/8)$, or as a hypergeometric function:
\begin{eqnarray}
\label{46}
&&{\frac { 2\,\sqrt {3}}{ (  (t+6)^{3}
 \, (t^3+18\,t^2+84\,t+24)^3 )^{1/12}}} \\
&& \qquad \quad \times \, \, 
 _2F_1 \Bigl([{{1} \over {12}}, {{5} \over {12}}];[1];  
\, 1728\,\, {\frac { (t+9)^2 \, (t+8)^3 \, t}{
 \,(t+6)^3 \, ({t}^{3}+18\,{t}^{2}+84\,t+24)^3}} \Bigr), 
\nonumber 
\end{eqnarray}
and that (\ref{46}) is solution of the 
order-two linear differential operator
(obtained from (\ref{lets}) by $\,t \, \rightarrow \, \, \, - t/9$) :
\begin{eqnarray}
\label{h6}
D_t^2 \, \,  + \Bigl(  {{1} \over { t+8 }} \, 
+ {{1} \over {t}} \, +{{1} \over { t+9 }}  \bigr) \cdot  D_{t} \, 
 +{\frac {t+6}{ \left( t+8 \right) \left( t+9 \right)\cdot  t}}.
\end{eqnarray}
Therefore, after some changes of variables, one can see the (selected)
solution of $\, Z_2$
as a hypergeometric function (up to a pull-back) corresponding to
{\em weight-1 modular form}\footnote[3]{The simplest weight-1 modular form
is $\, _2F_1([1/12,5/12],[1], \, {\hat J}) \, = \, 
\, 12^{1/4} \, \eta(\tau)^2 \, {\hat J}^{-1/12}$
where $\, \hat{J}$ is the Hauptmodul, $\, \eta$ 
is the Dedekind eta function, 
and $\, \tau$ is the ratio of periods (see (4.6) 
in~\cite{Maier4}). It can also be expressed as a linear combination 
of $\, _2F_1([1/12,5/12],[1/2], \, 1\, -{\hat J}) $ and
 $\, _2F_1([7/12,11/12],[3/2], \,1 \, - {\hat J})
 \cdot (1 \, - {\hat J})^{1/2} $. } ($h_6$ in~\cite{Maier}). 
To sum-up $\, {\cal H}_{x}$, given by (\ref{Heunx}), has 
the following solution:
\begin{eqnarray}
\label{bingo}
&& {\cal S} \, = \, \, \Bigl( \Omega \cdot  {\cal M}_x \Bigr)^{1/12} 
\times  \, \, \,
 _2F_1 \Bigl([{{1} \over {12}}, {{5} \over {12}}];[1]; \, \, {\cal M}_x), 
\quad \, \, \, \,\,\,\, \,\hbox{where:}  \\
&&\Omega \, = \, \, 
{\frac {1}{1728}}\,{\frac { (1-4\,x)^6 
\, (1-x)^{6}} {x \cdot (1 \,+ 3\,x\, +4\,x^2)^2 \,
 (1+ 2\,x)^6}}, \nonumber \\
&&  {\cal M}_x\, = \, \,  \, 
1728\,{\frac {x \cdot (1 \, +3\,x\, +4\, x^2)^2
 \, (1\, +2\,x)^{6}
 \, (1-4\,x) ^{6}
 \, (1-x) ^{6}}{ 
 (1\, +7\,x+4\,{x}^2) ^{3} \cdot  P^{3}  }},  \nonumber \\
&& P \, = \, \,  \,1\, +237\,x\, +1455\,{x}^{2}+ 4183\,{x}^{3}
+5820\,{x}^{4}+3792\,{x}^{5} +64\,{x}^{6}. 
\nonumber 
\end{eqnarray}
Recalling (\ref{40}) and (\ref{conj}), 
the solution of the operator $\, Z_2$ in terms of hypergeometric functions
 then corresponds to the action of
the intertwinner $\, U_1$ on the solution
  $ \, {\cal S}/h(x)$
of  $\,  M_2$. The global nilpotence of $\, Z_2$ can now be understood 
from this hypergeometric function (up to a  modular invariant 
pull-back) structure.

\subsection{Atkin's modular functions}
\label{atkin}

The modulus $\,{\cal M}_x$ in the argument of $\, _2F_1$ 
actually corresponds~\cite{Maier}
 to the (genus zero) modular curve 
which amounts to multiplying, 
or dividing, the ratio of the two periods of the elliptic
curve by $\, 6$:
\begin{eqnarray}
\label{sixmodu} 
\Phi_6(j, \, j') \, = \, \, \Phi_6(j', \, j)\, = \, \, 0, 
\end{eqnarray}
obtained from the elimination of $\, z$ between:
\begin{eqnarray} 
\label{j6} 
&&j \, = \, \, j_6(z) \, = \, \, 
{\frac { \left( z+6 \right)^{3} 
\, (z^3+18\,z^2+84\,z+24)^3}
{z \left( z+9 \right)^{2} \left( z+8 \right)^3}}, \\
&&j \,  \, = \, \,j_6(\Bigl( {{2^3 \cdot 3^2 } \over { z }}  \Bigr) 
 \, = \, \, {\frac { (z+12)^{3} \, (z^3+252\,z^2
+3888\,z+15552)^3}{z^6 
\, (z+8)^{2} \, (z+9)^3}}. 
\nonumber 
\end{eqnarray}
Actually, similarly to (\ref{simplechange}), if one introduces:
\begin{eqnarray} 
\label{changer}
z \, = \, \, \,{\frac {72  \, \,\, x}{ (1-x) 
 \, (1-4\,x) }}, \quad \quad   \quad  \hbox{and:} 
\quad  \quad  \quad 
 {\cal M}_z\, = \, \,  {{12^3} \over {j_6(z)}} 
\end{eqnarray}
one finds immediately that $\, {\cal M}_x$ in (\ref{bingo}) 
is {\em nothing but the Hauptmodul} $\,{\cal M}_z$.

The singularities of the linear ODE of $\, \chi^{(3)}$ 
(or $\, \Phi^{(3)}_H$)
correspond to the singularities of $\, j(z)$:
\begin{eqnarray} 
&&z \left( z+9 \right)  \left( z+8 \right)
\, = \, \,\, \\
&&\qquad \qquad 5184\,\, {\frac {x \cdot  (1 +3\,x +4\,{x}^2) 
 \, (1+\, 2\,x)^2}
{ (1-x)^3 \, (1-4\,x)^3}} \nonumber 
\end{eqnarray}
Similarly the ``Atkin-dual''~\cite{Atkin} change 
of variables (see \ref{atkin}):
\begin{eqnarray} 
\label{dualchanger}
z' \, = \, \,{{72} \over {z}}
 \, = \, \, {\frac { (1-x)  \, (1-4\,x) }{x}},
\end{eqnarray}
gives:
\begin{eqnarray} 
&&z' \left( z'+9 \right)  \left( z'+8 \right) \, = \, \,\, \\
&&\qquad \, = \, \,\, {\frac { (1-x)  \, (1-4\,x) 
 (1+2\,x)^2 \, (1+ 3\,x\, +4\,x^2) } {x^3}} \nonumber 
\end{eqnarray}

Recalling the {\em modular Atkin-polynomial}\footnote[2]{In the case
of  modular Atkin-polynomial of degree 
one in $\, j$, like (\ref{modularAtkin}),
 one sees immediately the rational parametrization
(\ref{j6}). The introduction of the modular Atkin-polynomial becomes necessary
 when one does not have a rational parametrization
like (\ref{j6}) anymore, because the genus of the modular
 curves is no longer zero.}~\cite{Atkin} 
for (\ref{sixmodu}) (see \ref{atkin}):
\begin{eqnarray} 
\label{modularAtkin}
z \cdot (z+9)^2 \, (z+8)^3 \cdot j \,\,\, 
-(z+6)^{3} \, ({z}^{3}+18\,{z}^{2}+84\,z+24)^3, 
\end{eqnarray}
one finds out that the singularities of the linear ODE 
of $\, \chi^{(3)}$, are obtained 
from this {\em modular Atkin-polynomial}~\cite{Atkin}
 (\ref{modularAtkin})
{\em together with the covering} (\ref{simplechange}) (or (\ref{changer})). 

 These results strongly suggest that {\em all the singularities, and the
associated polynomials with integer coefficients we obtained in
previous papers (from some involved Landau singularities 
analysis}~\cite{Landau,Singul}),
{\em should have an interpretation as singularities of an (absolute)
 Klein modular invariant $\, j(N\cdot \tau)$, or equivalently of the
modular Atkin-polynomials}~\cite{Atkin},  for higher values
of $\, N$, {\em when rewritten in the $\, w$ variable.}

The difficult part, here, corresponds to 
find the well-suited covering 
(\ref{simplechange}), or (\ref{changer}), ``wrapping'' the seven
singularities in $\, \tilde{Z}_2$
onto the three singularities $\, 0$, $\, 1$, $\, \infty$ 
of the hypergeometric function $\, _2F_1$. 
Noticeably, the well-suited covering 
(\ref{simplechange}) (or (\ref{changer}))
{\em does not} correspond to a partition according 
to the critical exponents (\ref{indici})
(see section (\ref{revisitglob})), but to the following  selected
partition:
\begin{eqnarray}
\label{covering}
&&w=\, 0, \,\,\quad \, \,w=\, \infty 
\qquad \rightarrow \qquad  \, 0, \nonumber \\
&&w \, = \,-1/2, \,\,\,   \,\,\, {{-3\, \pm i\sqrt {7}} \over {8}}
\qquad \rightarrow \qquad  \, 1,  \\
&&w\,=\,\,1, \,\quad \,\, w \, = \,1/4, 
\qquad  \rightarrow \qquad  \, \infty. 
\nonumber
\end{eqnarray}
To find~\cite{Delaunay} this covering (\ref{covering}) 
among the $\, \sum_{p=2}^{6} \, {7 \choose p}
 \cdot (2^{p}-2) \, = \,$   $ \, 3^7 -3 \cdot (2^7-2) \, -3 \,\,\, = \, $
$\,\, 1806 \, $ possible ones,
one had to see the seven singularities 
on the {\em same footing}.

\vskip .3cm

\section{From $\, _2F_1$ with a pull-back to $\, _3F_2$ with a pull-back.}
\label{diffequtilde3}

\subsection{Linear differential equation for 
${\tilde \chi}_d^{(3)}(t)$.}
\label{kidiag}
For $\, \chi^{(3)}_d$, (see (\ref{d2n})),  we choose
 $x= \,t^{1/2}$ as
  our independent variable. We find 
that the linear differential operator for
  ${\tilde \chi}_d^{(3)}(x)$ is of order six, and has   
the {\em direct sum} decomposition
\begin{eqnarray}
{\cal L}_6^{(3)}\,= \,\,\, \, 
L_1^{(3)} \oplus L_2^{(3)} \oplus L_3^{(3)},
\label{ds3}
\end{eqnarray}
with
\begin{eqnarray}
L_1^{(3)}\, =\,\,\,\, Dx \,\, + \,\, {{d} \over {dx}} \ln(x-1), 
\nonumber
\end{eqnarray}
\begin{eqnarray}
L_2^{(3)} =\,\,\,
Dx^{2}\,\, +2\,{\frac { (1+2\,x)}{ \left( 1+x
 \right)  \left( x-1 \right) }} \cdot Dx
\,+{\frac {1+2\,x}{(1+x)  \, (x-1)\, x }},
\end{eqnarray}
\begin{eqnarray}
&& L_3^{(3)} =\,\,\,\, Dx^{3}\,  \nonumber  \\ 
&&\qquad \quad 
+  {{3} \over {2}}\,{\frac { \left( 8\,{x}^{6}
+36\,{x}^{5}+63\,{x}^{4}
+62\,{x}^{3}+21\,{x}^{2}-6\,x-4 \right)}{ (x+2)  \, (1+2\,x)  \, (1+x)  
\left(x-1 \right)  \, (1+x+x^2)\, x}} \cdot Dx^{2} \nonumber \\
&&\qquad \quad  +{\frac { n_1(x) }{ (x+2)  \left(1+2\,x \right)
  \left( 1+x \right)^{2} \, (1\, -x)^2 
\left(1+x+x^2 \right)\,  {x}^{2}}} \cdot  Dx \nonumber \\
&&\qquad \quad 
 -{\frac { n_0(x)}{ (x+2)  \,(1+2\,x)  \,(1-x)^3 \, (1+x+x^2) 
 \, (1+x)^2 \, x^{2}}},  
\end{eqnarray}
where:
\begin{eqnarray}
&&n_0(x) \, = \, \,
2\,{x}^{8}+8\,{x}^{7}-7\,{x}^{6}-13\,{x}^{5}
-58\,{x}^{4}-88\,{x}^{3}-52\,{x}^{2}-13\,x+5,   \nonumber \\
&&n_1(x) \, = \, \, 
14\,{x}^{8} +71\,{x}^{7}+146\,{x}^{6}+170\,{x}^{5}
 +38\,{x}^{4}  \nonumber \\
&&\quad \quad \quad \quad \quad
 -112\,{x}^{3}-94\,{x}^{2}-19\,x+2.   \nonumber
\end{eqnarray}

The linear differential operator of order two, $L_2^{(3)}$ 
is equivalent 
to the second order operator corresponding to 
the complete elliptic integral $\, E(x)$:
\begin{eqnarray}
\label{LEx}
{\cal L}_E\, = \,\, \, \, D_x^{2}\,  \,\,  + {{ D_x} \over {x}} \,\, 
 +{\frac {1}{ \left(1-x \right)  \left( 1+x \right) }}.
\end{eqnarray}

Consequently, this order-two linear differential operator
 $L_2^{(3)}$  is  globally nilpotent: actually we have
 calculated its $\, p$-curvature
and found that the corresponding characteristic
 polynomial (or minimal polynomial)
reads $\, T^2$. 

Similarly, the order-three linear differential
 operator $\, L_3^{(3)}$ is 
globally nilpotent: we have calculated its $\, p$-curvature
and found that the corresponding characteristic 
polynomial (or minimal polynomial)
reads $\, T^3$. 

\subsection{The solution of $\, L_3^{(3)}$ as a $\, \, \, _3F_2$
 with a Hauptmodul pull-back.}
\label{3F2pull}

The order-three linear differential operator $\, L_3^{(3)}$
corresponds to a  generalization
of the weight-1 modular form operators which have been discovered 
in the previous section (\ref{Z2}). 
Actually, introducing the rational function
\begin{eqnarray}
\rho(x) \, = \, \, \, {\frac { (1\, +2\,x)  \, (x+2) }
{ (1-x)  \, (1+x, +x^2) }},
\end{eqnarray}
one can find that the three
 solutions of  $\, L_3^{(3)}$
 are two MeijerG functions 
\begin{eqnarray}
&&\rho(x)
\cdot {\it MeijerG} 
\left( [[-1/2,1/3,2/3],[]],[[0,0],[0]],  \, Q \right), \\
&&\rho(x)
\cdot {\it MeijerG} 
\left( [[-1/2,1/3,2/3],[]],[[0,0,0],[]],  \, -Q \right),
\nonumber 
\end{eqnarray}
and a $\, \, _3F_2$ hypergeometric
 function\footnote[1]{Reminiscent of, for instance, 
$\, \, _2F_1([1/3, 2/3], [1];  \,  \, Q)\, $ in (23) 
of~\cite{Zudilin}.}:
\begin{eqnarray}
\label{3F2}
\rho(x)
\cdot \, _3F_2\Bigl( [1/3, 2/3, 3/2], [1, 1];  \, Q  \Bigr) 
\end{eqnarray}
where $\, Q$ is nothing but the Hauptmodul (the reciprocal
 of the Klein's invariant) 
with the (elliptic lambda function) $\, x$ changed into $\, -x$:
\begin{eqnarray}
\label{Q}
Q \,  = \, \,\, 
{\frac {27}{4}}\,{\frac { (1+x)^2\,x^2}{ ({x}^{2}+x+1)^3}}
\,  = \, \,\, \,{{ 1} \over {J(-x)}}  = \, \,\, 
{{ 12^3} \over {j(-x)}}. 
\end{eqnarray}

With (\ref{3F2}) we have a straight generalization
 of the previous weight-1 modular form 
where the expression for $\, h_6$ was given as a
 $\, _2F_1$ hypergeometric function 
with a pull-back corresponding to the Hauptmodul (\ref{changer}).
One has here a $\, _3F_2$ hypergeometric 
function\footnote[9]{Note that this hypergeometric function
is not a well-poised hypergeometric function.} 
with a pull-back corresponding to the  Hauptmodul (\ref{Q}).
If one seeks for a modular form interpretation
of the solutions of $\, L_3^{(3)}$ and, 
in particular, for (\ref{3F2}),
one would like to see $\, L_3^{(3)}$ as equivalent to the 
symmetric square of a $_2F_1$ operator with the same
Hauptmodul  pull-back (\ref{Q}). 
In this respect, it is worth
recalling the Clausen identity\footnote[5]{For such selected
arguments od the hypergeometric function we also have the highly remarkable
Gauss-Kummer quadratic 
relation: $\,_2F_1( [a,\, b], [a+b+1/2];  \, 4\, z\, (1-z))$
$\, = \,_2F_1( [2\,a,\, 2\, b], [a+b+1/2];  \, z)$. }:
\begin{eqnarray}
&&_2F_1\Bigl( [a,\, b], [a+b+1/2];  \, z \Bigr)^2
 \,\,\, = \, \, \,  \\
&&\qquad \qquad 
 _3F_2\Bigl( [2\, a,\,  a+b,\,  2\, b],
 [a+b+1/2,\, 2\, a\, + 2\, b];  \, z \Bigr), 
\nonumber
\end{eqnarray}
and in particular
\begin{eqnarray}
\label{fourth}
&&_2F_1\Bigl( [1/6,\, 1/3], [1];  \, z \Bigr)^2
 \,\,\, = \, \, \,  \\
&&\qquad \qquad 
 _3F_2\Bigl( [1/3,\,  2/3,\,  1/2], [1,\, 1];  \, z \Bigr), \nonumber
\end{eqnarray}
which is one of the four classes\footnote[2]{Four
 arithmetic triangle subgroups, 
commensurable with the full modular group, yielding 
four hypergeometric representation
of periods of elliptic curves~\cite{Chud3}.}
of $_3F_2$ found by Ramanujan~\cite{Chud} 
that are squares of  $_2F_1$ representations
of complete elliptic integrals. The symmetric square
of the second order operator
\begin{eqnarray}
\label{G1}
W_1 \, = \, \,  D_x^2\,
+{{1} \over {2}} \cdot 
{\frac { (2\,x+1)  \, ({x}^{2}+x+2) }{x
 \, (1+x)  \, ({x}^{2}+x+1) }}
\cdot D_x \, \, 
-{{3} \over {2 \, \, ({x}^{2}+x+1)^2}},
 \nonumber 
\end{eqnarray}
annihilates the left-hand side
of (\ref{fourth}) where $z$ is taken to be equal 
to the ``Hauptmodul'' (\ref{Q}). We actually found 
out that this symmetric square
is equivalent to $\, L_3^{(3)}$ with two simple order-one
 intertwinners $\, V_1$ and $\, V_2$:
\begin{eqnarray}
\label{intertw}
&&V_2 \cdot Sym^2\Bigl(W_1 \Bigr) \, = \, \, \, L_3^{(3)}
\cdot V_1, \qquad \quad \hbox{where:} \\
&&V_1\, = \, \, \, 
{\frac {x \cdot (1+x) }{ (1-x)^{2}}} \cdot D_x
\, + \, \, {{1} \over {2}}
\cdot {\frac {2\,{x}^{2}+5\,x+2}{ ({x}^{2}+x+1)  \, (1-x) }}.
\nonumber
\end{eqnarray}
Equivalently, we could have used the following 
identity\footnote[1]{Other rewritings
involve the Heun functions $\, Heun(10,23/9,1/6,1/3,3/2,1,1-Q)$ and 
$\, Heun(10,-35/18,-1/3,-1/6,1/2,1,1-Q)\cdot (1-Q)^{-1/2}$. }
 to rewrite (\ref{3F2}):
\begin{eqnarray}
&&_3F_2\Bigl( [1/3, 2/3, 3/2], [1, 1];  \, Q  \Bigr) \, = \, \, \, \, 
_2F_1\Bigl( [1/6, 1/3], [1];  \, Q  \Bigr)^2\, \\
&&\quad \quad \quad  + \, \, {{2 \, Q} \over {9}} \cdot
\,  _2F_1\Bigl( [1/6, 1/3], [1];  \, Q  \Bigr)
 \cdot \,_2F_1\Bigl( [7/6, 4/3], [2];  \, Q  \Bigr) \nonumber \\
&& \hbox{where:} \quad  \quad 
 _2F_1\Bigl( [7/6, 4/3], [2];  \, Q  \Bigr)\, = \, \, \, 
18\,   {{d} \over {dQ}} _2F_1\Bigl( [1/6, 1/3], [1];  \, Q  \Bigr). \nonumber 
\end{eqnarray}
This generalization of the Clausen identity concludes our modular form 
interpretation~\cite{Chud7,Pi} of the third 
order operator $\, L_3^{(3)}$. 
 
\vskip .2cm

\section{Global nilpotence without integral representation:
 three-choice polygons, directed compact percolation, vicious walkers, \dots}
\label{notDFG}

Let us also give here  three more examples of global nilpotence
 that {\em do not correspond
to $n$-fold integrals},  emerging from the theory of the Ising model,
or  to $n$-fold integrals of the Ising class, but to enumerative
combinatorics. Their global nilpotence, or the fact that they could be DFG
cannot be immediately seen from a representation as an $\, n$-fold 
integral of an algebraic expression\footnote[5]{Recall that the integral of
a closed differential $\, n$-form 
 over a closed $\, n$-cycle is said to come
from algebraic geometry. In our case we have 
an algebraic $\, n$-form, so we  automatically 
have a closed $\, n$-form and our integration
 on hypercubes and not cycles, can be reduced 
to cycles because the global nilpotence property~\cite{Maxim} remains
 stable~\cite{Belkale} by the required 
extensions (desingularization, relative 
de Rham cohomology, etc.)[Y. Andr\'e e-mail].}. Of
 course one can always imagine to 
see this global nilpotence as a consequence of a detailed analysis 
of the corresponding series expansions,
 showing explicitely that they are probably
arithmetic Gevrey series and in fact {\em G}-functions.

 Along this line, it is worth noticing that, for instance,
 the series expansion of the $\, \tilde{\chi}^{(n)}$ 
in the variable $\, w$ are series expansions
 with {\em integer} coefficients (see (\ref{closedn})).
The fact that these series of holonomic
 functions are series with a finite radius
 of convergence and with integer coefficients,
 gives us a strong prejudice that
we are studying 
{\em G-functions}. For the various examples 
of enumerative combinatorics displayed 
in this section, we have a similar 
strong prejudice in favour of arithmetic 
Gevrey series: 
{\em G-functions}.   
Instead of such an arithmetic approach, we prefer to
consider directly the global nilpotence, 
 performing $\, p$-curvature calculations
modulo large set of primes, since this approach 
is {\em simple, algorithmic and effective}.

\subsection{Global nilpotence without (known) integral representation:
 three-choice polygons}
\label{threechoice}
An order-eight Fuchsian linear differential operator,
 was found for the perimeter
 generating function of the three-choice\footnote[3]{Note that there 
is a factor $\, 4$ in the definition of $\, x$ between 
the ODE displayed here and the one given
 in~\cite{threechoice}.} polygons~\cite{threechoice}. It is 
the direct sum of two order-one operators and
an order-six linear differential operator $\, M_6$.

This order-six linear differential operator $\, M_6$, is the
 product of an order-three,
 an order-two and an order-one linear differential operator, respectively
$\, M_3$, $\, M_2$ and $\, M_1$: 
\begin{eqnarray}
\label{M2M3first}
&&M_6 \, = \, \, M_3 \cdot M_2 \cdot M_1, \quad 
\quad \,\, \hbox{where:}\,\, \quad \quad  
M_1 \, = \, \, D_x \, \, - {{d } \over {dx}} \ln(1-x), \nonumber \\
&& M_2 \, = \, \, \,
 D_x^2 \, \,+\, {{P_{11} }\over {(x-1) \cdot P_{12} \cdot x}} \cdot D_x \, \,
+ \, {{ P_{21}} \over {4 \cdot P_{22}}}, \qquad  
\end{eqnarray}
where the two  order-three and  order-two operators
 are given in (\ref{M2M3})
and have, beyond apparent singularities,
  singularities at $\, x\, = \, 0$ and $\, x\, = \, 1$ and 
at the roots of the quadratic polynomial $\,  16+4\,x+7\,x^2$.
Let us consider the order-two linear differential operator
$\, M_2$.  The indicial polynomials of these singularities read:
\begin{eqnarray}
&&x\, = \, \, 0, \quad \rightarrow \quad r \cdot (r+2),
\,  \quad 16+4\,x+7\,{x}^{2}\, = \, \, 0, 
\quad \rightarrow \quad r \cdot (r-1),
\nonumber \\
&&x\, = \, \, 1, \quad \rightarrow \quad (2\, r \, +3)^2,\,
  \quad \quad \, \, 
x\, = \, \, \infty, \quad \rightarrow \quad
 (2\, r \, +1) \cdot (2\, r \, -1).
\nonumber 
\end{eqnarray}
All the formal series around these singularities have logarithms,
except $\, x\, = \, \, \infty$ {\em which has
 only square roots singularities} ($\, x^{-1/2}$).
Keeping in mind that the
wronskian of $\, M_2$ is a rational function,
one deduces~\cite{JAW} that
the differential Galois group of $\, M_2$ is $\, SL(2, \, C)$.

We have calculated the $\, p$-curvatures of the 
two linear differential operators of order three and two 
of (\ref{M2M3}). We found that $\, M_3$ (resp. $\, M_2$)
is globally nilpotent, the characteristic polynomial 
of its  $\, p$-curvature 
being $\, T^3$ (resp. $\, T^2$) for almost all primes. 

\subsection{Global nilpotence without (known) integral representation:
directed compact percolation}
\label{notDFG2}

Generating functions, associated with
 the mean cluster size and length  for the 
 {\em directed compact percolation} problem, satisfy various linear ODEs
that are displayed in~\cite{directpercol}. For instance, equation (12)
of~\cite{directpercol}, when rewritten 
in a homogeneous way, corresponds to the 
order-three linear differential operator which 
is the direct sum $\, L_1 \oplus L_2$,
of an order-one operator $\, L_1$  and
 an order-two operator $\, L_2$:
\begin{eqnarray}
\label{compact}
&&L_1 \,\, = \, \,\, D_t \,\, 
+ \, {{d} \over {dt}} \ln\Bigl( {\frac { (1-2\,t) 
 \, (1-t)^{3}}{2\,{t}^{4}+2
\,{t}^{3}-6\,{t}^{2}+4\,t-1}} \Bigr), \nonumber \\
&&L_2 \,\, = \, \,\, D_t^2 \, 
+ \, 2\,{\frac {12\,{t}^{4}-12\,{t}^{3}-11\,{t}^{2}
+8\,t-1}{ (1-2\,t)  \, (t-1)\, t \, (1+4\,t-4\,t^2) }} 
  \cdot D_t \, \nonumber \\
&& \qquad \qquad -\, 6\,{\frac {4\,{t}^{4}-12\,t^3
+7\,{t}^{2}-2\,t+1}{(1-2\,t)^2
 \,(1+4\,t-4\,t^2) 
 \, (t-1)^{2} \, t }}. 
\end{eqnarray}
Similarly equations (16), (18) and (21)
of~\cite{directpercol}, when rewritten in a homogeneous
 way correspond  to the 
order-three linear differential operator which 
is the direct sum $\, L'_1 \oplus L'_2$,
of an order-one operator $\, L'_1$  and an order-two operator $\, L'_2$.
 For instance for equation (16)
of~\cite{directpercol} we have:
\begin{eqnarray}
\label{compact2}
&&L'_1 \,\, = \, \,\, D_t \,\, + \, {{d} \over {dt}} \ln\Bigl(
{\frac {{t}^{3}}{4\,t^2 \, +8\,t \, -1}}
 \Bigr), 
 \nonumber \\
&&L'_2 \,\, = \, \,\, D_t^2 \,\,  
\, +{\frac {24\,{t}^{4}-72\,{t}^{3}+46\,{t}^{2}+8\,t-7}{(1-2\,t) 
 \left( t-1 \right) \,  (1 \, +4\,t \, -4\,{t}^{2}) \, t  }} 
 \cdot D_t \, 
\nonumber \\
&&\qquad \qquad  +\, 
{\frac {8\,{t}^{4}-24\,{t}^{3}+2\,{t}^{2}+28\,t-9}{
\, (t-1)  \, (1-2\,t)  \, (1 \, +4\,t \, -4\,{t}^{2})\, t^2 }}
\end{eqnarray}
All these order-two operators ($L_2$, $\, L'_2$, \ldots)  can be seen
 to be equivalent 
and are globally nilpotent.

\subsection{Global nilpotence without (known) integral representation:
 vicious walkers}
\label{notDFG3}
Another example of enumerative combinatorics corresponds to the
 vicious walkers and friendly walkers generating 
functions~\cite{viciousfriendly}.
Equation (4.34) of~\cite{viciousfriendly},  when rewritten 
in a homogeneous way corresponds to the 
order-three linear differential operator which
 is the direct sum $\, L_1 \oplus L_2$,
of an order-one operator $\, L_1$  and an order-two operator $\, L_2$:
\begin{eqnarray}
\label{compact5}
&&L_1 \,\, = \, \,\,\, D_t \,
+ \,   {{d} \over {dt}} \ln\Bigl({\frac {{t}^{3}}{1-t+3\,t^2}}\Bigr), 
\nonumber \\
&&L_2 \,\, = \, \,\, D_t^2 \, \, 
+2\,{\frac {16\,{t}^{2}+21\,t-4}{t \left( 8\,t-1 \right)  \, 
(t+1) }} \cdot D_t \, \, 
+ \, 4\,{\frac {4\,{t}^{2}+10\,t-3}{ (8\,t \, -1) \, (t+1) \, t^2 }}. 
\end{eqnarray}
Equation (4.38) of~\cite{viciousfriendly},  when 
rewritten in a homogeneous way,  corresponds to the 
order three linear differential operator which
 is the direct sum $\, L'_1 \oplus L'_2$,
of an order-one operator $\, L'_1$  and an order-two operator $\, L'_2$:
\begin{eqnarray}
\label{compact6}
&&L'_1 \,\, = \, \,\, D_t \,
+ \, {{d} \over {dt}} \ln\Bigl(
{\frac { (t+1)\, t^4}{ (t-1)  \, 
(1-2\,t+4\,{t}^{2}) }} \Bigr), \nonumber \\
&&L'_2 \,\, = \, \,\, D_t^2 \,  \,
 + 2\,{\frac {5-35\,t+16\,{t}^{2}}
{ \left( 8\,t-1 \right)  \, 
(t-1)\,  t}} \cdot D_t \,\nonumber \\
&& \qquad \qquad  + \, 4\,{\frac {5-18\,t-41\,{t}^{2}
+10\,{t}^{3}+6\,{t}^{4}-4\,{t}^{5}}{
 \, (1-8\,t)  \, (1- t)^{2} \, (1+ t)^2 \, t^{2} }}
\end{eqnarray}
where these last two operators ($L_2$ and $\, L'_2$)
 of order two are equivalent 
and actually correspond to a Heun function\footnote[3]{Note
that the notation in~\cite{viciousfriendly,vicious}
are Snow's notations which corresponds to a change 
of sign of the second argument
in the Heun functions compared to maple's notations. Also note
  a sign misprint in~\cite{viciousfriendly} ($\, t$ 
 in~\cite{viciousfriendly} instead of $\, -t$
 in ~\cite{vicious}). Do note that  changing $\, -1/4$
 into $\, +1/4$ in the previous Heun solution,
one gets $\, Heun(-1/8, \, +1/4, \, -1, -2, \, 2, \, -2; \, -t)/t^3$ 
which is solution of a second order 
operator almost identical to $\, L_2$
in (\ref{compact5}), where $\, 4\,{t}^{2}+10\,t-3$ is replaced by 
 $\, 4\,{t}^{2}+11\,t-3$. This
 new operator {\em is not globally nilpotent}, it differs from
 $\, L_2$ by the so-called~\cite{Messing,Splendid,Accessory2,Accessory}
 ``accessory parameters''.}. A solution of 
$\, L_1 \oplus L_2$ corresponds to a generating function 
with {\em integer} coefficients namely:
\begin{eqnarray}
&& \, {{1} \over {3 \, t^3}} \cdot 
 \Bigl(Heun(-1/8, \, -1/4; \, -1, -2, \, 2, \, -2; \, -t)
 \, \, \, -(1 -t +3\, t^2) \Bigr) \, \nonumber  \\
&& \qquad \, = \, \,\, \, 1\,+2\,t\,+6\,{t}^{2}\,+22\,{t}^{3}\,+92\,{t}^{4}
\,+422\,{t}^{5}\,\, + \,\, \cdots 
\end{eqnarray}
We have calculated the $\, p$-curvature 
of the corresponding linear
 differential operators and found that they are all globally nilpotent.

A large set of other enumerative combinatorics problems for which a linear
ODE exists for the generating function, but a 
representation of that generating function 
as a multiple integral has {\em not yet} been found,
 could be listed~\cite{convex,vicious,Self,directed,perimeter}. 

\section{Beyond order two and three ODEs: staircase polygons and Calabi-Yau
type ODEs.}
\label{beyondDwork}

Krammer's counterexample to the first Dwork's 
conjecture (reduction of global nilpotence to
$\, _2F_1$ hypergeometric function up to pull-back by a rational function)
 corresponded to a selected Heun 
function. Let us recall that some  $\, _3F_2$
 hypergeometric functions for 
selected values of the parameters~\cite{3F2} like
\begin{eqnarray}
_3F_2 \Bigl([\alpha, \, \beta, \, \gamma+1], 
[\delta, \, \gamma]; \,  x  \Bigr) 
\end{eqnarray}
have a Heun representation. 
We want, now, to explore globally nilpotent  higher order linear operators 
beyond the $\, _2F_1$, Heun and  $\, _3F_2$ framework.

In \ref{diag}, the linear differential operator for 
${\tilde \chi}_d^{(4)}(t)$ is displayed and factorized in direct sums 
of an operator of order one, an operator of order
 three and an operator of order
four. The operator of order three is seen to be {\em equivalent
to a symmetric square} of a second order operator corresponding to a
hypergeometric function $\, _2F_1$.
Similarly to the situation described 
in  section (\ref{3F2pull}), 
the order-four  linear differential operator  is not obviously
reducible to a symmetric cube of a $_2F_1$ (even with a pull-back),
 it may however be equivalent to  a symmetric
 cube of a $_2F_1$. It may well be a 
$\, _4F_3$ hypergeometric function with an 
involved (Hauptmodul, \ldots) 
pull-back, but, for the moment, we have not been able to decipher this
order-four operator. Let us display, 
in the following, a few more order-four 
(and more) globally nilpotent operators, related to 
staircase polygons and Calabi-Yau manifolds. 

\subsection{Staircase polygons.}
\label{beyondDwork1}

In an enumerative combinatorics framework
other Heun functions can be found for
 the generating function of the staircase polygons~\cite{Prell,Glasser}.

The Fuchsian differential equations corresponding to 
the {\em staircase polygons} generating functions
in $\, d$ dimensions, that we denote $\, {\cal Z}_d$ 
are given in~\cite{Prell}: 
\begin{eqnarray}
\label{stairc}
&&{\cal Z}_3 \, = \, \, \, 
{ D_{x}}^{2}\, \, +{\frac { (1-20\,x+27\,{x}^{2}) }{x \cdot (1-x) 
 \, (1-9\,x) }} \cdot D_x\, \,  \, 
-3\,{\frac {1-3\,x}{x \cdot (1-x)  \, (1-9\,x) }}, \nonumber \\
&&{\cal Z}_4 \, = \, \, \, D_{x}^{3} \, \, 
+3\,{\frac { (1-30\,x+128\,{x}^{2}) }
{x \cdot (1-4\,x)  \, (1-16\,x) }} \cdot D_x^2 \nonumber \\
&&\qquad  \, \, 
+{\frac { \left( 1-68\,x+448\,{x}^{2} \right)}
{{x}^2 \cdot  (1-4\,x)  \, (1-16\,x) } } \cdot D_x \, 
\, -4\,{\frac {1}{ x^2 \cdot (1-4\,x) }}, 
\nonumber
\end{eqnarray}
and we give $\, {\cal Z}_5$ and  $\, {\cal Z}_6$ in \ref{callZ5}.

All these linear operators (up to ${\cal Z}_7$ in~\cite{Prell})
 are globally nilpotent: 
this is a simple consequence of the
fact that the generating functions 
of the staircase polygons are expressed
as $\, n$-fold integrals of (very) simple algebraic integrands, and are 
therefore DFG.
We calculated systematically the $\, p$-curvatures of all these 
Fuchsian linear differential operators (considered modulo 
 the first thousand primes) and found  that 
their characteristic polynomial is
$\, T^{d-1}$, where $\, d-1$ is the order
 of the operator $\, {\cal Z}_d$.

The first two staircase linear differential 
operators $\, {\cal Z}_3$ and
 $\, {\cal Z}_4$ actually correspond to 
Heun functions. The solution of the Heun linear differential operator
$\, {\cal Z}_3$ and its series expansion has integer coefficients:
\begin{eqnarray}
&&Heun(1/9, 1/3, 1, 1, 1, 1, x) \, = \, \,
 Heun(9,3,1,1,1,1,\, 9\, x)\, = \, \,
\nonumber \\
&& \qquad  \, = \, \, 
1+3\,x +15\,{x}^{2} +93\,x^3 +639\,x^4 +4653\,x^5 \, 
+35169\,x^6  \, + \cdots
 \nonumber
\end{eqnarray}
The  $\, Heun(1/9, 1/3, 1, 1, 1, 1, x)$ solution
 can also be written in terms of a 
modular hypergeometric function (corresponding to 
the weight-1   modular form $\, h_6$ in~\cite{Maier}):
\begin{eqnarray}
\label{z3sol}
&&Heun(1/9, 1/3, 1, 1, 1, 1, x) \, = \, \,  \\
&&\Bigl( (3\,x-1)  \cdot (243\,{x}^{3}
-243\,{x}^{2}+9\,x-1)\Bigr)^{-1/4} \cdot \, 
_2F_1([1/12, 5/12],[1], M_6)  \nonumber \\
&&\quad \hbox{with:} \quad \quad \quad M_6 \, = \, \, \, 
1728\,{\frac { (x-1) \, (9\,x-1)^3  \cdot  x^2}
{ (3\,x-1)^3 \cdot (243\,{x}^{3}-243\,{x}^{2}+9\,x -1)^3}}.
 \nonumber 
\end{eqnarray}

Joyce has shown~\cite{Joyce4} that the square of
 this Heun function\footnote[2]{See
also equation (36) in~\cite{viciousfriendly} 
for the generating function of a watermelon 
counting (union of friendly walkers).}
 is related (up to quite involved 
algebraic transformations 
of the arguments see also equations (23), (24) in~\cite{Prell})
to the simple-cubic
Green function where its DFG nature becomes clear:
\begin{eqnarray}
\label{Green}
P(z) \, = \, \, 
{{1} \over {\pi^3}} \cdot \int \,  \int \, \int_0^{\pi} \, 
{{dx_1 \, dx_2 \, dx_3} \over {
1 \, -z/3 \cdot (\cos(x_1) \, + \cos(x_2)\, + \cos(x_3)) }}.
\nonumber 
\end{eqnarray}

This Heun function $Heun(1/9, 1/3, 1, 1, 1, 1, x)$,
we now see as a weight-1 modular form,
 had been seen by Guttmann and Prellberg
 to be equal to the product of two
 complete elliptic integrals of the first 
 kind $\, K$ (up to slightly involved algebraic transformations
 namely equations\footnote[3]{Remarkably, 
 this is also true for 
 $\, Heun(1/4,1/8,1/2,1/2,1,1/2,\, \, 4\,t)^2$
 (see below) (these are equations (25) and (28) in~\cite{Prell}).} 
(25) and (28) in~\cite{Prell}). As a (curious) byproduct
one thus deduces a new 
relation between a weight-1 modular form 
and a product of two complete elliptic integrals.

The function $\, Heun(1/4,1/8,1/2,1/2,1,1/2,\, \, 4\,x)$
 corresponds to the Heun linear  differential operator:
\begin{eqnarray}
\label{A4}
A_4 \, = \, \, D_x^2 \,  
+ \Bigl(   {{1} \over {x} } \, 
+  \, {{2} \over { 4\, x -1}} \,
 +  \,{{8} \over {16\, x -1}}   \Bigr) \cdot D_x \, 
 + \,{\frac {2 \cdot (8\,x-1) }{ (4\,x-1)  \, (16\,x -1)\, x }}.
\nonumber 
\end{eqnarray}

Its series expansion has {\em integer} coefficients:
\begin{eqnarray}
&& Heun(1/4,1/8,1/2,1/2,1,1/2,\, \, 4\,x)
 \, = \, \,  \nonumber \\
&& \qquad \, = \, \, Heun(4,1/2,1/2,1/2,1,1/2,\, \, 16\, x)
 \, = \, \,  \nonumber \\
&& \qquad  \, = \, \, 
1+2\,x +12\,{x}^{2} +104\,{x}^{3}+1078\,{x}^{4}
+12348\,{x}^{5} \, +150528\,{x}^{6}\, + \cdots  \nonumber 
\end{eqnarray}
One easily verifies that the 
{\em symmetric square} of the previous Heun 
linear differential operator $\, A_4$ is {\em nothing 
but} $\, {\cal Z}_4$. Actually
 $\, Heun(1/4,1/8,1/2,1/2,1,1/2,\, \, 4\,x)^2$
 is solution of $\, {\cal Z}_4$. 

To our knowledge  this 
Heun function $\, Heun(1/4,1/8,1/2,1/2,1,1/2,\, \, 4\,x)$
does not have an interpretation as a weight-1
 modular form  (see 
also~\cite{Maier,Maier1,Maier2,Maier4}).  With these two Heun
 functions we are thus still
in a ``Dworkian'' framework (hypergeometric functions 
up to a pull-back, selected Heun functions, etc.).

The very nature of $\, {\cal Z}_5$,  $\, {\cal Z}_6$,  $\, {\cal Z}_7$,
 \dots ,  is less clear. 
In particular, it is far from clear that 
they can be written in terms of $\, _2F_1$ 
hypergeometric functions. Let us introduce some kind of
 ``multi-singular'' (beyond the four singularities of a Heun ODE)
second order linear ODE:
\begin{eqnarray}
&&z_5 \, = \, \,  D_x^2\, 
 + \Bigl( {{1} \over {x}} - {{1} \over {3 (1-x)}} 
 - {{3} \over {1 \, -9\, x}} \, 
- {{25} \over { 3\, (1 \, -25\, x)}}  \Bigr)\cdot D_x \nonumber \\
&&\qquad \quad -{{19} \over {270\, (1-x)}}
\, -{{171} \over {40\, (1\, -9\, x)}}\, 
-{{5875} \over {216\, (1- 25\, x)}}\, 
-{{49} \over {30\, x}}\nonumber \\
&& \quad \quad \quad \quad 
+{{1} \over {90\, (1-x)^2}} \, 
+{{9} \over {10\, (1\, -9\, x)^2}}
\, +{{125} \over {18\, (1\, -25\, x)^2}}
\end{eqnarray}
One finds that the explicit expression of $\, {\cal Z}_5$ is 
  ``close'' to be a symmetric cube 
of this  ``multi-singular'' (beyond the four 
singularities of a Heun ODE) linear differential operator:
\begin{eqnarray}
{\cal Z}_5 \, = \, \, Sym^3(z_5) \, + \, \, 
{\frac {1}{100}}\,{\frac {P(x)}{ (1-x)^4 \, 
(1- 9\,x)^{4} \, (1-25\,x)^4 \cdot x^3}}, \nonumber 
 \end{eqnarray}
where $\, P(x)$ is a polynomial of degree
 eleven with integer numbers.
In fact this order four operator is 
associated\footnote[1]{A. J. Guttmann, private communication.} with 
 Calabi-Yau manifolds~\cite{private} (see also next section).

Similarly, introducing the 
 ``multi-singular'' (beyond the four singularities 
of a Heun ODE) second order 
operator
\begin{eqnarray}
\label{superHeun2}
&&z_6 \, = \, \,  D_x^2\, + \Bigl( {{1} \over {x}}
 - {{1} \over { 1\,-4\, x}}  - {{4} \over {1 \, -16\, x}} \, 
- {{9} \over { 1 \, -36\, x}}  \Bigr)\cdot D_x \nonumber \\
&& \quad \quad \quad \quad 
+ {{1} \over { 5}} \cdot {\frac {p(x)}{x \cdot (1-4\,x)^2 
\left(1-16\,x \right)^2 \left(1-36\,x \right)^2}},  \\
&& \quad p(x)\, = \, \, -7+696\,x-22224\,{x}^{2}
+298816\,{x}^{3}\nonumber \\
&&\quad \quad\quad  \quad \quad \quad 
 -1603584\,{x}^{4} +3068928\,{x}^{5}, \nonumber 
\end{eqnarray}
one finds that $\, {\cal Z}_6$ is (up to some rational
functions $\, A(x)$ and $\, B(x)$) ``close'' to be a
 symmetric fourth power
of this previous 
 ``multi-singular'' (beyond the four singularities 
of a Heun ODE) operator (\ref{superHeun2}):
\begin{eqnarray}
&&{\cal Z}_6 \, = \, \,  \,  \,  \, 
Sym^4(z_6) \, + \, \, A(x) \cdot D_x \, + \, B(x). 
\end{eqnarray}

Let us recall that, even for a second order operator, 
it is not so easy, and systematic,
 to see that a solution reduces to 
a hypergeometric function $\, _2F_1$ up to a possibly 
involved rational pull-back and up to a multiplication by 
some $\, N$-th root of a rational function
 (see (\ref{alghyp2}), (\ref{bingo})).
It has to be the case~\cite{Boucher} for zero-curvature
second order operators (and 
this is an important motivation to perform our
$\, p$-curvature calculations, to extract,
 very quickly, the zero-curvature situations
like (\ref{alghyp2})), when we have a basis
 of algebraic solutions, 
but this is far from clear, at first sight, for a globally
 nilpotent operator (see (\ref{bingo})). 

When one encounters a higher order irreducible 
linear differential operators like
these $\, {\cal Z}_d$ for $\, d \ge \, 5$,
 (or other order-three irreducible 
operators like $\, M_3$ in the three-choice
 polygon problem, see (\ref{M3})), 
it remains to see\footnote[5]{To check if 
an order $\, q$ operator is exactly
a symmetric power of, say, an order-two 
operator can be done systematically (pattern matching). 
To see if  an order $\, q$ operator is equivalent (homomorphic)
 to some symmetric power, is less easy and 
systematic~\cite{Person,Nguyen3,Nguyen2,vanHoej,vanHoej2}.}
 if these irreducible operators cannot be symmetric powers 
of a smaller order operator with an emphasis on second order operators 
simply reducing, or not simply reducing, to
 $\, _2F_1$ hypergeometric functions.

With these globally nilpotent staircase operators
 $\, {\cal Z}_5$,  $\, {\cal Z}_6$,  $\, {\cal Z}_7$, \ldots
we are clearly in a win-win situation.
In a first scenario, the 
very simple and regular accumulation of singularities
 ($x\, = \, 1/n^2$, for $\, n$ even or odd)
 cannot be wrapped onto three,
or four, canonical singularities
 $\, 0$, $\, 1$, $\, a$, $\, \infty$, and 
one gets brand new examples of 
selected\footnote[3]{Do recall that, for instance, Heun operators
{\em are not generically globally nilpotent} (see (\ref{nonnilp})).} 
``multi-singular'' (beyond the four
 singularities of a Heun ODE)  globally nilpotent
 operators of arbitrary order $\, d-1$,
and fascinating algebraic geometry interpretations 
(similar to the one
Krammer gave for the Heun ODE like (\ref{Kram}), with 
Shimura curves) which remain to be found.
In a second scenario
these staircase generating functions can be expressed 
as products of complete elliptic integrals (up 
to involved algebraic transformations)
 or $_2F_1$,  $\, \,  _3F_2$, with 
some (Hauptmodul \ldots) pullback,
and this would provide a set of highly
 non-trivial effective algebraic geometry 
results associated with these $\, n$-fold integrals and 
 these enumeration of staircase polygons.  

\subsection{Calabi-Yau type ODEs.}
\label{beyondDwork2}

Other non-trivial examples of non-trivial globally
 nilpotent high order operators 
can be found with the fourth order differential equations of the 
so-called {\em Calabi-Yau type}~\cite{Calabi,Calabi2}. We
 will not give any detail 
on the construction of these new type of Calabi-Yau
 manifolds using conifold transformations 
from {\em toric Calabi-Yau hypersurfaces}. We just display 
some of these fourth order differential
operators.

A first order-four differential
operator comes from  Kontsevich's observation~\cite{Calabi} 
that two selected 
matrices for a quintic and its mirror, actually 
correspond to monodromy matrices
of the Picard-Fuchs operator:
\begin{eqnarray}
\label{Konse}
&&\theta^4 \, - \, 5^2 \, x \cdot 
(\theta\, +{{1} \over {5}}) (\theta\, +{{2} \over {5}})(\theta\,
 +{{3} \over {5}})(\theta\, +{{4} \over {5}}), \\
&&\hbox{where:} \qquad \qquad \theta \, = \, \,
 x \cdot {{d} \over {dx}} \nonumber
\end{eqnarray}
having four solutions which can be expressed in terms
 of hypergeometric functions $\, _4F_3$:
\begin{eqnarray}
_4F_3 \Bigl( 
[{{n} \over {5}}, \,{{n} \over{ 5}}, \, {{n} \over {5}},
 \,{{n} \over {5}}], \, \,
[{{n+i_1} \over {5}}, \,{{n+i_2} \over {5}}, \, {{n+i_3} \over {5}}];\, \,
 {{1} \over {5^2 \, x}} \Bigr), 
\end{eqnarray}
where $\, n \, = \, \, 1, \, 2, \, 3, \, 4$ and 
$\, i_1, \, i_2, \, i_3$ are three integers 
in the list $[1, \, 2, \, 3, \, 4]$.
Another example of order-four differential
operator which can be found in~\cite{Calabi} reads:
\begin{eqnarray}
\label{Riemann}
&&\theta^4 \, - \, \, x \cdot 
(65 \, \theta^4 \, + \, 130 \, \theta^3 \,
 + \, 105 \, \theta^2 \, + \, 40 \, \theta \, + \, 6)
\nonumber \\
&&\qquad \, \,\,
 + \, 4\, x^2  \, (4\, \theta+3)\, ( \theta+1)^2 \, ( 4\, \theta+5).
\end{eqnarray}
The critical exponents of (\ref{Riemann}) 
are given in~\cite{Calabi} in  P-Riemann function
notations. They are very simple integer or 
rational numbers ($3/4, \, 5/4$).

Associated with the diffeomorphisms~\cite{Calabi2}
  $\, X^{45}_{144,120}$ another order-four differential
operator reads:
\begin{eqnarray}
\label{X45a}
&& \theta^4 \, - \, \, 2\, x \, (102 \, \theta^4 \, + 204 \, \theta^3 \, 
+ \, 155\, \theta^2\, + \, 53 \, \theta\, +7) \nonumber \\
&& \quad \quad +\, 4  \, x^2 \, (\theta+1)^2 \cdot 
(396 \, \theta^2 \, + \, 792 \,\theta \, + 311) \, \\
&& \quad \quad
-784 \, x^3 \, (\theta+1)( \theta+2)( 2\,\theta+1)( 2\, \theta+5), 
\nonumber 
\end{eqnarray}
which, simply written in $\, x$, has a form very similar to the 
staircase operators $\, {\cal Z}_5$
given by (\ref{caZ5}) in \ref{callZ5}, namely : 
\begin{eqnarray}
\label{X45}
&&D_x^4 \, + \,  6\,{\frac {1568\,{x}^{2}-268\,x+1}
{x \, (1-4\,x) 
 \, (1-196\,x) }} \cdot D_x^3\, \nonumber \\
&& \quad\quad + 
{\frac { 7 -2962\,x\, +39260\,{x}^{2} -116816\,{x}^{3}}
{ (1-196\,x) 
 \, (1-4\,x)^2 \, x^2}} \cdot D_x^2 \, \nonumber \\
&& \quad \quad
+ {\frac {1 -1028\,x +22740\,{x}^{2} -90944\,{x}^{3}}{ (1-196\,x) 
 \, (1-4\,x)^2 \, x^3}} \cdot D_x \, \\
&& \quad \quad
 - 2\,{\frac {7+3920\,{x}^{2}-622\,x}{{x}^{3} \cdot (1-196\,x) 
 \, (1-4\,x)^2}}.\nonumber
\end{eqnarray}

Associated with the diffeomorphisms~\cite{Calabi2}
  $\, X^{51}_{200,140}$
 another order-four linear differential
operator reads:
\begin{eqnarray}
\label{X51}
&& \theta^4 \, - \, \,\, x \, (113 \, \theta^4 \, + 226 \, \theta^3 \, 
+ \, 173\, \theta^2\, + \, 60 \, \theta\, +8) \nonumber \\
&& \quad \quad  -\, 8  \, x^2 \, (\theta+1)^2 
\cdot (119 \, \theta^2 \, + \, 238 \,\theta \, + 92) \, \nonumber \\
&& \quad \quad 
-484 \, x^3 \, (\theta+1)( \theta+2)( 2\,\theta+1)( 2\, \theta+5),
\nonumber
\end{eqnarray}
which, simply written in $\, x$, has, again, a form very similar to the 
staircase operator $\, {\cal Z}_5$ given
by (\ref{caZ5}) in \ref{callZ5}, 
or to the previous order four operator (\ref{X45}), 
but with the denominators $\,  (1-196\,x) 
 \, (1-4\,x)^r $ of  (\ref{X45})
changed into  $\,(1-121\,x) \,(1+4\,x)^{r} $. 
\vskip .1cm 
We have calculated the $\, p$-curvature of
 these four order-four linear differential 
operators (\ref{Konse}), (\ref{Riemann}), (\ref{X45})
and (\ref{X51}) (modulo the first
thousand primes) and found that the characteristic
 polynomial, as well as the minimal polynomial,
read $\, T^4$, and that the Jordan block reduction 
of the $\, 4 \times 4$ $\, p$-curvature
matrix reads: 
\begin{eqnarray}
\label{Jordan}
\left[ \begin {array}{cccc} 
0&1&0&0\\
\noalign{\medskip}0&0&1&0 \\
\noalign{\medskip}0&0&0&1\\
\noalign{\medskip}0&0&0&0
\end {array} \right].
\end{eqnarray}

\section{Beyond holonomic functions: ratio of holonomic functions.}
\label{beyondholo}

In~\cite{PRE,Annals} it has been shown that the enumeration
 of three-dimensional convex polygons 
can be written as {\em ratio of holonomic functions}. 

For instance the equation
in Proposition 4.12 in~\cite{Annals}, or the equation 
just before the conclusion in~\cite{PRE},
gave the perimeter generating function
for three-dimensional oriented convex polygons as the ratio 
\begin{eqnarray}
\label{ratio}
{{ N_1 } \over {S_3(x)}}, \qquad \hbox{with:} \qquad 
 N_1 = \, \, A_1(u) \cdot S_3(x)  + \, A_2(u),
\end{eqnarray}
where $\, u$ denotes the square root of $\, x$, 
where $\, A_1(x)$ and  $\, A_2(x)$ 
denote algebraic expressions of  $\, u$
 and where $\, S_3(x)$ denotes the solution
of the previous staircase operator $\, {\cal Z}_3$ given 
by (\ref{stairc}). Such a  numerator $\, N_1$,
 is also of the type considered in section (\ref{notDFG}), namely
functions for which an integral representation has not
 yet been found. Do note that {\em ratio}
 of holonomic functions are far from
 being holonomic\footnote[8]{In contrast
the product of two holonomic functions
 is holonomic.}, as can be seen 
on the solutions of the Chazy III non-linear ODE
 (see for instance (3.37) in~\cite{Kruskal},
 and page 1878 in~\cite{ze-bo-ha-ma-05}). 
In terms of $\, x$ and not its square root $\, u$, 
the algebraic expression $\, A_2$
 is solution of a linear differential operator
of order four, $\, L_4$ (direct sum of an operator of order three
and an operator of order one). The algebraic expression
 $\, A_1$ is solution of a linear differential operator
of order five, $\, L_5$ (direct sum of an operator of order two
and three operators of order one), the
 product  $\, A_1(u) \cdot S_3(x)$ being 
 solution of a linear differential operator 
of order ten $\, L_{10}$ (symmetric product
of  $\, {\cal Z}_3$ and $\, L_5$).
The numerator $\, N_1$ is solution of an order fourteen linear 
operator that can be obtained as the LCLM of $\, L_{10}$ and  $\, L_4$.
We have calculated\footnote[5]{Note that these
 $\, p$-curvature calculations are very quickly performed, when the
 factorization (or LCLM-factorization) of such order fourteen operators
is easily reaching the limits of our 32-Gigas computer facilities.} 
the $\, p$-curvature for the {\em first two hundred primes},
 of this order fourteen operator
 (the degree of the polynomial coefficients is 320) and found that 
its characteristic polynomial is $\, T^{14}$, its
 minimal polynomial being $\, T^{2}$. The global nilpotence 
of this order-fourteen operator is a straight
 consequence of the expression of
$\, N_1$ in (\ref{ratio}). Examples like
the enumeration of three-dimensional convex polygon 
suggest to seek for new classes of solutions that are, not
 only ratio of holonomic functions, but {\em ratio of solutions 
of two globally nilpotent operators}
 ``algebraically equivalent'' in the sense
of relation (\ref{ratio}), such ratio of periods
 probably having a (modular)
 interpretation as
$\, \tau$ functions and Painlev\'e-like
 Picard-Fuchs deformations~\cite{Harnad}.

\section{Beyond global nilpotence:  linear differential operators 
with irregular singularities}
\label{beyondglob}

We have encountered in previous publications~\cite{Holo}
$\, n$-fold integrals annihilated by (minimal)
 linear differential operators which are obviously globally nilpotent,
namely the two-point correlation functions 
and the form factors of the off-critical lattice Ising model.
The  linear differential operators of the form factors have a
nice ``russian-doll'' structure
 (see  (\ref{F7531})). The linear 
differential operators $\, F_{j}(N)$ 
occurring~\cite{Holo} in these factorised russian-doll form 
were seen to be equivalent 
to symmetric powers 
of the second order linear differential $\, L_E$ corresponding 
to the complete elliptic integral of the second kind $\, E$. 
Consequently, they are obviously globally nipotent.
The {\em scaling limit} of these linear
 differential operators also exhibit a 
``russian-doll'' structure, but they are not globally nilpotent.

The scaling limit of the $ \, f^{(n)}_{N,N}$'s amounts, on the functions,
and on the corresponding differential operators, to
 taking the limit $\, N \, \rightarrow \, \infty$
and $\, t \, \rightarrow \ 1$, keeping the limit 
$ \, x \, = \, \, N\cdot (1-t)$ finite, or in other words, to
 performing the change of variables  $\, t=1-x/N$,
keeping only the leading term in $\, N$. 
Performing these straightforward calculations, the linear
differential operators in $\, t$
for the $ \, f^{(n)}_{N,N}$'s where $\, N$ was a parameter,
 become linear differential operators in the  
scaling variable $\, x$. 

Calling $\, F^{scal}_j$ the scaling
 limit of the operator $\, F_{j}(N)$ we
found~\cite{Holo} for $j$ odd, that
\begin{eqnarray}
\label{f7scal}
&&F_{1}^{scal}  \, = \, \,   L_2^{scal}, \nonumber \\
&&F_{3}^{scal}  \, = \, \,  L_4^{scal} \cdot  L_2^{scal},  
\nonumber \\
&&F_{5}^{scal}  \, = \, \,L_6^{scal} \cdot
  L_4^{scal} \cdot  L_2^{scal},  \quad   \quad  \cdots \nonumber
\end{eqnarray}
where  
\begin{eqnarray}
&&L_4^{scal}  \, = \, \,16\,x^4\,  D_x^4 \, +96\,x^3\,  D_x^3
+ 40\, \left( 2-x^2 \right)\,  x^2\,  D_x^2 \nonumber \\
&&\quad \quad \quad \quad +8\, ({x}^{2}-2) \,x \,  D_x\,\, 
+9\,x^{4} -8\,{x}^{2}+16,\nonumber  \\
&&L_2^{scal}  \, = \, \, \,
4\,x \cdot   D_x^{2}\,\, +4\, D_x \, \, -x, 
\end{eqnarray}
and $\, L_{10}^{scal} $,  $\, L_8^{scal} $,  $\, L_6^{scal} $ 
are given in~\cite{Holo}. Similar relations 
occur for $\, j$ even~\cite{Holo}.
Thus, we see that the scaled operators 
$\, F_j^{scal}$ have a ``russian doll''
structure inherited from the lattice operators $F_j(N)$.

Consider the linear differential operator
 corresponding to the modified
Bessel function $Bessel(n,x/2)$ for $\, n=0$, namely:
\begin{eqnarray}
\label{modBessel}
 B\, = \, \, \,\, D_x^2\,\, +{\frac {D_x}{x}}\,\, - {{1} \over {4}}.
\end{eqnarray}
We recognize, in this linear differential
 operator, the exact identification with
the scaled differential operator $F_1^{scal}=L_2^{scal}$. 
We find  that the symmetric square of the linear
 differential operator $\, B$, and the scaled 
operator $\, L_3^{scal} $ {\em are equivalent},
 the symmetric third power of the
 linear differential operator $\, B$,
 and  the scaled operator $\, L_4^{scal} $ are equivalent, 
and, more generally, the symmetric $j$-th power  of
 (\ref{modBessel}) and the scaled operator
 $\, L_{j+1}^{scal} $ {\em are equivalent}, 
 $\,L_{j+1}^{scal} \,\, \, \simeq \,\, \,\,  \,  {\rm Sym}^j(B)$. 
 
Global nilpotence implies fuchsianity. The scaling limit 
generates a {\em confluence
 of the regular singular points}~\cite{Ramis,Ramis2} 
we had on the lattice, yielding  linear differential operators,
 which are {\em not} Fuchsian anymore 
because of an {\em irregular
singular point at infinity}: we are leaving 
the universe of {\em G}-functions
for the universe of ``Hamburger'' functions~\cite{Hamburg}. 

 Let us explore, however, 
the $\, p$-curvatures of the previous {\em non-Fuchsian}
linear differential operators 
which correspond to scaling limits of 
globally nilpotent linear differential operators.
The calculations give, modulo the prime $\, p$, 
the following  characteristic 
polynomial for  $\, L_2^{scal}$:
\begin{eqnarray}
 \Bigl(T + {{p-1} \over {2}}\Bigr)\cdot 
\Bigl(T + {{p+1} \over {2}}\Bigr) 
\end{eqnarray}
and the following  characteristic 
polynomial
for  $\, L_4^{scal}$:
\begin{eqnarray}
\Bigl(T + {{p-3} \over {2}}\Bigr) \cdot
 \Bigl(T + {{p-1} \over {2}}\Bigr)\cdot 
\Bigl(T + {{p+1} \over {2}}\Bigr) \cdot
 \Bigl(T + {{p+3} \over {2}} \Bigr).
\end{eqnarray}
We have also calculated the $\, p$-curvatures 
of  $\, L_3^{scal}$  and 
the corresponding characteristic 
polynomials. For almost all primes, 
 this  characteristic 
polynomial has a very simple expression one can see 
as a {\em simple deformation} of the characteristic 
polynomial of a nilpotent operator:   $\,  T^3 \, - T\, =$
$ \, \, (T-1)\cdot T \cdot (T+1)$.
Similar calculations performed for $\, L_5^{scal}$, 
$\, L_6^{scal}$,  $\, L_7^{scal}$,
up to $\, L_{10}^{scal}$ give the following results for the
characteristic 
polynomial of the corresponding $\, p$-curvature:
\begin{eqnarray}
&& \Bigl(T - {{n-1 } \over {2}}\Bigr)\,\, \cdots \,\,
\Bigl(T - 2\Bigr)\, \,\cdots \, \,  \Bigl(T - 1\Bigr) 
 \times \\
&& \qquad T  \cdot \Bigl(T + 1\Bigr)\,\, \cdots \,\,
\Bigl(T + 2\Bigr)  \,\,
\cdots \,\, \Bigl(T + {{n-1 } \over {2}}\Bigr) \nonumber 
\end{eqnarray}
for  $\, L_n^{scal}$ with $\, n$ odd
and 
\begin{eqnarray}
\prod_{i \, =\, 1\, -n/2 }^{i \, =\,n/2 }
 \Bigl(T + {{p-1} \over {2}} \, +i\Bigr),
\end{eqnarray}
for  $\,  L_n^{scal}$ with $\, n$ even.
All these calculations have been performed 
for all the primes $\, p\, < \, 100$. 

A remarkable structure ``beyond global nilpotence'' 
clearly remains to be discovered by mathematicians
for the functions of ``Hamburger\footnote[2]{To sum up 
quite brutally the situation, one 
may say~\cite{Andre3,Andre4} that
 almost all the special functions occurring
 in theoretical physics
 are either $\, G$-functions for Fuchsian ODEs
or ``Hamburger'' functions~\cite{Hamburg}
 when an irregular singularity occurs.} type'' 
(one irregular singular point at infinity)
that occur in field theory, or, more simply, 
in the scaling limit of DFG holonomic functions
of lattice problems.

\section{Conclusion}
\label{concl}

One can probably conjecture that when the generating functions of the 
 various problems of enumerative combinatorics
 are found to be solutions
of Fuchsian ODEs, quite systematically the
 corresponding linear differential operators are
globally nilpotent, these holonomic functions
 being ``DFG'', their rewriting in terms
of $\, n$-fold integrals being just a question of time,
 work and/or stamina. The
generating function of the perimeter 
three-choice polygon, of the directed compact percolation
or of the vicious walkers are such examples.
 In this paper, we have studied a quite
 large number of $\, n$-fold integrals 
of algebraic integrands and their
 corresponding Fuchsian ODEs. In particular, 
we looked at the $\, p$-curvatures of
 their factors, not to see if these linear
differential factors were globally 
nilpotent\footnote[5]{This question has been
solved by mathematicians: these $\, n$-fold integrals
are holonomic functions with rational critical exponents,
and are even DFG.}, but to understand
{\em how} these differential factors 
``succeed'' to be globally nilpotent. One must keep in mind
that the Fuchsian ODEs for the $\, n$-fold integrals
of theoretical physics 
(Feynman diagrams~\cite{Kreimer2000},  \ldots) are generically
 of quite large orders  (as an example the minimal order ODE 
for $\chi^{(5)}$  is of order 33, see~\cite{Experimental}).
Since the corresponding  minimal order ODEs are 
necessarily globally nilpotent because 
they are DFG, the question one can ask is
{\em how} an order $23$, $33$, $50$ linear differential
operator succeeds to be globally nilpotent?  Throughout all the
examples displayed in this paper we have seen that the 
(minimal order) linear differential operators of quite large order
actually factorise into products (sometimes direct 
sums and products) of linear  differential operators
of smaller orders (one, two, sometimes three 
and four). The global nilpotence
of the order-one operators just corresponds to wronskians that
are $N$-th roots of rational functions, most of the 
order-two linear differential operators
and, for instance, the order-three and four operators occurring in the
factorizations of $\chi^{(3)}$
and  $\chi^{(4)}$, having a typically ``Dworkian'' interpretation
since they correspond to either
the second order linear differential operator
associated with the complete elliptic integrals 
(of the first or second kind),
or equivalently to $_2F_1$ hypergeometric function or 
to symmetric powers (square and cube) 
of these second order linear differential operators.
More remarkably other second order linear differential operators
were found to correspond not only to globally nilpotent 
operators but to zero curvature operators. The solutions are  algebraic 
functions corresponding to selected algebraic curves. 
For instance, we encountered {\em genus six and genus five} algebraic curves
(for $\, \Phi_D^{(5)}$ and $\, \Phi_D^{(6)}$ respectively),
their roots being expressed as complete elliptic integrals
of the third kind with a ``characteristic'' 
corresponding to {\em genus three} curves.
As far as algebraic curves are concerned, we  also gave many examples of 
zero curvature  linear differential operators of different orders 
corresponding exactly to  algebraic 
functions associated to {\em modular curves}, namely 
the $\, \lambda$-extensions $\, C(N, \, N;\,  \lambda)$
for some selected values~\cite{Holo} of  $\, \lambda$. 
From a ``Dworkian viewpoint'' let us recall that these algebraic 
functions  can be expressed as pull-backs of 
 $_2F_1$ hypergeometric functions. We also gave many examples of 
zero curvature  linear differential operators of different orders 
corresponding exactly to  algebraic 
functions associated to {\em modular curves}. 
As far as second order linear differential operators are concerned,
the most spectacular example came from the
 linear differential operator $\, Z_2$
occurring in the factorization of the  linear differential operator
for $\, \chi^{(3)}$.  The global nilpotence of that operator $\, Z_2$
was seen to correspond to a highly non-trivial pull-back of 
 $_2F_1$ hypergeometric function, namely 
the {\em weight-$\, 1$ modular forms} $\, h_6$ !
Most of these examples correspond to $n$-fold integrals 
 associated with the Ising models or more generally
$n$-fold integrals of the so-called Ising class. 

Among all the globally nilpotent operators we displayed in
this paper, other $n$-fold examples came from enumerative combinatorics
and others from Picard-Fuchs (Gauss-Manin connection~\cite{man})
constructions on (mirror) Calabi-Yau hypersurfaces with 
conifold singularities. In the first case (enumerative combinatorics)
only two examples were clearly ``Dworkian'' the $n$-fold integral 
 being expressed alternatively as a Heun function or as a pull-back of 
 a $_2F_1$ hypergeometric function. All the other examples
seem to go beyond a strict  hypergeometric ``Dworkian'' framework. 
One seems to explore, similarly to Krammer counterexamples,
Heun functions that cannot be reduced to
  $_2F_1$ hypergeometric functions,
and, more generally, holonomic functions corresponding to
linear differential 
operators of order two, three, four, \dots ,  with, at first sight,
 {\em many more singularities} than $0$, $\, 1$, and $\, \infty$. 

The details and the richness of the situations, and structures,
encountered with our physical examples were certainly not
obviously expected from the DFG diagnostic.   

Understanding {\em how} linear differential
operators are globally nilpotent led us to discover different 
structures on various algebraic varieties
(elliptic curves and complete elliptic integrals,
algebraic curves that are
 {\em modular curves, weight-$1$ modular forms}, \ldots) 
that provide a deeper understanding of the
underlying mathematical structures ``hidden'' in the 
physics problems we study. When we see that,
 to sum up things brutally, the
global nilpotence of the linear differential operator for $\chi^{(3)}$
is inherited from the global nilpotence of $\, Z_2$
which corresponds to the weight-$1$ modular forms $\, h_6$, one 
understands the ``complexity'' of the holonomic
 function $\chi^{(3)}$, totally, and utterly, differently. 
An interesting generalization of the previous weight-$1$ modular form
was found with $\tilde{\chi}_d^{(3)}$ and equation (\ref{3F2}),
with the occurence of a $\, _3F_2$ hypergeometric 
function with a {\em Hauptmodul pull-back}. 
This reinforces
the viewpoint we tried to promote that a (serious) theory of the
Ising model embed all the theory of elliptic curves (modular curves, 
modular forms,  \ldots). The last Calabi-Yau examples confirm
 that viewpoint: discovering the underlying algebraic varieties 
(or projective spaces minus some singular 
sets~\cite{Sakai1,Sakai2}), is fundamental. 
What are these algebraic varieties curves, surfaces,  and higher 
dimensional algebraic varieties? Going a bit further beyond the simple
mantra ``it is derived from geometry'', we tried,
and often succeeded, to find {\em explicitly} the 
 structures of effective algebraic geometry that are the 
``deus ex machina''  of our theoretical physics problems. 

One can now propose the following systematic, and quite algorithmic, 
study of every $\, n$-fold 
integral of algebraic expression encountered in theoretical physics:
 first generate large series expansions of these  $\, n$-fold 
integrals to find out the linear differential operators that 
annihilate these series, then get the minimal 
order differential operators,
then factorize and LCLM-factorise, as much as possible,
these minimal order differential 
operators, then examine the irreducible factors to 
see if they are not equivalent to 
symmetric powers of smaller order operators, 
then calculate the corresponding
 $\, p$-curvatures\footnote[9]{For operators of 
quite large order and degree of the polynomial coefficients, 
the $\, p$-curvature calculations
are much quicker than the
 factorization and LCLM-factorisation.}
of  all these smaller order irreducible operators
to see if they have zero curvature, or if they are globally nilpotent,
and, finally, examine all these smaller order irreducible operators to 
find out if they correspond to $_2F_1$, $_3F_2$, $_4F_3$,  \ldots 
hypergeometric functions up to a rational pull-back.

Saying that theoretical physics should eventually reduce
 to classification of
 singular varieties~\cite{Sakai1,Sakai2},
thus reducing most of it to effective algebraic geometry, 
is certainly a too drastic simplification. With section (\ref{beyondglob}) 
we see that some nice generalisation
of the notion of global nilpotence does exist, and needs to be explored,
 for the Hamburger functions~\cite{Hamburg} that typically occur in physics
(linear ODEs with one irregular singularity, 
for instance at $\, \infty$).  Let us just say that
an effective algebraic geometry viewpoint of lattice statistical
mechanics, enumerative combinatorics, particle physics,
solid state physics, theoretical physics, hopefully yielding the emergence
of a new ``Algebraic Statistical Mechanics'' is certainly a step in 
the good direction.

\vskip .1cm 

\ack
We would like to thank Y. Andr\'e and F. Beukers 
for many illuminating explanations.
We would like to thank A.J. Guttmann for providing 
many useful references and precious comments.  
We would like to thank I. Jensen for useful comments. 
We would like to thank  M. Rybowicz for providing some 
closed formulae for the first $\, \Phi^{(n)}_D$'s.
The first author was supported in part by an ANR ``Gecko''
and by the Microsoft Research-INRIA Joint Centre.
Part of the computations described in this article have been
performed on the server venus[at]unilim.fr.
This work is partially supported by a PICS/CRNS grant.
As far as physicists authors of this paper are concerned,
this work has been performed without any ANR or ERC or Egide support.

\vskip .1cm

\vskip .1cm 

\appendix

\vskip .1cm 

\section{Factorisations of multiple integrals linked to $\, \zeta(3)$}
\label{fin}

In Ap\'ery's proof of the irrationality 
of $\, \zeta(3)$ a crucial role is played
 by the linear differential operator~\cite{Beukers}: 
\begin{eqnarray}
\label{Apery}
&& (t^2 \, -34\, t \, +1)\cdot t^2 \cdot  D_t^3 \, 
+ \, (6\, t^2 \, -153\, t \, +3)\cdot t \cdot  D_t^2   \\
&& \qquad \qquad  \qquad \qquad \, 
+ \, (7\, t^2 -112\, t \, +1) \cdot   D_t 
\, \,  \,  + (t-5),  \nonumber 
\end{eqnarray}
this operator being linked to the modularity of the algebraic variety:
\begin{eqnarray}
 x\, +\,{{1} \over {x}}\,   +\,y\, +\,{{1} \over {y}}\,  
 +\,z\, +\,{{1} \over {z}}\,   +\,w \, +\,{{1} \over {w}} \,= \, \, \, 0.  
\nonumber
\end{eqnarray}
Operator (\ref{Apery})
 is, in fact, the symmetric square of the 
second order operator~\cite{Frits}:
\begin{eqnarray}
\label{Apery21}
&&  4 \cdot (t^2 \, -34\, t \, +1) \cdot t \cdot  D_t^2 \,   \\
&& \qquad \qquad  \quad \quad \, 
+ \, 4 \cdot (2\, t^2 \, -51\, t \, +1) \cdot  D_t  
+ \, (t \, -10),  \nonumber 
\end{eqnarray}
If one introduces the change of variable
\begin{eqnarray}
t \,\,  = \, \, \, \, {{ x \cdot (1\, - \,9 \, x)} \over {1-x}},
\end{eqnarray}
the second order operator (\ref{Apery21}) becomes
\begin{eqnarray}
\label{Apery22}
 D_x^2 \,  \,  \,
+ \, {{1 \, -18\, x } \over { (1 \, -9\, x ) \, x}} \cdot  D_x    \,  \, 
- \,  {{ 1} \over {4}} \cdot   {{ 10 \, -11\, x \, + \, 9 \, x^2 }
 \over { (1 \, -9\, x ) (1 \, -\, x )^2\, x}},  
\end{eqnarray}
Considering the solution up to a multiplicative
square root  $\, (1-x)^{1/2}$, amounts to perform
the symmetric product of this operator (\ref{Apery22}) with
the order-one operator 
$ \, D_x \, +1/2/(x-1)$ and transforms (\ref{Apery22}) into the 
second order operator $L_x$:
\begin{eqnarray}
\label{Apery23}
  D_x^2 \,  \,
+ \, {{1 \, -20\, x \, +27 \, x^2} \over {
 (1 \, -9\, x ) \, (1 \, -\, x )\, x}} \cdot  D_x \, 
- \,  3 \cdot   {{ 1 \, -3\, x  }
 \over { (1 \, -9\, x ) (1 \, -\, x ) \, x}},  \nonumber 
\end{eqnarray}
which is nothing but  
$\, {\cal Z}_3$
(see section (\ref{beyondDwork1}),
 or equivalent to the operators $Z_2$ and  $M_2$ of $\, \Phi^{(3)}_H$)
that occurred in the staircase polygons, and has 
the Heun function solution
 $\, (1-x)^{1/2} \cdot Heun(1/9, \, 1/3; 1, \, 1, \, 1, \, 1; \, x)$
that already occured in $Z_2$ and  $M_2$. Again 
we have here~\cite{Frits} the occurence 
of a Picard-Fuchs equation\footnote[5]{Similarly~\cite{Coster} one also
has Gauss-Manin systems of Shimura families of abelian surfaces having
multiplication by a quaternion algebra
 over $\mathbb{Q}$.} for the modular
family of elliptic curves associated to $\, \Gamma_1(6)$,
that is the weight-1 modular form $ \, h_6$.  

In~\cite{Singul} we also obtained an order-four Fuchsian linear
differential equation (also related to the analyzis of
$\, \zeta(3)$)  which factorizes\footnote[2]{Note
 a misprint in~\cite{Singul} one should read
  $\ln A_i$, instead of  $\, A_i$,
in the equations defining the $\, A_i$ after equation
 (H.2) in~\cite{Kiev,Landau}.} 
in four order-one differential operators 
($Dx$ denotes $d/dx$):
\begin{eqnarray}
\label{operat}
&&L_n\, =\, \,\,\,Dx^4 \,\,
+\frac{2 \, (3\, x\, -1)}{(x-1)\, x} \cdot Dx^3 \, \, 
\nonumber \\ 
&&\qquad 
+\frac{\left(7\, x^2\, +(n^2+n-5)\, x\, 
-2\,n\,(n+1)\right)}{(x-1)^2 \,x^2 } \cdot Dx^2
\nonumber \\ 
&&\qquad + \frac{\left(x^2\,+2\,n\,(n+1)\right)}
{ (x-1)^2 \,x^3 } \cdot Dx \, \, \\
&& \qquad   +\frac{n \cdot  (n+1) \cdot 
\left((n^2+n+1)\,x\,+(n-1)\,(n+2)\right)}{(x-1)^2\, x^4}  \,
 \nonumber \\
&& \qquad \, = \, \, \, 
\Bigl(Dx \, + \, {{d \ln(A_1)} \over {dx}}  \Bigr) 
\cdot \Bigl(Dx \, + \, {{d \ln(A_2)} \over {dx}}  \Bigr) 
 \nonumber \\
&& \qquad \qquad 
\times  \Bigl(Dx \, + \, {{d \ln(A_3)} \over {dx}}  \Bigr) 
\cdot \Bigl(Dx \, + \, {{d \ln(A_4)} \over {dx}}  \Bigr),
 \nonumber
\end{eqnarray}
These order-one linear differential operators 
{\em have rational solutions}. Such factorization into  
 order-one linear differential operators with rational solutions
is characteristic of factorization of globally nilpotent operators 
when they are considered modulo a prime. Here, remarkably, such a
 factorization takes place for the
 exact operators over $\mathbb{Q}$
and {\em not only mod prime} !
The Jordan block reduction of the 
$\, p$-curvature of the operator (\ref{Apery})
read respectively:
\begin{eqnarray}
\label{Jordan2}
\left[ \begin {array}{ccc} 
0&1&0\\
\noalign{\medskip}0&0&1 \\
\noalign{\medskip}0&0&0
\end {array} \right],
\end{eqnarray}
when it gives matrix (\ref{Jordan}) for (\ref{operat}) 
for any integer $\, n$. For any integer $\, n$ the characteristic 
polynomial, as well as the minimal
polynomial of the $\, p$-curvature of (\ref{operat}) 
reads $\, T^4$. For $\, n$ a non integer parameter
the characteristic polynomial reads
\begin{eqnarray}
&& {T}^{4}\, 
+ \, (x-2)^p  \cdot U \cdot {T}^{2}\, 
+ \, \Bigl((x+1)\, (x-1)^2\Bigr)^p \cdot U^2, \\
&& \qquad  \hbox{with:} \qquad \quad  \quad U \, = \, \,
  A_n^2 \cdot 
 \Bigl( {{ (x-1)^2 \, x^6 } \over { (n-1)^4 \, n^8 }}\Bigr)^p,
 \qquad \quad \quad \hbox{where:}  \nonumber \\
&& \qquad A_n \, = \, \,\, 
(n-1) \cdot n \cdot (n+1) \cdot  (n+2) \, \cdots \, (n+p-2),
\nonumber
\end{eqnarray}
where $\, A_n$, as it should, vanishes modulo the prime $\, p$,
when $\, n$ is an integer.

\section{Display of miscellaneous Fuchsian linear operators of the paper.}
\label{display}

\subsection{Operators $\, M_2$ (resp.  $\, M_3$) for
 three-choice polygons}
\label{threechoibis}
The order-two (resp. three) operator $\, M_2$ (resp.  $\, M_3$)
occurring in the factorisation (\ref{M2M3first})  
of the order-six linear differential operator $\, M_6$  associated
with the three-choice polygons generating 
function (see section (\ref{threechoice}))
reads
\begin{eqnarray}
\label{M2M3}
&& M_2 \, = \, \, \,
 D_x^2 \, \,+\, {{P_{11} }\over { (x-1) \cdot P_{12} \cdot x}} \cdot D_x \, \,
+ \, {{ P_{21}} \over {4 \cdot P_{22}}}, \qquad  
\end{eqnarray}
with:
\begin{eqnarray}
&&P_{11} \, = \,\,  3437\,{x}^{7}-1341\,{x}^{6}+4188\,{x}^{5}
-24160\,{x}^{4}+38400\,{x}^{3} \nonumber \\
&&\qquad +10752\,{x}^{2}-34816\,x+12288, \nonumber \\
&&P_{12} \, = \, \,  \, 3437\,{x}^{6}-5826\,{x}^{5}
+5280\,{x}^{4}-7360\,{x}^{3}\nonumber \\
&& \qquad +7680\,{x}^{2}+3072\,x-4096, \nonumber \\
&&P_{21} \, = \, 24059\,{x}^{9}+35756\,{x}^{8}\, 
+116792\,{x}^{7}-480784\,{x}^{6}
+693824\,{x}^{5} \nonumber \\
&&\qquad -2361856\,{x}^{4}+2886656\,{x}^{3}-3739648\,{x}^{2}
+3670016\,x-1376256,\nonumber \\
&&P_{22} \, = \, \,
 -(16+4\,x+7\,x^2)  \cdot (1-x)^2 \cdot  x \cdot P_{12},  \nonumber 
\end{eqnarray}
and:
\begin{eqnarray}
\label{M3}
&&M_3 \, \,= \, \,\, P(x) \cdot D_x^3 \,\, + \, \cdots 
\end{eqnarray}
where $\, P(x)$ denotes the (head) polynomial:
\begin{eqnarray}
&&P(x) \, = \, \, 4\, x^3 \, (1+x)  \, (1-x)^{2}
 \, (4+x^2) \nonumber \\
&& \quad \quad \quad \times \,(16+4\,x+7\,x^2)
\cdot P_{12}(x)^3 \cdot Q_(x) ,\nonumber \\
&&Q_(x) \, = \, \,116620\,{x}^{12} \, 
-39739\,{x}^{11}+2816770\,{x}^{10}
-4827228\,{x}^{9} \nonumber \\
&&\qquad -5350720\,{x}^{8}+12343408\,{x}^{7}
+473056\,{x}^{6}-13436096\,{x}^{5}
\nonumber \\
&&\qquad +9007872\,{x}^{4}+1064960\,{x}^{3} \,
 -1421312\,{x}^{2}-327680\,x-65536.
\nonumber
\end{eqnarray}

\subsection{The order-two operator $\, M_2$ in $\, L_{\Phi^{(3)}_H}$}
  \label{m2}
The order-two operator $\, M_2$ occurring in the factorization 
$L_{\Phi^{(3)}_H} \, = \,  \, \,  M_3 \cdot M_2$ reads:
\begin{eqnarray}
\label{Phi3Hrevi}
&&p_2 \cdot M_2 \, =\, \, \, 
p_2 \cdot Dx^2 \, + \, (x-1) \cdot p_1 \cdot Dx\, + \, 3 \cdot p_0, 
 \nonumber 
\end{eqnarray}
with:
\begin{eqnarray}
&&p_2 \, =\, \, \, (x-4)  \, (x-2)  \, (x-1)^{2} \cdot 
 (2+x)  \, (4+3\,x+x^2) \cdot x,  \nonumber \\
&&p_1 \, =\, \, \,
\, -64+128\,x+196\,{x}^{2}+20\,{x}^{3}-57\,{x}^{4}
-14\,{x}^{5}+7\,{x}^{6}, \nonumber \\
&&p_0 \, =\, \, \,33\,{x}^{3}-16-20\,x+44\,{x}^{2}
-11\,{x}^{4}-9\,{x}^{5}+3\,{x}^{6}, 
\nonumber 
\end{eqnarray}
and:
\begin{eqnarray}
&& M_3 \, = \, \, q_3 \cdot Dx^3 \, + \, \cdots \nonumber \\
&& q_3 \, =\, \, \, \, (x-2)^2 \, (x-1) \cdot x^{2} \cdot (1+x)^2
\cdot p_2 \cdot r_3 ,  
\qquad \qquad \hbox{where:} \nonumber \\
&& r_3 \, =\, \, \, 1280-1344\,x-6848\,x^2
-21456\,{x}^{3}+82416\,{x}^{4}
+74876\,{x}^{5}\nonumber \\
&&\qquad -44684\,{x}^{6}-48873\,{x}^{7}+32112\,{x}^{8}
+25252\,{x}^{9}+1728\,{x}^{10}
\nonumber \\
&&\qquad -1918\,{x}^{11} +648\,{x}^{12}
 +120\,{x}^{13}+4\,{x}^{14}-{x}^{15}.\nonumber
\end{eqnarray}
Note that $\, x$ here is actually $\, 4 \, w$. The 
set of singularities have a 
$\, w \, \leftrightarrow \, 1/4/w$ covariance, that is a
$\, x \, \leftrightarrow \, 4/x$ 
 covariance:
\begin{eqnarray}
p_2\Bigl({{4} \over {x}}\Bigr) \, = \, \,  \,  \, 
 256\cdot {\frac {4-x}{ (1-x) \cdot {x}^{9}}}  \cdot  p_2(x)
 \nonumber 
\end{eqnarray}

\subsection{The order-four operator $\, M_4$ in  $\, L_{\Phi^{(4)}_H}$}
  \label{m4}
The order-four operator $\, M_4$ occurring in the factorization 
$L_{\Phi^{(4)}_H} \, = \,  \, \, 
M_4 \cdot K_1 \cdot Z_1$ reads:
\begin{eqnarray}
&&M_4  \, = \, \,  \, q_4 \cdot D_x^4 \, +  q_3 \cdot D_x^3 \, 
+ q_2 \cdot D_x^2 \, + q_1 \cdot D_x \, + q_0, 
\qquad \quad  \hbox{with:}
\nonumber \\
&&q_4 \, = \, \, 16\, (4-x)  \, (1-x)^{4} \, 
{x}^{4} \cdot Q_4, \nonumber \\
&&Q_4 \, = \, \,  -128-2233\,x+2847\,{x}^{2}-3143\,{x}^{3}
+3601\,{x}^{4}-144\,{x}^{5}+64\,{x}^{6} ,
\nonumber \\
&&q_3 \, = \, \,32\, x^3 \, (1-x)^{3} \cdot Q_3, \nonumber \\
&&Q_3 \, = \, \,768\, x^8 -4712\,{x}^{7}+54621\,x^6
-226585\,{x}^{5}+271255\,x^4 \nonumber \\
&& \qquad \quad  -253247\,{x}^{3}+190228\,x^2\, 
-45848\,x-3328, \nonumber \\
&&q_2 \, = \, \,-8\, x^2 \, (1-x)^2 \cdot Q_2, \nonumber \\
&&Q_2 \, = \, \,
  23360\, x^9 -140752\,{x}^{8}
+1814065\,{x}^{7}-7479930\,{x}^{6}+10944040\,{x}^{5}\nonumber \\
&& \quad \quad -11262286\,{x}^{4}
+9445431\,{x}^{3}-4048776\,{x}^{2}+419280\,x+47104, \nonumber \\
&&q_1 \, = \, \,8\,x \, (1-x) \cdot Q_1, \nonumber \\
&&Q_1 \, = \, \,  64640\,{x}^{10}-382600\,{x}^{9}
+5520835\,{x}^{8}-22754401\, x^7\nonumber \\
&& \quad \quad  +38212402\,{x}^{6} -43444138\,x^5
+39867319\,{x}^{4}-22329197\,{x}^{3}\nonumber \\
&& \quad \quad +5075028\,x^2 -21248\,x -47104, \nonumber \\
&&q_0 \, = \, \,
-65536+1444096\,x+4876704\,{x}^{2}
-79483588\,{x}^{3}+250389985\,{x}^{4}\nonumber \\
&&\quad \quad +382946518\,x^6 -307163242\,x^7
-380545497\,{x}^{5}+165955737\,{x}^{8}\nonumber \\
&&\quad \quad 
-40044089\,x^9+2440592\,x^{10}\, -419904\,x^{11}.
 \nonumber 
\end{eqnarray}

\subsection{The second order operator $\, N_1$ in  $\, L_{\Phi^{(5)}_D}$}
  \label{n1}
The second order operator $\, N_1$ occurring in the factorization
of  $\, L_{\Phi^{(5)}_D}$ reads:
\begin{eqnarray}
&&N_1\, = \, \, \, P_2  \cdot  x \cdot D_x^2 \,
+ \, \,  P_1 \cdot D_x \, + \, x^2 \cdot P_0,  \nonumber \\
&&P_2\, = \, \, \,-2\, (1+8\,x+20\,x^2+15\,x^3+4\,x^4)  
(1-x-3\,{x}^{2}+4\,x^3)  \, \nonumber \\
&& \quad \quad\quad  \times \,
  (1+2\,x-4\,{x}^2) \,(1-3\,x+x^2)  \, (1+2\,x)  \, 
(x-1)  \, (1+x)  \cdot  p_2 \nonumber \\
&&p_2\, = \, \, \,  608\,x^{16}+88\,x^{15}-3092\,{x}^{14}
-7329\,x^{13}+2156\,x^{12}+15088\,x^{11} \nonumber \\
&& \qquad +7054\,{x}^{10}-
3476\,x^9 -15856\,x^8 -6198\,x^7
+7668\,{x}^{6}+4064\,{x}^{5}\nonumber \\
&& \quad \quad 
-1014\,x^4 -732\,x^3 +36\,x^2 +63\,x +8, \nonumber \\
&&P_1 \, = \, \, -16-188\,x -592\,x^2 -384\,{x}^{3} 
+48895500\,x^{11}-24850920\,x^{12} \nonumber \\
&& \quad \quad -216496824\,{x}^{13}
 +526808\,x^7 
-24630720\,{x}^{14}  -203254764\,x^{23} \nonumber \\
&& \quad \quad 
+526770352\,x^{15}+243268011\,x^{16}+34759480\,x^{24}
+91207968\,x^{25}\nonumber \\
&& \quad \quad +16933696\,x^{26}
-13056000\,x^{27}-5926400\,x^{28}
+112640\,x^{29}\nonumber \\
&& \quad \quad 
+622592\,x^{30}+80791756\,{x}^{21}
-255450647\,x^{22}-7415\,x^4 \nonumber \\
&& \quad \quad -1427696\,x^8 -6112060\,x^9
+10047555\,x^{10}-686378692\,x^{17}\nonumber \\
&& \quad \quad 
+511457760\,x^{20} -507535276\,x^{18}
+380400540\,x^{19} -33404\,x^5 \nonumber \\
&&  \quad \quad +113836\,x^6, 
 \nonumber \\
&&P_0 \, = \, \, -144-1344\,x -4416004\,x^{23}-7073308\,x^{11}
+223379312\,x^{12}\nonumber \\
&& \quad \quad +107333356\,x^{13}+66480\,x^3
+1927564\,{x}^{7}+2688\,x^2\nonumber \\
&& \quad \quad -364322176\,x^{14}
-265131272\,x^{15}+279826283\,x^{16} \\
&& \quad \quad -10176392\,x^{24}
-2082592\,{x}^{25}+67584\,x^{26}+311296\,x^{27}\nonumber \\
&& \quad \quad +39019332\,x^{21}
+45440155\,x^{22}+176313\,x^4 +14440640\,x^8\nonumber \\
&& \quad \quad -6309980\,x^9
-73169703\,x^{10} +282707420\,x^{17}
-61145056\,x^{20}\nonumber \\
&& \quad \quad -44575908\,x^{18}
-149321200\,x^{19}-446364\,x^5 -2110212\,x^6. \nonumber 
\end{eqnarray}

\subsection{The order four and five operators for the staircase polygons.}
  \label{callZ5}

Let us give the order four and five operators
 $\, {\cal Z}_5$ and  for $\, {\cal Z}_6$ 
 the staircase polygons generating functions~\cite{Prell}:
\begin{eqnarray}
\label{caZ5}
&&{\cal Z}_5 \, = \, \, \, D_x^4 \, 
+2\,{\frac { (3-140\,x+1295\,{x}^{2}-1350\,{x}^3) }{x \, (1-x)
 \, (1-9\,x)  \, (1-25\,x) }} \cdot D_x^3 \nonumber \\
&&\qquad  
\, +{\frac { (7-518\,x+6501\,{x}^{2}-8550\,{x}^3)
 }{{x}^{2} \cdot (1-x) \, 
 \, (1-9\,x)  \, (1-25\,x) }} \cdot D_x^2
\nonumber \\
&&\quad \quad \quad 
+{\frac { (1-196\,x+3963\,{x}^{2}
-7200\,{x}^{3}) }{{x}^{3} \, (1-x)  \, (1-9\,x)
 \, (1-25\,x) }} \cdot D_x  \, \nonumber \\
&&\qquad  
\, -5\,{\frac {1-57\,x+180\,{x}^2}{ x^3 \, (1-x) 
 \, (1-9\,x)  \, (1-25\,x) }},
\nonumber \\
&&{\cal Z}_6 \, = \, \, \, D_x^5\, 
+10\,{\frac {1-70\,x+1176\,x^2 -4032\,x^3}{x
 \, (1-4\,x)  \, (1-16\,x) 
 \, (1-36\,x) }} \cdot D_x^4\nonumber \\
&&\quad\quad  +{\frac {25-2408\,x+51196\,{x}^{2}
-211968\,{x}^{3}}{{x}^{2} 
\, (1-4\,x)  \, (1-16\,x) 
 \, (1-36\,x) }} \cdot D_x^3 
\nonumber \\
&&\quad\quad  +3\,{\frac {5 -812\,x +23992\,x^2
-126720\,{x}^{3}}{ x^3 
\, (1-4\,x)  \, (1-16\,x) 
 \, (1-36\,x) }}\cdot D_x^2 
\nonumber \\
&& \quad \quad +{\frac {1-516\,x+25956\,x^2 -193536\,x^3}{ x^4
 \, (1-4\,x)  \, (1-16\,x)  \, (1-36\,x) }}
\cdot D_x \nonumber \\
&&\quad \quad  -6\,{\frac {1-170\,x+2304\, x^2}{ x^4 \, (1-4\,x)
 \, (1 -16\,x)  \, (1 -36\,x) }}. 
\end{eqnarray}

\section{Exponents of Fuchsian linear ODEs
 are generically algebraic  numbers, not rational numbers}
\label{fuchindi}

\subsection{``Lattice'' Fuchsian ODEs.}
\label{latticeFuchs}

Keeping in mind some mainstream~\cite{Todorov} conformal theory prejudice,
the fact that critical exponents (for 
the ferromagnetic/antiferromagnetic
 critical points) are rational numbers
is too often taken for granted. However, here, we have
 {\em a much stronger result}:
in all our previous calculations, the critical
 exponents {\em all} appear 
to be {\em rational numbers
for all the singularities of these ODEs} (in our $\, n$-fold integrals
of the Ising class, the exponents are even half-integers). 
Considering the polynomial coefficients in front of the successive
derivatives in all our Fuchsian linear ODEs,
one could wrongly imagine that this rational number exponent result
 is a straight consequence of the fact
that these polynomial
have, themselves, {\em integer coefficients}, this property 
being straightforwardly inherited from 
 the enumerative combinatorics nature of the lattice problem.
In that respect the simplest example of a Fuchsian linear 
differential operator certainly
 corresponds to the 
Gauss hypergeometric second-order differential operator
($D_x$ denotes the derivative $d/dx$)
\begin{eqnarray}
\label{Gauss}
x  \cdot (1-x) \cdot  D_x^{2} \,\,
 + \,( c\, - (a+b+1)\cdot  x) \cdot  D_x\, \,-a\,b,
\end{eqnarray}
which has the following indicial polynomials  for the regular
singular points $x\, = \, 0, \, 1, \, \infty$:
\begin{eqnarray}
r \cdot (r-1+c), \qquad  \quad   \, r \cdot (r\, +a\, b/c),  
\qquad  \quad   r \cdot (r+1-c).
\end{eqnarray}
On these indicial polynomials one sees clearly that
 exponents being rational numbers is straightforwardly 
inherited from the rational character of the coefficients 
of (\ref{Gauss}). More generally for Fuchsian linear ODEs 
of arbitrary order it can easily be shown, 
  for a singular point 
which is a rational number\footnote[2]{And we 
have many examples~\cite{Experimental},
 in $\, s$ or $\, w$ or $\, x$, of rational number
singular point, namely $\, w \, = \, \pm 1/4$, 
$\, w \, = \, \pm 1/2$, $\, w \, = \, \pm 1$, 
 $\, x \, = \,  1/16$,  $\, x \, = \,  1$, $\, x \, = \,  1/4$,
 $\, x \, = \,  1/9$,  $\, x \, = \,  1/25$,
  $\, x \, = \,  1/8$, \ldots} 
 and {\em occurring in the ODE with multiplicity one},
that its exponents are necessarily rational numbers. 

However, it is important to underline 
that Fuchsian linear ODEs with integer 
coefficients {\em do  not have 
 necessarily rational number exponents}.
Generically exponents of such integer coefficients Fuchsian linear ODE
 are {\em algebraic numbers} (algebraic over $\, \mathbb{Q}$) 
{\em not rational numbers}. 

\subsection{``Lattice'' Fuchsian ODEs with
algebraic numbers but not rational exponents.}
\label{latticeFuchs}

Let us consider, for instance, the order-four linear differential
 operator:
\begin{eqnarray}
\label{B3}
&& (x-1)^{2} \, x^2 \cdot  D_x^4 
+\, 4 \cdot (1-2\,x) \, (1-x) \cdot  x \cdot  D_x^3 \\
&&\qquad \quad + \, (1-6\,x +6\,x^2) \cdot  D_x^2
+\, 6 \cdot  (1-2\,x) \cdot D_x \,\,  +12. \nonumber 
\end{eqnarray}
This  linear differential operator is Fuchsian. It
 has three regular singular points
$x \, = \, 0$, $\, x\, = \, 1$ and
 $\, x\, = \, \infty$, some of its exponents 
being algebraic numbers simply expressed in terms
 of the golden number as can be seen on the 
indicial polynomial corresponding
 respectively to $x\, = \, 0, \, 1, \, \infty$:
\begin{eqnarray}
\label{index}
&& r \cdot (r-1) \cdot (r^2-r-1), \, \qquad  
 r \cdot (r-1) \cdot (r^2-r-1), \nonumber \\
&&\qquad \qquad  
 (r-3)\,(r-2)\,(r+1)\,(r+2) , \nonumber 
\end{eqnarray}
as well as on 
the solutions of this linear operator:
\begin{eqnarray}
\label{B5}
&&\left( x-1 \right)^{-1/2\,\sqrt {5}+1/2}   \cdot 
{x}^{1/2+1/2\,\sqrt {5}} \cdot 
\left( -\sqrt {5}-3+6\,x \right), \nonumber \\
&& {x}^{-1/2\cdot \sqrt {5}+1/2} \,
 (x-1)^{1/2+1/2\,\sqrt {5}} 
\cdot  \left( -3+6\,x+\sqrt {5} \right), \\
&&\left( 3\,x -1 \right) \cdot x, 
\qquad \quad \quad   1-6\,{x}^{2}.\nonumber
\end{eqnarray}

It is interesting to calculate the $\, p$-curvature of this
Fuchsian operator with a rational wronskian 
$1/((x-1)^{4} \cdot x^{4})$, but {\em with non-rational critical 
exponents} (which cannot, therefore, be globally nilpotent). 
For an infinite number of primes the $\, p$-curvature
is not nilpotent (a fortiori zero). However, 
for a subset of prime numbers (for instance, 
$\, p \, = \, 11, \, 19,\,$
$  29,\,  31, \, 41,\,  59, \, \cdots, \, 281, \, 311, \, \cdots$)
one finds a zero $\, p$-curvature. A heuristic 
interpretation is that, for 
some primes such that the golden number is ``like'' a rational
number\footnote[9]{Namely the primes $\, p$ 
such that $\, r^2-r-1$ factorize in $\, F_p$. For instance 
$\, r^2-r-1$ factorizes into $\, (r+7)\, (r+3)$ mod. 11,
$\, (r+4)\, (r+14)$ mod. 19, 
 \ldots ,  $\, (r+58)\, (r+252)$ mod. 311.},  an expression,
 like the first two ones in (\ref{B5}),
 can be seen as an algebraic one.
Of course this is not true for almost all primes. 
 
Similarly, the order four linear differential  operator 
\begin{eqnarray}
\label{B6}
&&(1 -x)^{2} \,{x}^{2} \cdot D_x^4
+\, 4\cdot (1 -2\,x) \,  
\, (1 -x) \cdot x \cdot  D_x^3 \\
&&\qquad  \quad \, + \,(1-14\,x +14\,{x}^2) \cdot  D_x^2
\,\,  -2 \cdot  (1- 2\,x) \cdot  D_x, \nonumber 
\end{eqnarray}
is also Fuchsian with solutions that can be expressed
 in terms of hypergeometric functions 
$_3F_2$ and has quadratic number exponents 
as can be seen on the indicial polynomials corresponding
 respectively to $\, x\, = \, 0, \, 1$,  
coincide with (\ref{index}) and for $\, x= \,\infty$ 
 with $\, r^2\, (r-1)^2$. Again calculating 
the $\, p$ curvature of this
Fuchsian operator with the same rational wronskian 
$1/((x-1)^{4} \cdot x^{4})$, with algebraic 
 {\em but not rational numbers} critical 
exponents (thus excluding global nilpotence) one finds,  
for a {\em subset of the prime numbers},
 a nilpotent characteristic polynomial, namely
$\, T^4$ (with a minimal polynomial $\, T^2$).
Generically the  characteristic polynomial rules out the 
nilpotence, since it reads:
\begin{eqnarray}
\label{generalcurvature}
T^4 \, + \, \, {{p-5} \over {x^{4\, p} \,
 + \, (p-2)x^{3\, p} \, +x^{2\, p} }} \cdot T^2. 
\end{eqnarray}

Another simple example is the order three 
linear differential  operator
\begin{eqnarray}
\label{B7}
x^{2} \cdot  (1 -x) \cdot  D_x^{3}\, 
+\, x \cdot (2 -3\,x)  \cdot   D_x^{2}\, 
+ \, (1 +2\,x) \cdot   D_x\, \, -1,
\end{eqnarray}
which has exponents that can simply be expressed
 in terms of third root of unity and golden number,
as can be seen on the indicial polynomial corresponding
 respectively to $x\, = \, 0, \, 1, \, \infty$:
\begin{eqnarray}
r \cdot (r^2-\,r+1), \,\,\,\, \,\,\,\,  
\quad \,\,\,\, r \cdot (r-1)^2, 
\, \,\,\,\,\,\,\,  \quad \,\,\,\,
 r \cdot (r^2\, -3\, r+1). \nonumber
\end{eqnarray}
Again calculating the $\, p$-curvature of this
Fuchsian operator with a rational wronskian 
$1/((x-1) \cdot x^{2})$ but {\em non rational} critical 
exponents (thus excluding global nilpotence), one finds  
for a subset of the prime numbers
 ($\cdots, \, 109, \, 163, \, 181, \, 199, \, \cdots$)
a characteristic polynomial $\, T^3$ 
with a minimal polynomial $\, T^2$,
and for a smaller set of primes
 ($\cdots, \, 73, \, 271, \, \cdots$)
a characteristic polynomial $\, T^3$ with a 
minimal polynomial $\, T^3$.

The miscellaneous examples we have displayed
 actually correspond to the
generic situation of Fuchsian linear ODEs
 with {\em integer coefficients}
 (the proper framework we expect,
 at first sight, for lattice statistical 
mechanics quantities satisfying a linear ODE). 
However they are not globally nilpotent 
and thus are not ``derived from geometry'':
generically\footnote[5]{In a mathematical 
perspective. In contrast
 physics seems to favour
the DFG framework for the minimal order ODEs \ldots} a 
Fuchsian linear ODE does not 
have solutions that can be expressed as 
$\, n$-fold integrals of algebraic integrands.

\section{Linear differential equation for 
${\tilde \chi}_d^{(4)}(t)$.} 
\label{diag}

The  linear differential operator for ${\tilde \chi}^{(4)}_d(t)$ 
is of order eight, and has the {\em direct sum} decomposition
\begin{eqnarray}
\label{ds4}
{\cal L}_8^{(4)}\,=\,\,\, \,
  L_1^{(4)} \oplus L_3^{(4)} \oplus L_4^{(4)},
\quad \quad \, \,\hbox{with:} 
\quad \, \,
L_1^{(4)}\,=\,\,\, D_t \, \,
 + \,\, {{d} \over {dt}} \ln({{t-1} \over {t}}), 
\nonumber
\end{eqnarray}
\begin{eqnarray}
&&L_3^{(4)}\, = \,\,   D_t^{3}\, 
+{\frac { \left( 5\,{t}^{2}+6\,t-1 \right) }
{ \left( 1+t \right)  \, (t-1)\, t }} \cdot D_t^{2}
\,\, +{\frac { \left( 3\,{t}^{3}+6\,{t}^{2}-2\,t-1 \right)}
{ \left( 1+t \right) {t}^{2}
 \left( t-1 \right)^{2}}}\cdot  D_t \nonumber\\
&&\qquad \qquad \qquad -{\frac {3}{ 2\, (1+t) \, (t-1)\, t^2 }}, 
\end{eqnarray}
where  $\, L_4^{(4)}$ is an order-four linear differential operator 
with apparent singularities 
$\, {t}^{2}-10\,t +1 \, = \, 0$, that will not be displayed here.
 Introducing the 
order-one operator: 
\begin{eqnarray}
G_1 \, = \, \, D_t \, + \,  \,\, 
  {{d} \over {dt}} \ln\Bigl(  {{ (t^2-10\, t+1) \, (t-1)^6 \cdot t^4 }
 \over { (t+1)^{8}}}  \Bigr)
\end{eqnarray}
one get rid of these apparent singularities (desingularization),
 and obtains an order-five 
Fuchsian linear differential operator $\, G_1 \cdot L_4^{(4)}$
which, after simple conjugaisons, can be simply written as:
\begin{eqnarray}
&& {t}^{4} \, (t-1)^{3} \, (t+1) ^{2} \cdot G_1 \cdot L_4^{(4)} 
\, = \, \,  \, \,  
 (t-1) ^{3} \, (t+1)^{2} \, t^4  \cdot  D_t^5 \,  \nonumber \\
&& \qquad \qquad + \, (-9+11\,{t}^{2}+26\,t)
 \, (t-1)^{2} \, (t+1)\,  t^3 \cdot   D_t^4 \,  \\
&& \qquad \qquad  \, + \, 
(31\,t^4 +172\,t^3 +126\,t^2 -140\,t+19)  \, (t-1)\,  t^2 \, \cdot 
 D_t^3\,  \nonumber \\
&&\qquad  \qquad 
+2\, (11\,t^5 +107\,t^4
+179\,t^3 -271\,t^2 +74\,t -4) \, t \cdot 
\, D_t^2 \,  \nonumber \\
&& \qquad \qquad  \,
+ \, (2\,{t}^{4}+43\,{t}^{3}+327\,{t}^{2}-199\,t+19)
 \,t \cdot    D_t\,
+3\, \, (t+1)^{3} .\nonumber 
\end{eqnarray}

The linear differential operator of order three, $\, L_3^{(4)}$ 
is {\em actually equivalent} 
to the symmetric square $\, Sym^2(L_E)$ of the second
 order operator corresponding to 
the complete elliptic integral $\, E(x^{1/2})$ 
(see (\ref{LE})). This order-three linear differential operator 
$\, L_3^{(4)}$ is  therefore globally nilpotent. 
Actually, we have calculated its $\, p$-curvature of  $\, L_3^{(4)}$
and found that the corresponding characteristic 
polynomial (or minimal polynomial)
reads $\, T^3$. 

The order-four linear differential
 operator  $\, L_4^{(4)}$ is also
globally nilpotent: we have calculated the $\, p$-curvature
and found that the corresponding characteristic 
polynomial (or minimal polynomial)
reads $\, T^4$. For the moment we have not been able to write
one of his four solutions as a $\, _4F_3$ hypergeometric function
up to a pull-back (trying to generalize subsection (\ref{3F2pull})). 

\vskip .4cm 

\section{Revisiting the global nilpotence of
 $\, \Phi^{(n)}_D$ for $\, n\, = \, 3, \, 4, \, 6$.}
\label{revin}

\subsection{Revisiting the global nilpotence of $\, \Phi^{(3)}_D$.}
\label{revi}

The global nilpotence of $\, \Phi^{(3)}_D$ can
 be understood from the factorisation
of the corresponding linear differential operator
which can be seen as the direct sum of 
an operator of order three and 
of $\, D_x$:
\begin{eqnarray}
&&D_x \oplus L_3, \qquad \quad  \hbox{where:}
 \qquad \quad L_3 \, = \, \, z_2 \cdot L_1,
\qquad \quad \hbox{with:} \\
&& L_1 \, = \, \,\, D_x \, \, 
 + \, {{1} \over {2}} \,  {{d} \over {dx}}
 \ln\Bigl( {{(1+3\, x \, +4\, x^2) \, 
(1 \, +2\, x) \, (x-1)} \over {(1+x)^2}} \Bigr),
 \nonumber \\
&&q_2 \cdot z_2 \, = \, \, q_2 \cdot  D_x^2 \, 
 +  2\cdot (1+x) \cdot q_1 \cdot  D_x \,
 + 4 \cdot q_0, \qquad \quad   \hbox{where:} \qquad
\nonumber 
\end{eqnarray}
\begin{eqnarray}
&&q_2 \, = \, \, x \,  (1-x)  \, (1+ 4\,x) 
 \left( 1+2\,x \right)
 \left( 1-4\,x \right)  \left( 1+3\,x+4\,{x}^{2} \right)  
\left( 1+x \right)^{2} \cdot Q_2,  \nonumber \\
&&Q_2 \, = \, \, 3264
\, x^8 +56\,x^7 -862\,x^6 +3641\,x^5
+1873\,{x}^{4}\nonumber \\
&& \qquad +149\,{x}^{3}
-23\,{x}^{2}+2, \nonumber \\
&&q_1 \, = \, \,1253376\,x^{15} +1330688\,x^{14}\,
-492800\,x^{13}+1432064\,x^{12} \nonumber \\
&&\quad \quad +3680288\,x^{11}+1249562\,{x}^{10}\,
-1192677\,x^9 -1051887\,x^8 \nonumber \\
&&\quad \quad -317269\,x^7 -47698\,x^6
-8120\,x^5 -2801\,x^4\,
-693\,x^3 \nonumber \\
&&\quad \quad -50\,{x}^{2}+15\,x+2, \nonumber \\
&&q_0 \, = \, \,626688\,x^{15}+1237248\,x^{14}\,
+237504\,x^{13}+898720\,x^{12} \nonumber \\
&&\quad \quad +3726900\,x^{11}+3657589\,x^{10}
+1424484\,x^9 +315618\,x^8\nonumber \\
&&\quad \quad
+122103\,x^7  +24147\,x^6 -21786\,x^5 -14389\,x^4\,
-3444\,x^3\nonumber \\
&&\quad \quad -375\,x^2 -9\,x +2.  \nonumber 
\end{eqnarray}
The order-two linear differential operator $\, z_2$ can be seen to 
be homomorph to $\, Q_E$, defined in
(\ref{32}), corresponding to the complete elliptic integral
$\, {\it E}(4\, x)$:
\begin{eqnarray}
\label{w1}
&&  z_2 \cdot W_1 \, = \,  \, \,  W_2 \cdot Q_E, 
\end{eqnarray}
 where $\, W_1$ and $\, W_2$ are two linear operators of order one.
From the explicit expression of $\, W_1$
 one easily finds the following solution for $\, z_2$:
\begin{eqnarray}
&&-\,{\frac { (12\, x^3 +7\, x^2 +x-2)
\cdot  {\it E}(4\,x) }{ (1+3\,x +4\,x^2) \, (1- 4\,x)  
\, (1+2\,x) \, (1-x) \, x  }} \nonumber \\
&&\qquad \qquad - \,{\frac { (34\, x^4
+11\,x^3 +6\, x^2 +7\,x +2) \cdot  {\it K}(4\,x) }
{ (1+2\,x)  \, (1+x) \, 
 \, (1+3\,x+4\,x^2) \, (1-x) \, x  }}. \nonumber 
\end{eqnarray}

\subsection{Revisiting the global nilpotence of $\, \Phi^{(4)}_D$.}
\label{revi2}

The global nilpotence of $\, \Phi^{(4)}_D$ can
 be understood from the factorisation
of the corresponding linear differential operator
which can be seen as the direct sum of
 a linear operator of order-three and 
of $\, D_x$:
\begin{eqnarray}
&&D_x \oplus L_3, \qquad \quad  \hbox{where:} \qquad
\quad  L_3 \, = \, \, L_2 \cdot M_1, 
\qquad \quad  \hbox{with:} \nonumber \\
&& M_1 \, = \, \,\, D_x  \,\,
+ \, {{1} \over {2}} \, {{d} \over {dx}}
 \ln \Bigl( {{4\, x\, -1} \over {(x-1)^2 }}  \Bigr), 
\nonumber \\
&&L_2 \, = \, \, q_2 \cdot  D_x^2 \, 
 +  \, (x-1)\cdot q_1 \cdot  D_x \,
 +  q_0,  \qquad \quad \quad  \hbox{where:} 
\end{eqnarray}
\begin{eqnarray}
&&q_2 \, = \, \,  \, (1-16\,x)  \, (1- 4\,x) 
 \, (1024\,{x}^{3}+28\,{x}^{2}-42\,x+1) 
 \, (1-x)^2 \cdot x, \nonumber \\
&&q_1 \, = \, \, 262144\,x^6 -228608\,x^5\,
-4496\,x^4 +19420\,x^3 \,\nonumber \\
&&\qquad \quad \quad -3088\,x^2 +125\,x-2, \nonumber \\
&&q_0 \, = \, \, 147456\,x^6 -242624\,x^5\,
+13376\,x^4 +49864\,x^3 \nonumber \\
&&\qquad \quad \quad -14530\,x^2 +961\,x +2.
 \nonumber 
\end{eqnarray}

The  order-two linear differential operator 
\begin{eqnarray}
\label{QE}
\tilde{Q}_E \,\, \, =\, \, \,\,{ D_x}^{2}\,
+\,{\frac {D_x}{x}}\,
+  \,{\frac {4}{ (1-16\,x)\, x}}.
\end{eqnarray}
corresponding to the complete elliptic integral
$\, {\it E}(4\, \sqrt{x})$, is equivalent to 
the linear differential operator $\, L_2$ 
\begin{eqnarray}
\tilde{Q}_E \cdot z_1 \, \, = \, \, \, s_1 \cdot L_2, 
\end{eqnarray}
where $\, z_1$ and $\, s_1$ are two order-one 
linear differential operators.

\subsection{Revisiting the global nilpotence of $\, \Phi^{(6)}_D$.}
\label{revi2}

The linear differential operator for $\, \Phi^{(6)}_D$
is an order-five Fuchsian linear operator which is the direct sum
of $\, D_x$ (here $\, x\, = \, \, w^2$) 
and of an order-four operator
which factorises as a product of 
two order-two operators 
\begin{eqnarray}
L_{\Phi^{(6)}_D} \, = \, \, D_x \oplus L_4,  \, 
\qquad \hbox{where:} \qquad  L_4
\, = \, \,  \,  M_2 \cdot  L_2,
\end{eqnarray}
where $\, M_2$ is a pretty large order-two linear 
differential operator
 (with a rational wronskian) and 
\begin{eqnarray}
L_2 \, = \,\,  D_x ^2 \,\, - \, 2 \cdot (1\,-4\, x) \cdot {{P_1 } 
\over {P_2}} \cdot D_x \, \,
-\, 2 \, {{ P_0} \over {P_2}}, 
\qquad \quad \quad \hbox{with:} \nonumber 
\end{eqnarray}
\begin{eqnarray}
&& P_2\, = \,\,\, \, (1-4\,x)^{2} \, (1-x)
  \, (1-9\,x)  \, ( 1-10\,x+29\,{x}^{2})\nonumber \\
&& \qquad \times 
 \, (1722\,{x}^{6}-3306\,{x}^{5}+2973\,{x}^{4}
-1548\,{x}^{3}+403\,{x}^{2}-46\,x+2),
\nonumber \\
&& P_1\, = \, \,\, 898884\,x^{10} -2797104\,x^9
+4902117\,x^8 -5573337\,x^7\, \nonumber \\
&& \qquad 
+3999969\,x^6 -1764005\,x^5 +477136\,x^4 -79113\,x^3
\nonumber \\
&& \qquad +7883\,{x}^{2}-441\,x+11, \,\,  \\
&& P_0\, = \,\,\, 898884\,x^{10} -2559756\,x^9
+3491100\, x^8 -2205501\,x^7\,
+556746\, x^6\nonumber \\
&& \qquad
  +92091\,x^5 -92841\,x^4 +23740\,x^3
-3081\,x^2 +226\,x -8.
\nonumber 
\end{eqnarray}
The square of the wronskian of $\, L_2$
is a simple rational function. The Fuchsian
 linear operator $\, L_2$ is
 such that the $\, p$-curvatures 
{\em are zero for almost all primes}, and therefore it has
 a basis of {\em  algebraic solutions}. Note that 
 the differential Galois group of $\, L_2$ is 
isomorphic to the {\em group
of quaternions} (eight elements).
Its algebraic solutions correspond to an algebraic curve
of genus $\, g  \,= \,\, 5$. 
The equation of that algebraic curve reads:
\begin{eqnarray}
&& 21025 \cdot (2-36\,x+218\,{x}^{2}-558\,{x}^{3}+553\,{x}^{4}\,
-106\,{x}^{5}+27\,{x}^{6})^4 \nonumber \\
&& \quad - 58 \cdot p_2 \cdot (1-4\,x)^{2} \, (1-x)^2
  \, (1-9\,x)^2  \, ( 1-10\,x+29\,{x}^{2})^2 \cdot Z^2
\nonumber \\  
&& \quad +(1-4\,x)^{4} \, (1-x)^4
  \, (1-9\,x)^4  \, ( 1-10\,x+29\,{x}^{2})^4 
\cdot Z^4 \, = \, \, 0, 
\end{eqnarray}
with:
\begin{eqnarray}
&&p_2 \, = \,  \, \, 1053\,x^{12} +46836\,x^{11} -429262\,x^{10}
+520760\,x^9 +1315505\,x^8\nonumber \\  
&&\quad \quad  -3318300\,x^7 +3056140\,x^6
-1518520\,x^5 +448000\,x^4 -80280\,x^3\,\nonumber \\  
&&\quad \quad  +8552\,x^2 -496\,x +12.\nonumber 
\end{eqnarray}
Again, these algebraic functions, roots of a {\em genus five}
 algebraic curve, 
can be expressed as
linear combinations of complete elliptic
integrals of the third kind with a ``characteristic'' (first argument
of the complete elliptic integral of the third kind) 
associated with a {\em genus three}
curve (see \ref{towards1}).

\vskip .1cm 
 We have also calculated the
 $\, p$-curvature of the (quite large)
order-two Fuchsian linear differential operator $\, M_2$
(for primes $ < 400$) and found that all these $\, p$-curvatures
are nilpotent. One can actually prove that $\, M_2$
is equivalent to the previous second order
operator (\ref{QE}),
associated with $\, {\it E}(4 \sqrt{ x})$.

\vskip .3cm 

\vskip .4cm 

\section{Towards a geometrical interpretation of global nilpotence }
\label{towards}

\subsection{Towards an interpretation as periods of algebraic
 varieties: closed formula for $\, \Phi_D^{(n)}$}
\label{towards1}

The integrals  $\, \Phi_D^{(n)}$ can 
all be expressed
as sums of complete elliptic integrals of
 the third kind $\,  {\it \Pi} (y(w), \, w)$, where the
 characteristic\footnote[5]{The first argument in
a complete elliptic integral of
 the third kind is called the characteristic.} 
$\, y= \, y(w)$ corresponds to some algebraic curves:
\begin{eqnarray}
\label{closed}
\Phi_D^{(n)} \, = \, \, 
\sum_{i}  A_i(w) \cdot \Pi(y_i(w), \, w), \qquad \quad 
 P_n(y_i, \, w)\, = \, \, 0,
\end{eqnarray}
where $\, A_i(w)$ are algebraic expressions
 of\footnote[8]{In the following 
subsections the calculations are expressed
in terms of a variable $\, x$ that is equal to $w$ for $\, n$ odd and 
to $w^2$ for $\, n$ even.} $w$
 and $\, P_n$ are simple polynomials 
of $y_i$ and $w$ with integer coefficients. 
 
\subsubsection{Towards an interpretation as periods of algebraic
 varieties: closed formula for $\,  \Phi_D^{(3)}$.}
\label{towards12}

Let us give an exact expression for the integral
$\, \Phi_D^{(3)}$ (see (\ref{chinaked})) in terms of 
complete elliptic integrals of
 the third kind\footnote[2]{We thank M. Rybowicz for kindly providing 
to us other closed expressions in terms of 
complete elliptic integrals of
 the third kind.}.
Let us introduce $\, f_1$ and $\, f_2$ solutions of:
\begin{eqnarray}
\label{cccu1}
&& (1-x-4\,{x}^{2})^{2} \cdot (f_{1}^{2}\, -1)\, 
 -2\,{x}^{2} \cdot ( 4\,x+1)^{2} \cdot (f_{1}-1)\, \nonumber \\
&&\qquad \qquad  -2\, (1+2\,x)  \cdot ( 4\,x-1)  \cdot (f_{1}+1)
\,\, = \, \,\, 0,  \\
\label{cccu2}
&&(1 +3\,x +4\,{x}^{2})  \cdot (f_2^2 -1) \, 
+ \,(1+ 2\,x)  \, (1- 4\,x)  \cdot (f_2 +1) \, 
\nonumber \\
&&\qquad \qquad 
- \, (1 +4\,x)^{2} \cdot(f_2 -1)\,\, = \, \,\, 0, 
\end{eqnarray}
which are {\em rational} curves
 that can be parametrized as follows:
\begin{eqnarray}
\label{ratparam}
x \, = \, \,\,  {\frac {4 +{u}^{2}}{2(8 -u^2)}},
 \quad \, \, \, \, 
f_1 \, = \, \, \, {\frac { (u^2+6\,u-8)^{2}}
{ ({u}^{2}-6\,u-8)^{2}}}, \quad \, \, \, \, 
f_2 \, = \, \, {\frac {u^2 +6\,u +16}{u^2 -6\,u +16}}.
\nonumber 
\end{eqnarray}
Let us introduce the three {\em involutive} birational transformations: 
\begin{eqnarray}
 J(f_i)\, = \, \, 
{\frac { (1+x)\,  -(1-\, x) \cdot f_i}
{(1-\,x) \, -(1+x) \cdot f_i}}   , \quad  
H(f_i)\, = \, {{1} \over {f_i}}, \quad 
I(f_i) \, = \, -f_i. \nonumber 
\end{eqnarray}
The non-involutive birational transformation 
$\, I  \circ J$ maps (\ref{cccu1}) onto
 (\ref{cccu2}), and, of course,
 $\, J \circ  I$ maps back (\ref{cccu2}) onto
 (\ref{cccu1}). These two rational curves 
are invariant under the (Hadamard) involution $\, H$: 
$f_1 \rightarrow 1/ f_1$ and 
$f_2 \rightarrow 1/ f_2$. Note that $\, H$ and 
$\, I \circ  J$ commute.

Let us introduce the expression:
\begin{eqnarray}
R(f, \, 4\, x) \, = \, \, 
{{2 \cdot {\it \Pi} (4\, x\, f, \, 4\, x)} \over {\pi}} \cdot 
\sqrt {{\frac { (4\,x\, f\, -1)  \, (4\,x\, -f) }{f}}},
\end{eqnarray}
one then easily finds that $\, R(f, \, 4\, x)$ is such that
\begin{eqnarray}
\label{identity}
&& R(f, \, 4\, x) \, + \, R(1/f, \, 4\, x) \,= \,  \, \, \nonumber \\
&& \qquad \,= \,  \, \, 1 \, + \, \, 
 {{2} \over {\pi}} \cdot K(4\, x) \cdot 
\sqrt {{\frac { (4\,x\, f\, -1)  \, (4\,x\, -f) }{f}}}, \\
&&{{d} \over {dx}} R(f(x), \, 4\, x) \,= \,
 Q(x) \cdot {{2} \over {\pi}} \cdot E(4\, x)
\, + \, P(x) \cdot {{2} \over {\pi}} \cdot K(4\, x),
\nonumber
\end{eqnarray}
for some $\, Q(x)$ and $\, P(x)$ that can be deduced from $\, f=\, y(x)$.
These identities are valid for any $\, y(x)$ and do not require
$\, y(x)$ to be, for instance, a rational function of $\, x$.
The occurence of several operators equivalent to $\, L_E$ 
(associated to the complete elliptic integral $\, E$) 
in the factorisation of the
linear differential operators
corresponding to
the $\, \Phi^{(n)}_D$, 
 can be seen as a 
consequence of that identity (\ref{identity}) 
on the complete elliptic integral of the third kind for arbitrary 
characteristic $\, y(x)$.

Introducing a combination of four complete elliptic
 integrals of the third kind,
with characteristics satisfying the
 rational curves (\ref{cccu1}), (\ref{cccu2}):
\begin{eqnarray}
\label{sigm}
\Sigma \, = \,  \,   
R(f_1, \, 4\, x) \, -  R(1/f_1, \, 4\, x)  \,
 +  R(f_2, \, 4\, x)\,   - R(1/f_2, \, 4\, x), 
\quad \quad 
\end{eqnarray}
$\, \Phi^{(3)}$ can then be simply written in terms $\Sigma$ as:
\begin{eqnarray}
&&\Phi^{(3)}_D\, \, = \,\,  \, \, {{1} \over {8}} \,  \,  \, 
+  \,{{1}\over{12\,\pi}} \cdot {\frac { (1+4\,x) 
 \, (1-9\,{x}^{2}-12\,{x}^{3}) }
{ (1-x-4\,{x}^{2})  \, (1+3\,x+4\,{x}^{2}) }} \cdot K(4\,x) \, 
+\, C \cdot \Sigma, \nonumber
\end{eqnarray}
where:
\begin{eqnarray}
C\, = \, \, \,  -{{i} \over {48}} \cdot 
{\frac { (1+ x) }
{\sqrt {1+2\,x}\sqrt {1-x} \sqrt {1+3\,x+4\,{x}^{2}}}}.
\end{eqnarray}

\subsubsection{Towards an interpretation as periods of algebraic
 varieties: closed formula for $\, \Phi_D^{(4)}$.}
\label{towards13}
\vskip .1cm 

For $\, \Phi^{(4)}_D$, and in a similar 
way as for $\, \Phi^{(3)}_D$, 
one has now the two {\em rational} curves:
\begin{eqnarray}
\label{FF1}
&&3 \cdot  (1 +4\,x)\,  (f_{1}+1)^2 
+ \, (1\,-4\,{x}) \,(f_{1}-1)^2\,  = \, \, 0,  \\
\label{FF2}
&&27 \cdot  (1 +4\,x)\,x^4 \, (f_{2}+1)^{2} 
+ \, (1 -4\,x) \, ( 2-5\,{x}^{2} )^2
 \, (f_{2}-1)^2  \, = \, \, 0. 
\end{eqnarray}
A simple parametrization reads:
\begin{eqnarray}
x \, = \, \, {{1} \over {4}} \cdot {{u^2+3} \over {u^2-3}}, \quad 
f_1 \, = \, \, {{1\, + \, u} \over {1\, - \, u}}, \quad 
f_2 \, = \, \, \Bigl({\frac {1 \,-\, u}{1\,+\, u}} \Bigr)\cdot 
\Bigl( {\frac {  {u}^{2}-4\, u\, -9}{ u^{2}+4\, u\, -9}}\Bigr)^2.
\nonumber
\end{eqnarray}
Let us introduce the three {\em involutive} 
birational transformations: 
\begin{eqnarray}
 J(f_i)\, = \, \, 
{\frac { (1-x^2)\,  -(1-4\, x^2) \cdot f_i}
{(1-4\,x^2) \, -(1-x^2) \cdot f_i}}, \quad  
H(f_i)\, = \, {{1} \over {f_i}}, \quad 
I(f_i) \, = \, -f_i. \nonumber 
\end{eqnarray}
the non-involutive birational transformation $\, J \circ  I$ 
maps the first curve (\ref{FF1}) 
onto the second curve  (\ref{FF2}), and of course 
its inverse $\, I \circ J$ maps back  (\ref{FF2}) 
onto (\ref{FF1}). Note that these two rational curves 
are actually invariant under the (Hadamard) involution $\,H$:  
$ f_1  \rightarrow  1/ f_1$ and $f_2  \rightarrow  1/ f_2$.
The rational curve (\ref{FF1}) has, of course,
 many other involutive birational automorphisms:
\begin{eqnarray}
(x, \, f_1) \, \rightarrow \, \, 
\Bigl(-{\frac {1 +5\,x}{5 +16\,x}}, \, -f_1\Bigr). \nonumber
\end{eqnarray}
Using the same form for $\Sigma$ as in (\ref{sigm}), 
the solution $\, \Phi_D^{(4)}$ can now be written simply as:
\begin{eqnarray}
&&\Phi_D^{(4)} \, = \, \, \, {{1} \over {32}} \, \, 
+ {{1} \over {144\pi}} \cdot E(4\,x) \,\nonumber \\
&& \quad \quad \quad  
+ \, {{8\, x^4 \,+ 9\, x^3 \, - 11\, x^2 \, -1} 
\over { 72\, \pi \, \, (2\, x\, +1) \, (x^2 \, +3\, x\, -1 )} }
 \cdot K(4\,x) \, + \, C \cdot \Sigma, \nonumber
\end{eqnarray}
where:
\begin{eqnarray}
C \, = \, \, {{i \, \sqrt{3} } \over {864}}
\cdot {{ x^2-1} \over { \sqrt{1-4\, x^2} }} \nonumber 
\end{eqnarray}

\subsubsection{Towards an interpretation as periods of algebraic
 varieties: closed formula for $\, \Phi_D^{(5)}$, 
$\, \Phi_D^{(6)}$, $\, \Phi_D^{(7)}$
and $\, \Phi_D^{(8)}$.}
\label{towards13}
\vskip .1cm 
For higher values of $\, n$ ($n \, \ge \, 5$) 
the algebraic curves $\, P_n(y_i, \, w)\, = \, \, 0$,
which occur in the closed formula 
(\ref{closed}) for the $\, \Phi_D^{(n)}$
are no longer genus zero curves
 but {\em  higher genus} curves.  
For instance for the $\, \Phi_D^{(5)}$ and
 $\, \Phi_D^{(6)}$, 
these curves are {\em genus three} curves. In terms of 
$\, q \, = \,\, 4\, w\, y $ they read respectively:
\begin{eqnarray}
&&{q}^{4}\, -4\,{q}^{3} \, 
-4\, (4\,{w}^{2}-3) \cdot  {q}^{2}\, 
+16\, (2\,{w}^{2}-1) \cdot  q \, \nonumber \\
&&\qquad \qquad +8\, \left( w+1 \right)
  \, (4\,{w}^{3}-3\,{w}^{2}-w+1)
\, \, = \, \, \, 0. \nonumber 
\end{eqnarray}
and:
\begin{eqnarray}
&& 10\,{q}^{4}\, 
-40\,{q}^{3}
\, -5\, (3\,{w}^{2}-2)\cdot  {q}^{2}\, 
+80\, \left( 3\,{w}^{2}-1 \right) \cdot q\, \nonumber \\
&&\qquad \qquad 
+32\, \left( w-1 \right)  \left( 2\,w+1 \right) 
 \, (2\,w-1)  \, (w+1)
\, = \, \, 0. \nonumber
\end{eqnarray}

For  $\, \Phi_D^{(7)}$ and $\, \Phi_D^{(8)}$
 these curves are {\em genus ten} curves. 

For $\, \Phi_D^{(7)}$, in terms of 
$\, q \, = \,\, 4\, w\, y$, one gets:
\begin{eqnarray}
&&q^6\, -6\,q^5 \, 
-6\, (-5+4\,w^2) {q}^{4}\, 
+16\, (6\,w^2-5) \cdot  {q}^{3}\nonumber \\
&&\qquad 
\, +24\, (6\,{w}^{4}+5-12\,w^2) {q}^{2}\, 
-96\, (w-1)  \, (w+1)  \, (3\,w^2 -1)\cdot  q\,
 \nonumber \\
&&\qquad 
+32-128\,w^6 -192\,w^2 +288\,w^4 +32\,w^5
\, = \, \, 0. \nonumber 
\end{eqnarray}

The vanishing conditions $\, \delta_n\, = \, 0$ of the
discriminants in $\, q$ of these
 polynomials, associated with $\,  \Phi_D^{(n)}$,
read respectively:
\begin{eqnarray}
&&\delta_5 \, = \, \, (1-2\,w)  \, (1\, +4\,w) 
 \, (1 \, -2\,w\,  -4\,w^2) 
 \, (w-1)^{2} \, \nonumber \\
&&\qquad \times (1\, +w \,-3\,w^2 \, -4\,w^3)^2  \, = \, \, 0,
\nonumber \\
&& \delta_6 \, = \, \,(1+4\,w)  \, (1+2\,w)  \, (1-2\,w) 
 \, (1-4\,w)  \, (1-10\,w^2 \,+29\,w^4)^2
 \, = \, \, 0,
\nonumber \\
&&\delta_7 \, = \, \,(1\, +4\,w)  \, (1 -2\,w \, -4\,w^2) 
 \, (1\, -2\,w\, -8\,w^2\, +8\,w^3)  \, (1-w)^2 \nonumber \\
&&\qquad 
\,(w^3-w^2-2\,w+1)^2 \cdot d_7^2 \, = \, \, 0\nonumber \\
&& d_7 \, = \, \, 
 1 +3\,w-10\,w^2-35\,{w}^{3}+5\,w^{4}
+62\,w^5 +17\,w^6-32\,w^7 -16\,w^8.
\nonumber 
\end{eqnarray}
With the exception of $\, (1\, +4\,w)\,(1\, -4\,w)\, = \, 0$ 
in $\, \delta_6\, = \, \, 0$, 
{\em these polynomial conditions are, respectively, the 
singularities of the linear ODEs
for  $\, \Phi_D^{(5)}$, $\, \Phi_D^{(6)}$ and $\, \Phi_D^{(7)}$,
 where $\, w$ has been changed into $\,(- w)$}.
 
\vskip .3cm 

\section{Atkin's modular curves and Weber's modular functions}
\label{atkin}
The classical modular curve~\cite{Hanna} which 
corresponds to the duplication
of the ratio of periods of the 
elliptic curves~\cite{KleinAbsolute}
 $\, j \, = \, \, j(\tau)\, \, \rightarrow 
\, \, j' \, = \, \, j(2\, \tau)$:
\begin{eqnarray}
\label{2tau}
&&j^2 \cdot j'^2 -(j+j') \cdot
 (j^2+1487 \cdot j\, j' \, +j'^2) \nonumber \\
&& \qquad  \,  
 +3 \cdot 15^3 \cdot (16\, j^2\, -4027\, j\, j' \, +16\, j'^2) \\
 &&   \qquad \,  -12\cdot 30^6\cdot (j+j') \, +8\cdot 30^9
\, = \, \, \, \, 0 , \nonumber
\end{eqnarray}
of course symmetric by  $\, j \leftrightarrow \, j'$,
 is well-known to be a genus zero curve with the 
rational parameterization:
\begin{eqnarray}
\label{duplication}
j \, = \, \, \, j_2(z) \, = \, \,
 {\frac { (z+16)^{3}}{z}}, \qquad \quad 
j' \, = \, \, {\frac { (z+256) ^{3}}{{z}^{2}}}
 \, = \, \, j_2\Bigl({{2^{12} } \over {z}} \Bigr). \quad  
\end{eqnarray}
This is the {\em duplication formula of the Klein's absolute invariant}.
The involution $\, z \, \, \rightarrow \, \, 2^{12}/z\, $ 
is the {\em  Atkin involution}~\cite{Atkin}.
Recall that the rational variable $\, z$ can simply be expressed
in term of the {\em Dedekind eta function}:
\begin{eqnarray}
z \, = \, \,\, 2^6 \cdot 
\Bigl( {{ \eta(2\, \tau) } \over {\eta(\, \tau) }} \Bigr)^{24}.
\end{eqnarray}
The modular curve (\ref{2tau}) can also be parameterized as:
\begin{eqnarray}
\label{duplik}
j \, = \, \,256\,{\frac { (1-{k}^{2}+{k}^{4})^3}{
 (1-\, {k}^{2})^{2} \, {k}^{4} }}, \quad \quad 
j' \, = \, \,16\,{\frac { (1+14\,{k}^{2}+{k}^{4})^3}{
 (1-{k}^{2})^{4} \,{k}^{2} }}
\, = \, \, j\Bigl({{ 2\, \sqrt{k}} \over {1+k}}\Bigr),
\nonumber 
\end{eqnarray}
where the occurrence of the Landen transformation becomes explicit,
or :
\begin{eqnarray}
\label{dupliw}
j \, = \, \,{\frac { (1-16\,w^2+16\,w^4)^3}
{ (1-16\,{w}^{2})\, \,w^8}}, \qquad 
j' \, = \, \, {\frac { (256\,w^4-16\,w^2+1)^3}
{ (1+4\,w)^2 \, (1-4\,w)^2\, w^4}}.
\nonumber 
\end{eqnarray}
Similarly the {\em triplication  formula 
of the Klein's absolute invariant}
can also be written rationally:
\begin{eqnarray}
\label{tripli}
j \, = \, \, \, j_3(z) \, = \,\,  \, 
{\frac { (z+27)  \, (z+3)^3}{z}}, \quad \, \, 
j' \, = \, \,  
 {\frac { (z+27)  \, (z+243)^3}{ z^3}} \, = \, \, 
 j_3 \Bigl({{ 3^6} \over { z }} \Bigr),
\nonumber 
\end{eqnarray}
where  $\, z \, \, \rightarrow \, \, 3^{6}/z\, $ 
is, again, the {\em  Atkin involution}.
The elimination of the rational variable $\, z$ yields 
the classical modular curve
 which corresponds to the triplication
of the ratio of periods of the elliptic curves
 $\, j \, = \, \, j(\tau)\, \, \rightarrow \, \,
 j' \, = \, \, j(3\, \tau)$:
\begin{eqnarray}
\label{modtri}
&&j^4+j'^4 \,  -j^3\cdot j'^3  \, +2232\cdot j'^2\, j^2\cdot (j+j')
-1069956\cdot (j\, j'^3 +j'\, j^3) \nonumber   \\
&&\quad +2587918086\, j^2\, j'^2 \, 
+36864000\cdot (j+j')\cdot (j^2 \, +241433\, j\, j' \, +j'^2) \nonumber   \\
&&\quad +16777216000000\cdot (27\, j^2+27\, j'^2\,  -45946\, j\, j') 
\nonumber \\ 
&&\quad +1855425871872000000000 \cdot (j+j')\, = \, \, 0.  
\end{eqnarray}
The genus zero  classical  modular curve (\ref{modtri}) is, of course, 
symmetric by  $\, j \leftrightarrow \, j'$.  
The rational variable $\, z$ 
can simply be expressed
in term of the {\em Dedekind eta function}~\cite{McKay,Martin}:
\begin{eqnarray}
\label{dedek}
z \, = \, \, \, 3^s \cdot 
\Bigl( {{ \eta(3\, \tau) } \over {\eta(\, \tau) }} \Bigr)^{2s},
 \qquad s\, =\, 6.
\end{eqnarray}

We will not write (though it is straightforward) 
the  classical modular curve (\ref{sixmodu})
 which corresponds to  $\, j \, = \, \, j(\tau)\, \, 
\rightarrow \, \, j' \, = \, \, j(6\, \tau)$.
 This  classical modular curve
is again a genus zero curve, corresponding to a polynomial relation
symmetric by $\, j \, \, \leftrightarrow \, \, j'$ 
with integer coefficients
and it can simply be obtained from the elimination of
 $\, z$ in the rational parameterization (\ref{j6})
with, again, an Atkin involution 
$\, z \, \rightarrow \, \,  2^3 \cdot 3^2/z$.

Note that the rational functions occurring 
in the rational parametrization of all these 
genus zero classical modular curves
are related together by some rational change of variables:
\begin{eqnarray}
\label{relationcompo}
 j_6(z)   \, = \, \, \, 
 j_2\Bigl( {\frac {z \cdot (z+8)^{3}}{z+9}}  \Bigr) 
 \, = \, \, \, 
 j_3\Bigl(  {\frac {z \cdot (z+9)^2}{z+8}} \Bigr). 
\end{eqnarray}

The integers $\, N$ such that the modular curves, 
$\, P_N(j(\tau), \, j(N\, \tau))\, = \, 0$, 
are genus zero is a highly selected set 
of integers corresponding to
the Monstruous Moonshine phenomenon~\cite{moon2,moon}.

\end{document}